\documentclass[prd,nofootinbib,reprint]{revtex4-1}
\usepackage[hyperindex,breaklinks]{hyperref}
\usepackage[letterpaper,hmargin=1in,vmargin=1in]{geometry}

\usepackage{graphicx,epstopdf,amsmath,amsfonts,amssymb,color,bm}

\def\be{\begin{equation}}
\def\ee{\end{equation}}
\def\ba{\begin{eqnarray}}
\def\ea{\end{eqnarray}}
\def\ge{\mathrel{\raise.3ex\hbox{$>$\kern-.75em\lower1ex\hbox{$\sim$}}}}
\def\la{\mathrel{\raise.3ex\hbox{$<$\kern-.75em\lower1ex\hbox{$\sim$}}}}

\def\simgt{\mathrel{\raise.3ex\hbox{$>$\kern-.75em\lower1ex\hbox{$\sim$}}}}
\def\simlt{\mathrel{\raise.3ex\hbox{$<$\kern-.75em\lower1ex\hbox{$\sim$}}}}

\newcommand{\nc}{\newcommand}

\nc{\gone}{\bar g_{\pi NN}^{(1)}}
\nc{\gzero}{\bar g_{\pi NN}^{(0)}}
\nc{\al}{\alpha}
\nc{\ga}{\gamma}
\nc{\de}{\delta}
\nc{\ep}{\epsilon}
\nc{\ze}{\zeta}
\nc{\et}{\eta}
\nc{\ka}{\kappa}
\nc{\rh}{\rho}
\nc{\si}{\sigma}
\nc{\ta}{\tau}
\nc{\up}{\upsilon}
\nc{\ph}{\phi}
\nc{\ch}{\chi}
\nc{\ps}{\psi}
\nc{\om}{\omega}
\nc{\Ga}{\Gamma}
\nc{\De}{\Delta}
\nc{\La}{\Lambda}
\nc{\Si}{\Sigma}
\nc{\Up}{\Upsilon}
\nc{\Ph}{\Phi}
\nc{\Ps}{\Psi}
\nc{\Om}{\Omega}
\nc{\ptl}{\partial}
\nc{\del}{\nabla}
\nc{\ov}{\overline}
\nc{\newcaption}[1]{\centerline{\parbox{15cm}{\caption{#1}}}}
\nc{\us}{U(1)$_S$}

\def\beq{\begin{equation}}
\def\eeq{\end{equation}}
\def\bmat{\begin{displaymath}}
\def\emat{\end{displaymath}}
\def\bear{\begin{eqnarray}}
\def\eear{\end{eqnarray}}
\def\ba{\begin{eqnarray}}
\def\ea{\end{eqnarray}}
\def\bery{\begin{array}}
\def\ery{\end{array}}
\def\bit{\begin{itemize}}
\def\eit{\end{itemize}}
\def\ben{\begin{enumerate}}
\def\een{\end{enumerate}}
\def\btab{\begin{tabular}}
\def\etab{\end{tabular}}
\def\btbl{\begin{table}}
\def\etbl{\end{table}}
\def\bfig{\begin{figure}[htb]}
\def\efig{\end{figure}}
\def\bpic{\begin{picture}}
\def\epic{\end{picture}}
\def\lsim{\mathrel{\rlap{\lower4pt\hbox{\hskip1pt$\sim$}}
    \raise1pt\hbox{$<$}}}                
\def\gsim{\mathrel{\rlap{\lower4pt\hbox{\hskip1pt$\sim$}}
    \raise1pt\hbox{$>$}}}                

\def\ga{\mathrel{\raise.3ex\hbox{$>$\kern-.75em\lower1ex\hbox{$\sim$}}}}
\def\la{\mathrel{\raise.3ex\hbox{$<$\kern-.75em\lower1ex\hbox{$\sim$}}}}
\def\gappeq{\mathrel{\rlap {\raise.5ex\hbox{$>$}}
{\lower.5ex\hbox{$\sim$}}}}
\def\lappeq{\mathrel{\rlap{\raise.5ex\hbox{$<$}}
{\lower.5ex\hbox{$\sim$}}}}

\def\gyr{{\rm \, G\kern-0.125em yr}}
\def\mev{{\rm \, Me\kern-0.125em V}}
\def\gev{{\rm \, Ge\kern-0.125em V}}
\def\tev{{\rm \, Te\kern-0.125em V}}

\def\slash#1{\rlap{\hbox{$\mskip 1 mu /$}}#1}%

\begin{document}

\title{Signatures of sub-GeV dark matter beams at neutrino experiments}

\author{Patrick deNiverville}
\email{pgdeniv@uvic.ca}
\affiliation{Department of Physics and Astronomy, University of Victoria, 
Victoria, BC V8P 5C2, Canada}

\author{David McKeen}
\email{mckeen@uvic.ca}
\affiliation{Department of Physics and Astronomy, University of Victoria, 
Victoria, BC V8P 5C2, Canada}

\author{Adam Ritz}
\email{aritz@uvic.ca}
\affiliation{Department of Physics and Astronomy, University of Victoria, 
Victoria, BC V8P 5C2, Canada}

\date{\today}
\begin{abstract}
\noindent

We study the high-luminosity fixed-target neutrino experiments at  MiniBooNE, MINOS and T2K and analyze their sensitivity to light stable states, focusing on MeV--GeV scale dark matter. Thermal relic dark matter scenarios in the sub-GeV mass range require the presence of light mediators, whose coupling to the Standard Model facilitates annihilation in the early universe and allows for the correct thermal relic abundance. The mediators in turn provide a production channel for dark matter at colliders or fixed targets, and as a consequence the neutrino beams generated at fixed targets may contain an additional beam of light dark matter. The signatures of this beam include elastic scattering off electrons or nucleons in the (near-)detector, which closely mimics the neutral current scattering of neutrinos. We determine the event rate at modern fixed target facilities and the ensuing sensitivity to sub-GeV dark matter.

\end{abstract}
\maketitle

\section{Introduction}

The existing gravitational evidence for dark matter provides limited information about its non-gravitational interactions, and many candidates are sufficiently non-relativistic and weakly interacting. The paradigm of a weak-scale thermal relic has the virtue of simplicity, with an abundance fixed without detailed knowledge of early-universe physics. However, direct detection experiments now impose stringent constraints on dark matter with a weak-scale mass; for example, spin-independent cross sections on nucleons must be at or below $10^{-45}$~cm$^2$. With this sensitivity now crossing the Higgs-mediation threshold, the minimal weakly-interacting massive particle (WIMP) paradigm may need generalization to allow new interaction channels, beyond the electroweak sector of the Standard Model (SM). This would position dark matter as part of a more complex hidden sector containing additional light states. The required relic density could then be achieved without either weak-scale 
interactions or a weak-scale mass \cite{Boehm,*light-chi,*Neil,*HZ,*AFSW,*PR,Fayet,*Fayet2,*Fayet3,BF,PRV}. 

This viewpoint has some interesting implications when one looks at the existing limits on direct WIMP scattering. The sensitivity of direct-detection experiments tends to fall rather sharply for masses below a few GeV, due to the recoil energy detection threshold. The GeV mass scale also happens to coincide with the Lee-Weinberg bound \cite{LW}, below which a thermal relic needs non-SM annihilation channels through light states to ensure the correct relic abundance. In combination, these observations naturally lead us to explore the use of new experimental tools to probe the sub-GeV mass range for thermal relic dark matter. 
The presence of light mediators coupled to the SM opens up the possibility of producing these
states directly in accelerators or fixed target facilities. This `dark force' phenomenology has been the focus of considerable interest in recent years. For example, a number of search strategies are based on the production of a GeV-scale vector mediator, with its subsequent decay to lepton pairs 
\cite{Fayet,*Fayet2,*Fayet3,Drees,tests,best,*others,*others2,*others3,slac,*Reece,BPR,bpr99c}. However, these search strategies are limited if, instead, the mediator is not the lightest hidden sector state and decays predominantly into the hidden sector, e.g. to dark matter. In this case, the scattering of those light states in a detector spatially separated from the production point represents perhaps the most efficient search strategy. Moreover, owing to the potentially large production rate, and the existence of large volume (near-)detectors, proton fixed-target facilities focusing on neutrino physics appear to be an ideal means for exploring these scenarios. 

In this paper, we analyze the sensitivity of neutrino facilities to a boosted light dark matter beam produced via the generation and subsequent decay of GeV-scale mediators. This extends our earlier analysis of MeV-scale dark matter  \cite{den2011,bpr99c} to the full sub-GeV range. We will find that high-luminosity
experiments such as MiniBooNE, MINOS and T2K have significant sensitivity to neutral current-like scattering of sub-GeV dark matter off nuclei in the (near-)detector. Although there is a long history of searches for exotics using fixed target facilities (see e.g. \cite{axions_exp,*axions_exp2,*neutrino_exp,*neutrino_exp2,*gluino_exp,*unstable, *losecco,Fayet,*Fayet2,*Fayet3,Drees,tests,best,*others,*others2,*others3,slac,*Reece,BPR,bpr99c}), neutrino experiments have the advantage that the large detector volume is sensitive to scattering signatures in addition to the products of SM decays. Since the recoil energy of sub-GeV halo dark matter  is generally
below threshold for underground direct detection experiments, and search channels at high energy colliders are less sensitive in the case of light mediators,
high-luminosity fixed-target experiments can play a complementary role in direct searches for dark matter.

In order to be as model-independent as possible, we parametrize the mediator interactions via the lowest dimension operators (portals) for a SM-neutral
hidden sector, ${\cal L}_{\rm int} = \sum {\cal O}_{\rm SM} {\cal O}_{HS}$, 
where ${\cal O}$ denotes SM and hidden sector (HS) operators. 
For light dark matter, fixed-target facilities have an advantage if the mediator can be produced on-shell, so we focus on the renormalizable
{\it vector}~\cite{portal1,*portal2} and {\it scalar} SM portals~\cite{pw,*higgsportal,*higgsportal2,*higgsportal3,*higgsportal4,*DMportal,*DMportal2,*DMportal3,*DMportal4}:
\be
\label{portals}
 {\cal L}_{\rm int} = {\cal L}_{\rm hid}(X,\ch) + \left\{ \begin{array}{cc} \kappa F^Y_{\mu\nu} V^{\mu\nu}, & \mbox{vector portal} \\  A S H^\dagger H, & \mbox{scalar portal} \end{array} \right.,
\ee
where $F^Y_{\mu\nu}$ and $H$ are the hypercharge field strength and the Higgs doublet, while ${\cal L}_{\rm hid}$ provides hidden sector couplings between
the mediator field $X=V^\mu$ or $S$ and the light dark matter candidate $\ch$.  We will limit attention to the kinematic regime
\be
 m_X > 2 m_\ch \sim {\cal O}({\rm MeV - GeV}),
\ee
so that with small portal couplings to the SM, the mediators predominantly decay into the hidden sector, Br$(X\rightarrow \ch\ch) \sim 1$. 
 
The rest of this paper is organized as follows. In Sec.~\ref{sec:DMProd}, we describe our model for production of the dark matter beam at MINOS, T2K and MiniBooNE using
both vector and scalar portals. In Sec.~\ref{sec:Con}, we discuss a number of existing constraints on sub-GeV dark matter, coupled to the SM via these portals, detail
the annihilation and scattering rates, and determine viable models which can be probed using neutrino facilities. In Sec.~\ref{sec:Sens}, we focus on the most viable
dark matter scenario, with scalar dark matter coupled via the vector portal, and analyze the sensitivity to the ensuing dark matter beam at MINOS, T2K and MiniBooNE.
We conclude in Sec.~\ref{sec:Conc}.

\section{Production of the dark matter beam}
\label{sec:DMProd}

\subsection{DM interactions}

The viability of thermal relic dark matter with a mass in the MeV--GeV range, well below the Lee-Weinberg bound,  seemingly rests on the presence of a light hidden sector with states that can mediate annihilation \cite{BF,Fayet,*Fayet2,*Fayet3,PRV}. Moreover, various phenomenological constraints \cite{PRV} suggest that the most viable scenarios are those in which the hidden sector is uncharged under Standard Model symmetries. This naturally leads us to the portal interactions (\ref{portals}) as the primary means of probing these sectors at low energies.

To keep our analysis as general as possible, we will consider both the vector and Higgs portals for production of the dark matter beam
in this section. These light mediators are necessary to allow for a viable annihilation channel in the early universe, but we will be agnostic about the 
precise choice of model. This will allow us to analyze the raw sensitivity of neutrino facilities to production of these light states, and we will turn to the 
model-dependent constraints on viable light dark matter scenarios in the following sections.

To fix the interactions, we use the simplest realizations for the vector and scalar portals, and moreover we will only need their low energy manifestations.
For the vector portal coupling, $F^Y_{\mu\nu} V^{\mu\nu}$, we have
\be
 {\cal L}_{V} = V_\mu \left( e\kappa J^\mu_{\rm em} + e' J^\mu_\ch\right) + {\cal L}_{\rm kin}(V,\ch) + \cdots
\ee
where we have used $\ptl_\mu  F^{\mu\nu} = eJ^\nu_{\rm em}$ in terms of the electromagnetic current $J^\mu_{\rm em} = \bar{q}\gamma^\mu q + \cdots$, 
while $J^\mu_\ch$ is the corresponding U(1) current for scalar (or Dirac fermion) dark matter, with gauge coupling $e'$,
\be
 J^\mu_\ch = \left\{ \begin{array}{cc} i\ch^\dagger \overleftrightarrow{\ptl^\mu} \ch + {\cal O}(V^\mu), & {\rm scalar} \\
       i \bar\ch \gamma^\mu \ch, & {\rm fermion} \end{array} \right..
\ee
 ${\cal L}_{\rm kin}(V,\ch)$ contains canonical kinetic and mass terms for $V$ and $\ch$, and 
higher order potential terms have 
not been written explicitly. 

For the trilinear scalar portal  coupling, $SH^\dagger H$, we have
\be
 {\cal L}_S = S \left( \theta J^m_{\rm EWSB}+ \beta J_\ch^m \right) + {\cal L}_{\rm kin}(S,\ch) + \cdots
\ee
where we have integrated out the SM Higgs, which induces a coupling $\theta \sim Av/m_h^2$ between $S$ and the SM fermions via 
$J^m_{\rm EWSB} = m_q\bar{q} q/v + \cdots$, while $J^m_{\ch}$  is an analogous mass current for scalar (or fermion) 
dark matter,
\be
 J^m_\ch = \left\{ \begin{array}{cc} m_\ch \ch^\dagger \ch, & {\rm scalar} \\
       i \bar\ch \ch, & {\rm fermion} \end{array} \right..
\ee
We have inserted a factor of $m_\ch$ in the scalar case, so that $\beta$ remains a dimensionless coupling.
As above,  ${\cal L}_{\rm kin}(V,\ch)$ contains the kinetic and mass terms for $S$ and scalar/fermion dark matter $\ch$.

We will refer to the mediator $V$ or $S$ generically as $X$, and the crucial kinematic assumption will be that $m_X > 2 m_\ch$, so that the
mediator can decay on-shell to dark matter. For small mixing via the portals, the hidden sector branching Br$(X\rightarrow \ch\ch) \sim 1$.

\subsection{Production mechanisms}

There are two viable production mechanisms for the mediator $X$ at proton fixed-target experiments:

\begin{itemize}
\item {\it Direct production:} This corresponds to hadron-level processes such as $pp(n) \rightarrow X^* \rightarrow \bar\ch \ch$ (or $\ch^\dagger\ch$) as shown in Figs.~1 and 2.
In practice, since $X$ can
decay to $\bar\ch\ch$, we will use the narrow width approximation so that $X$ is produced on-shell.  In this approximation, valid to ${\cal O}\left(e^{\prime 2},\beta^2\right)$, the cross section for the production of a DM pair can be written as
\begin{align}
&\sigma\left(pp(n)\to X^\ast\to\bar\ch\ch\right)
\nonumber
\\
&\quad\quad=\sigma\left(pp(n)\to X\right){\rm Br}\left(X\to\bar\chi\chi\right).
\end{align}

\begin{figure*}[t]
\centerline{\includegraphics[width=0.35\textwidth]{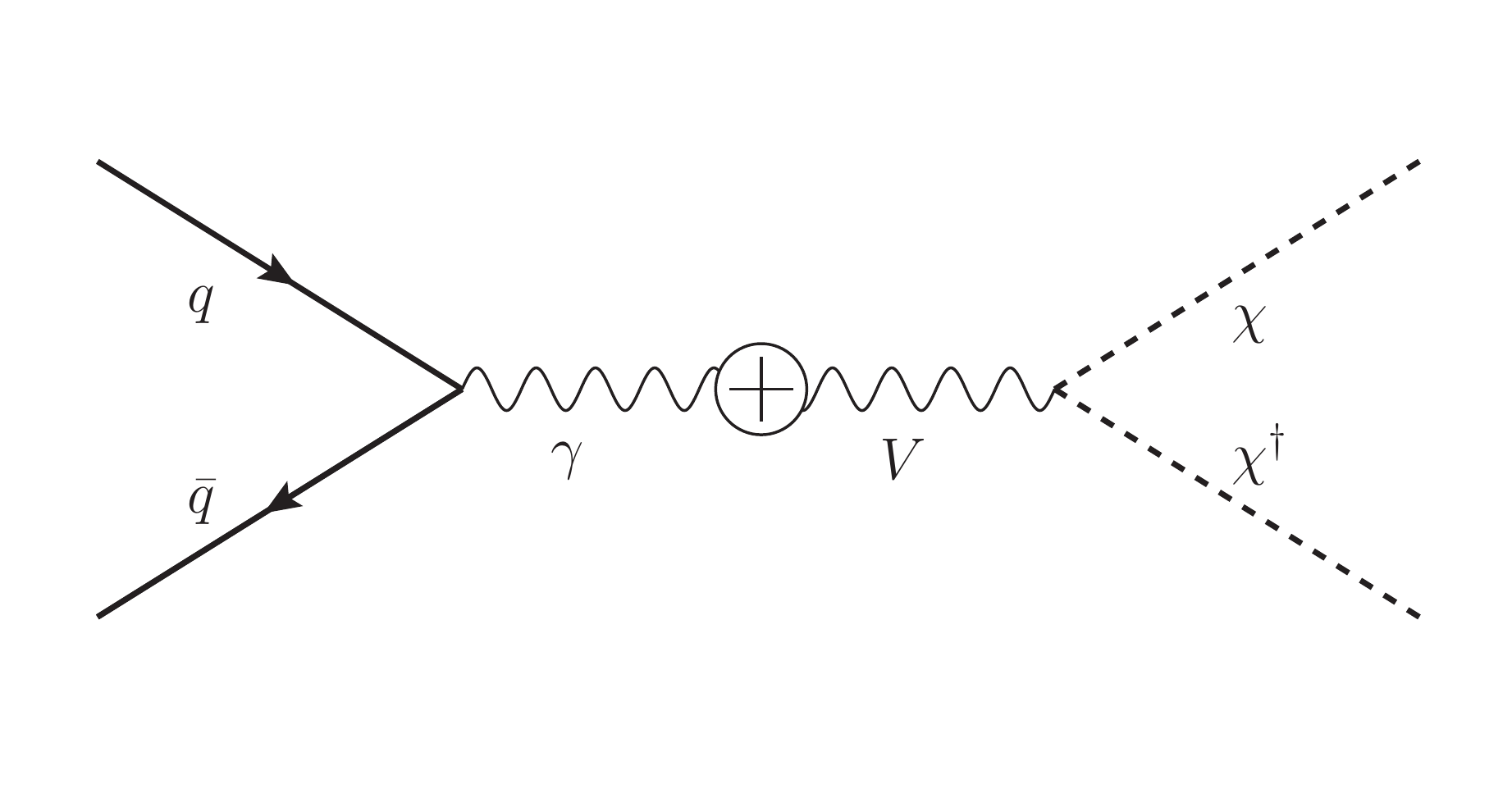}\includegraphics[width=0.35\textwidth]{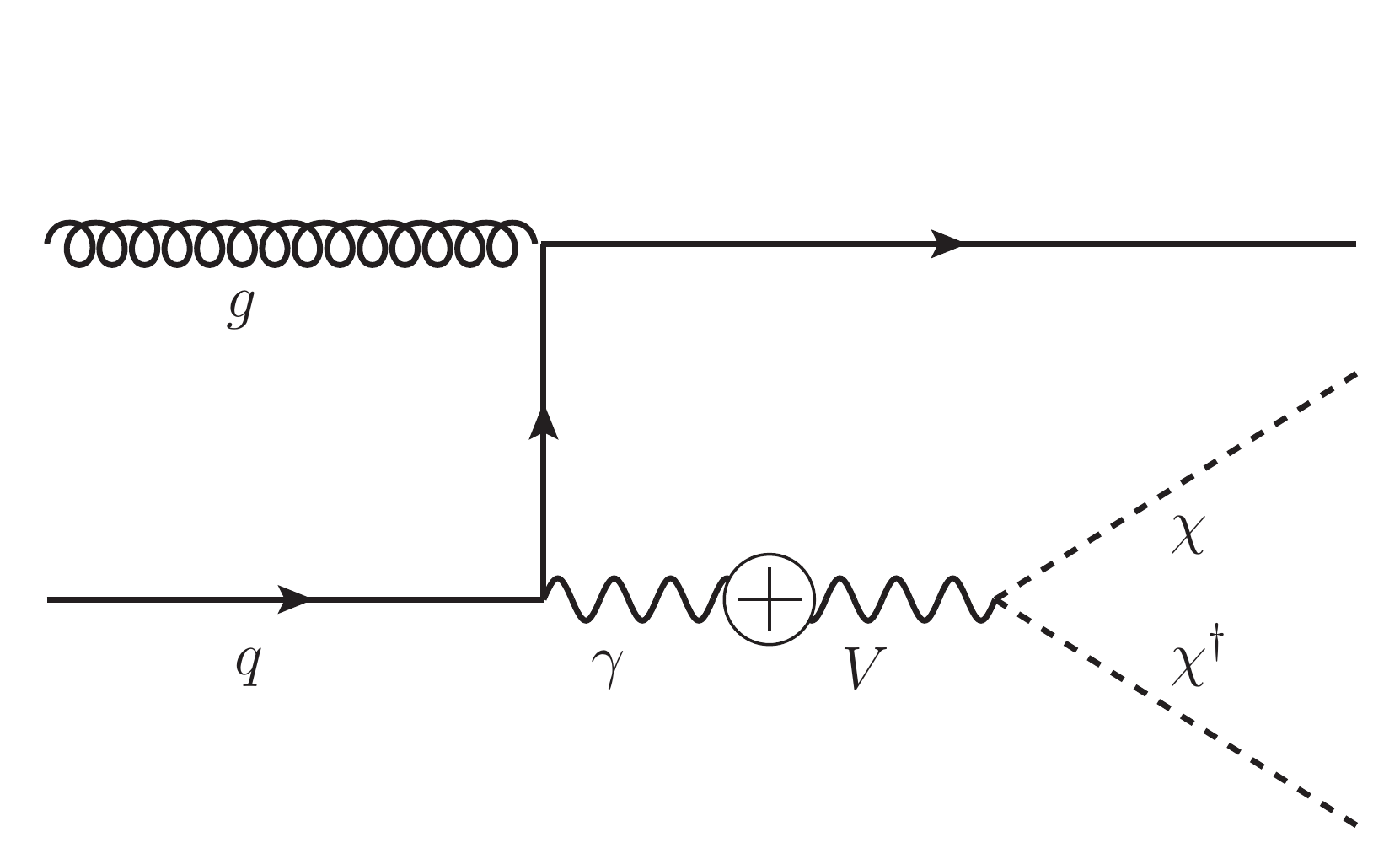}}
  \caption{\footnotesize Direct production of scalar dark matter via the vector portal. The leading-order process is shown on the left, which
  is helicity suppressed in the forward direction. The process on the right is higher order in $\al_s$, and also
  phase space suppressed, but has less helicity suppression in the forward direction.}
  \label{fig:Vprod}
\end{figure*} 

\begin{figure}[t]
\centerline{\includegraphics[width=0.35\textwidth]{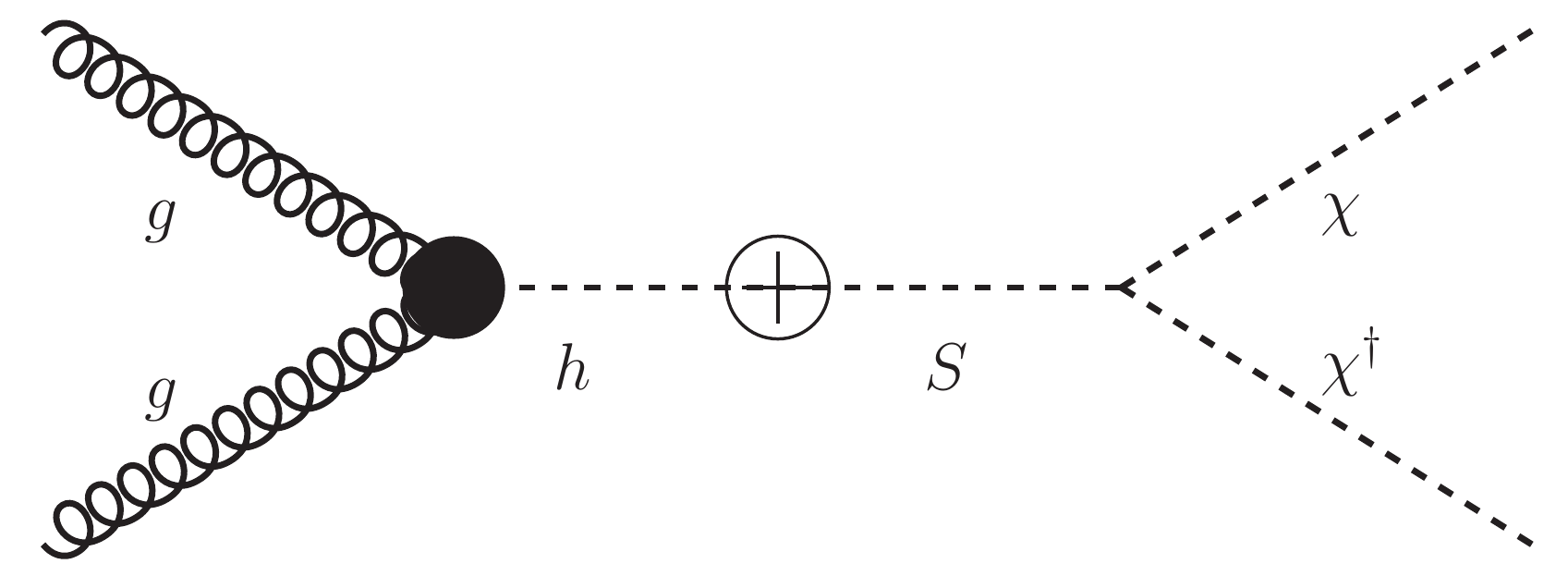}}
  \caption{\footnotesize Direct production of dark matter via the scalar portal. The solid gluon fusion $ggh$ vertex is generated at 1-loop. }
   \end{figure} 

The direct production cross section of a vector mediator is
\begin{widetext}
\begin{align}
\sigma\left(pp(n)\to V\right)&=\int_{\tau}^1dx\ \frac{d\sigma\left(pp(n)\to V\right)}{dx}
\nonumber
\\
&=\frac{4\pi^2\alpha\kappa^2}{m_V^2}\sum_q e_q^2\int_{\tau}^1 \frac{dx}{x}\ \tau\left[f_{q/p}\left(x\right)f_{\bar q/p(n)}\left(\frac{\tau}{x}\right)+f_{\bar q/p}\left(x\right)f_{q/p(n)}\left(\frac{\tau}{x}\right)\right],
\label{eq:sigmaV}
\end{align}
\end{widetext}
where $e_q$ is the charge of quark $q$ in units of the positron electric charge, $\tau=m_V^2/s$, and $\sqrt s$ is the hadron-level center-of-mass energy.  The parton distribution function (PDF) $f_{q/p(n)}\left(x\right)$ gives the probability of extracting the quark $q$ with momentum fraction $x$ from a proton (neutron) and similarly for $f_{\bar q/p(n)}\left(x\right)$.  We have omitted the scale, $Q$, at which the PDFs are evaluated.  To obtain estimates, we use CTEQ6.6 PDFs~\cite{cteq} and set $Q=m_V$; varying $Q$ in between $m_V/2$ and $2m_V$ resulted in an uncertainty in the production cross section of less than $\sim 30\%$ for $m_V>1~{\rm GeV}$ at T2K and MINOS beam energies.  Higher-order QCD corrections are large, introducing an error that can potentially be ${\cal O}(1)$.

\begin{figure}[t]
\centerline{\includegraphics[scale=.45]{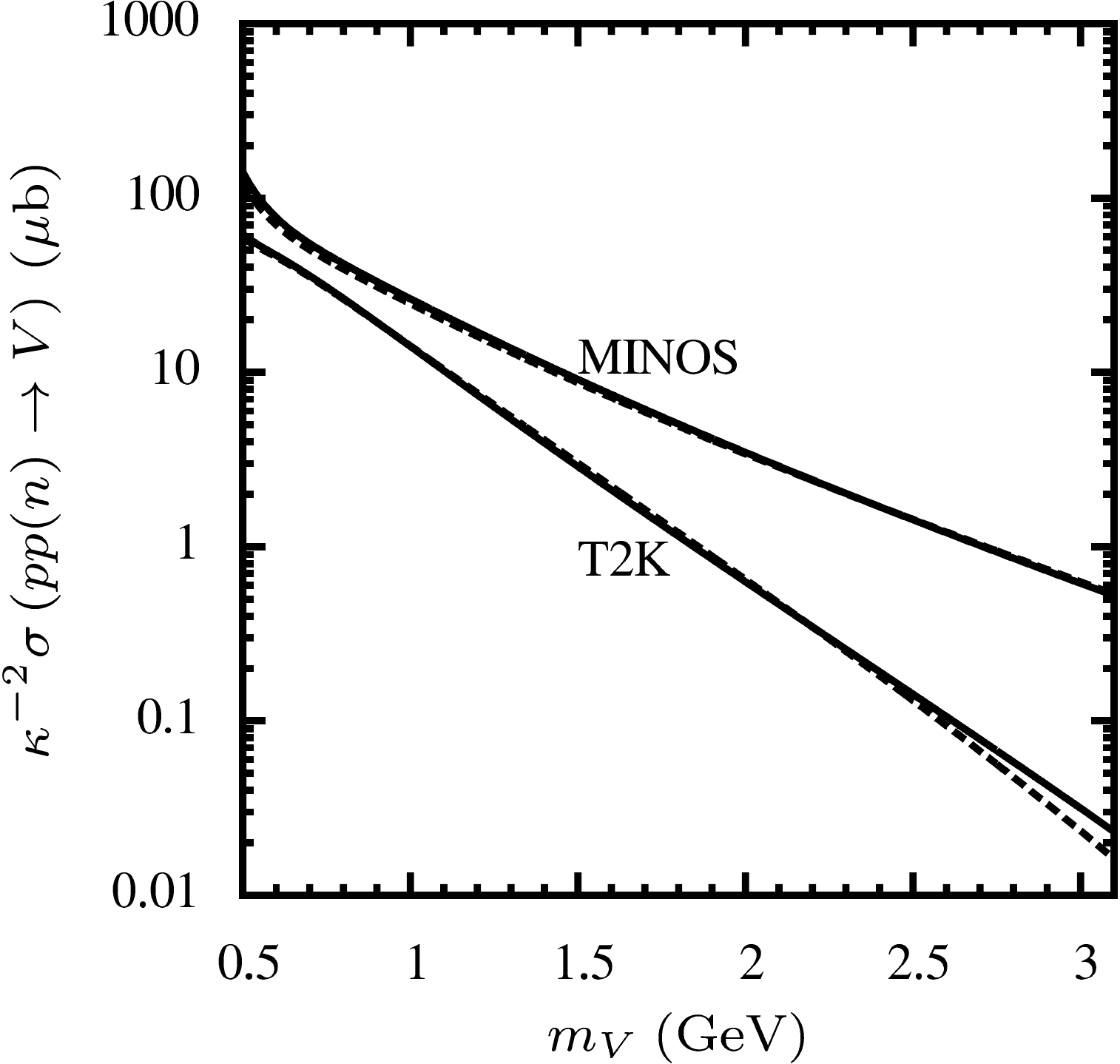}}
\caption{\footnotesize The total production cross section of a vector mediator at T2K and MINOS energies as a function of the mediator mass.  The solid and dashed curves indicate the cross sections for $pp\to V$ and $pn\to V$ respectively.  The PDF scale has been fixed to $Q=m_V$.}
\label{fig:xsec}
\end{figure}

The production cross section as a function of the DM lab frame energy, $E_\chi$, and the angle between its lab frame momentum and the beam direction, $\theta$, can be related to the differential cross section in Eq.~(\ref{eq:sigmaV}) through
\begin{align}
&\frac{d\sigma\left(pp(n)\to V\to\bar\chi\chi\right)}{dE_\chi d\cos\theta}=\left[\frac{\partial(x,\cos\hat\theta)}{\partial(E_\chi,\cos\theta)}\right]
\label{eq:com_to_lab}
\\
&\quad\times\frac{d\sigma\left(pp(n)\to V\right)}{dx}~{\rm Br}\left(V\to\bar\chi\chi\right)g\left(\cos\hat\theta\right),
\nonumber
\end{align}
where $\hat\theta$ is the angle between the momentum of $\chi$ and the beam in the $V$ rest frame and the quantity in square brackets is the Jacobian associated with this variable change.  The function $g$ describes the angular distribution of the DM in the $V$ rest frame.  For scalar DM produced through a vector mediator, this is
\begin{align}
g\left(\cos\hat\theta\right)=\frac{3}{4}\left(1-\cos^2\hat\theta\right).
\label{eq:sc_ang_dist}
\end{align}
If, instead, $\chi$ is a Dirac fermion, then
\begin{align}
g\left(\cos\hat\theta\right)=\frac{3}{8}\left(1+\cos^2\hat\theta\right).
\end{align}

We will find the distribution of $V$ momenta useful,
\begin{align}
f_V\left(p_V\right)&=\frac{1}{\sigma\left(pp(n)\to V\right)}\frac{d\sigma\left(pp(n)\to V\right)}{dp_V}
\\
&=\frac{1}{\sigma\left(pp(n)\to V\right)}\frac{dx}{dp_V}\frac{d\sigma\left(pp(n)\to V\right)}{dx},
\nonumber
\end{align}
with $p_V$ the momentum of $V$ in the lab frame which is related to $x$ through
\begin{align}
&p_V=\frac{\gamma p_Bm_T}{\sqrt s}
\\
&\quad\times\left[1+\beta\left(1+\frac{m_V^2s}{\left(x-\tau/x\right)^2p_B^2m_T^2}\right)^{1/2}\right],
\nonumber
\end{align}
where $m_T=m_{p,n}$ is the target mass, $p_B$ is the momentum of the beam, $\gamma\beta=p_B/\sqrt s$, and $\gamma=1/\sqrt{1-\beta^2}$.

\begin{figure*}[t]
\centerline{\includegraphics[scale=0.65]{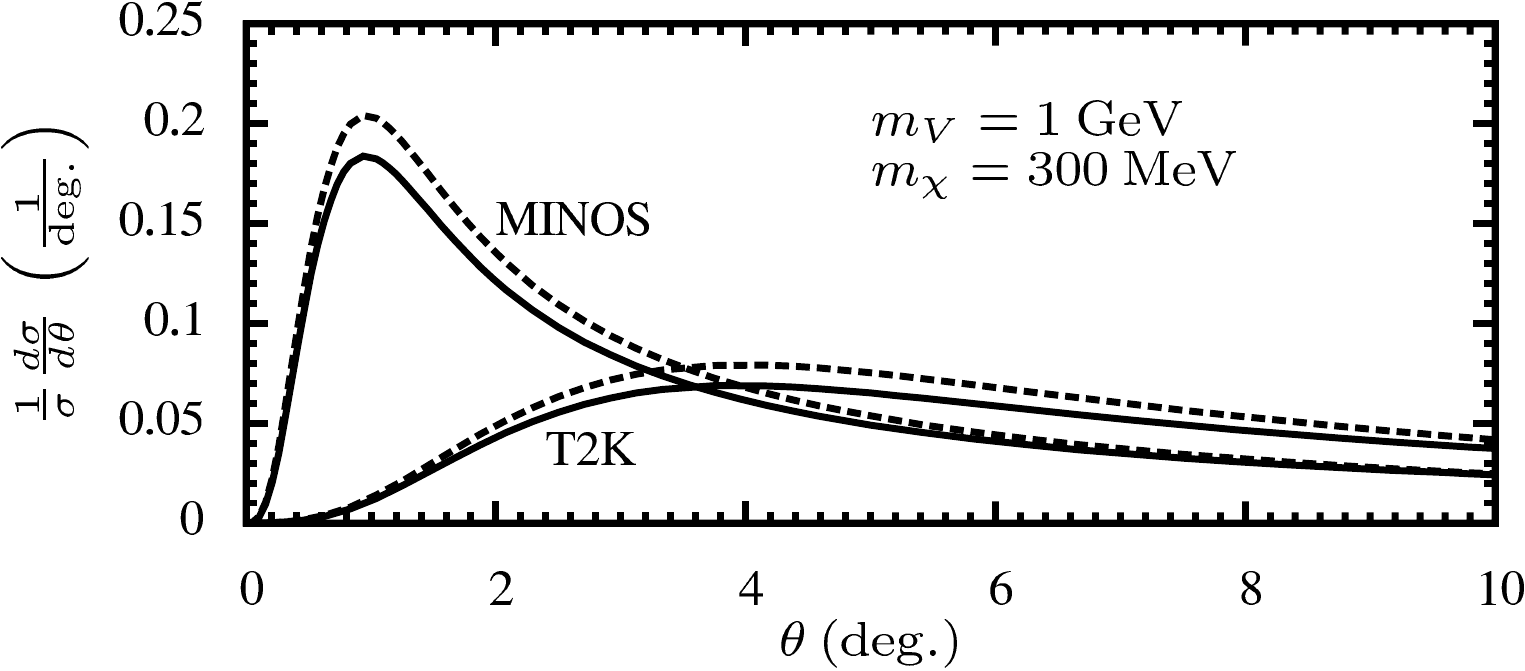}}
\centerline{\includegraphics[scale=0.55]{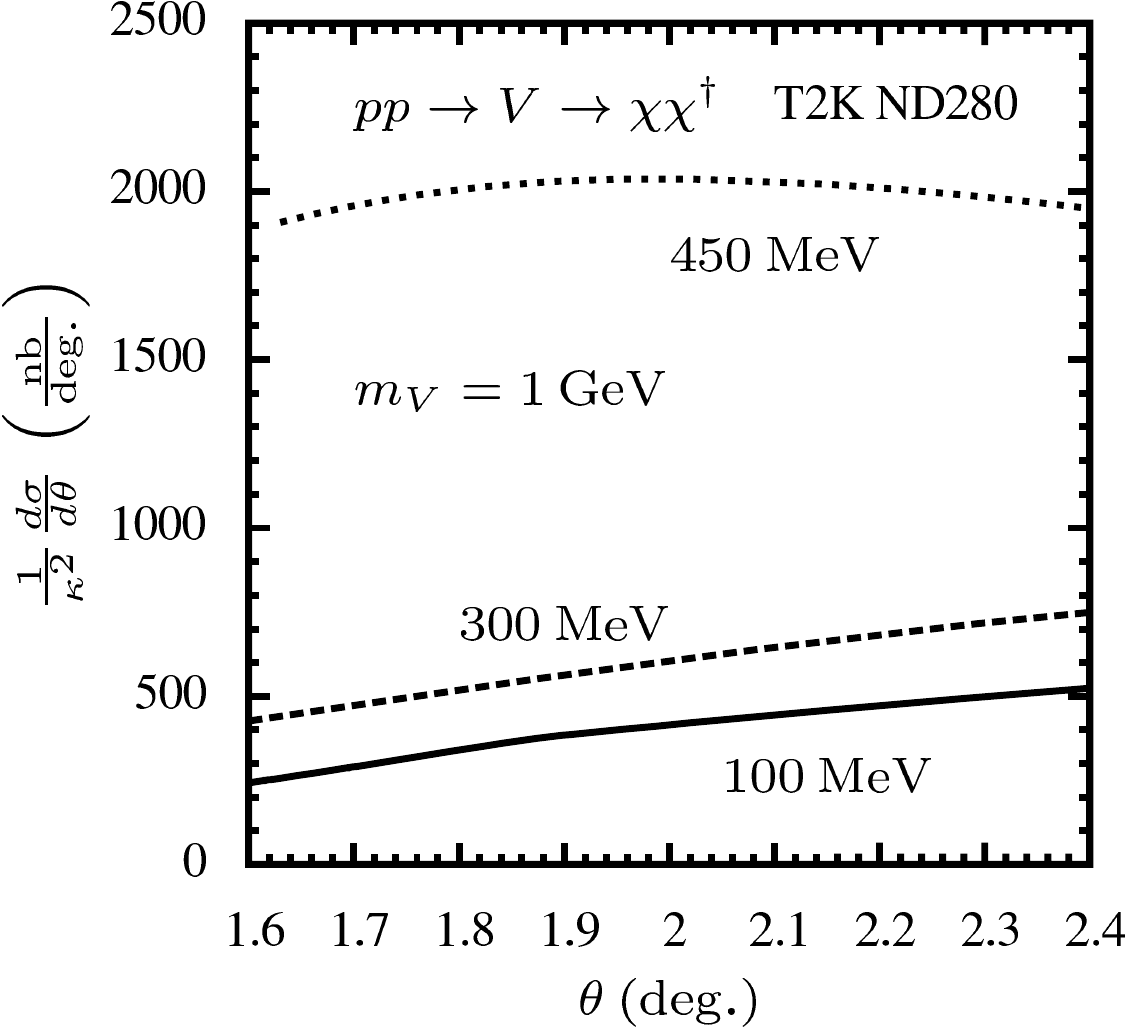}\hspace*{0.5cm}\includegraphics[scale=0.55]{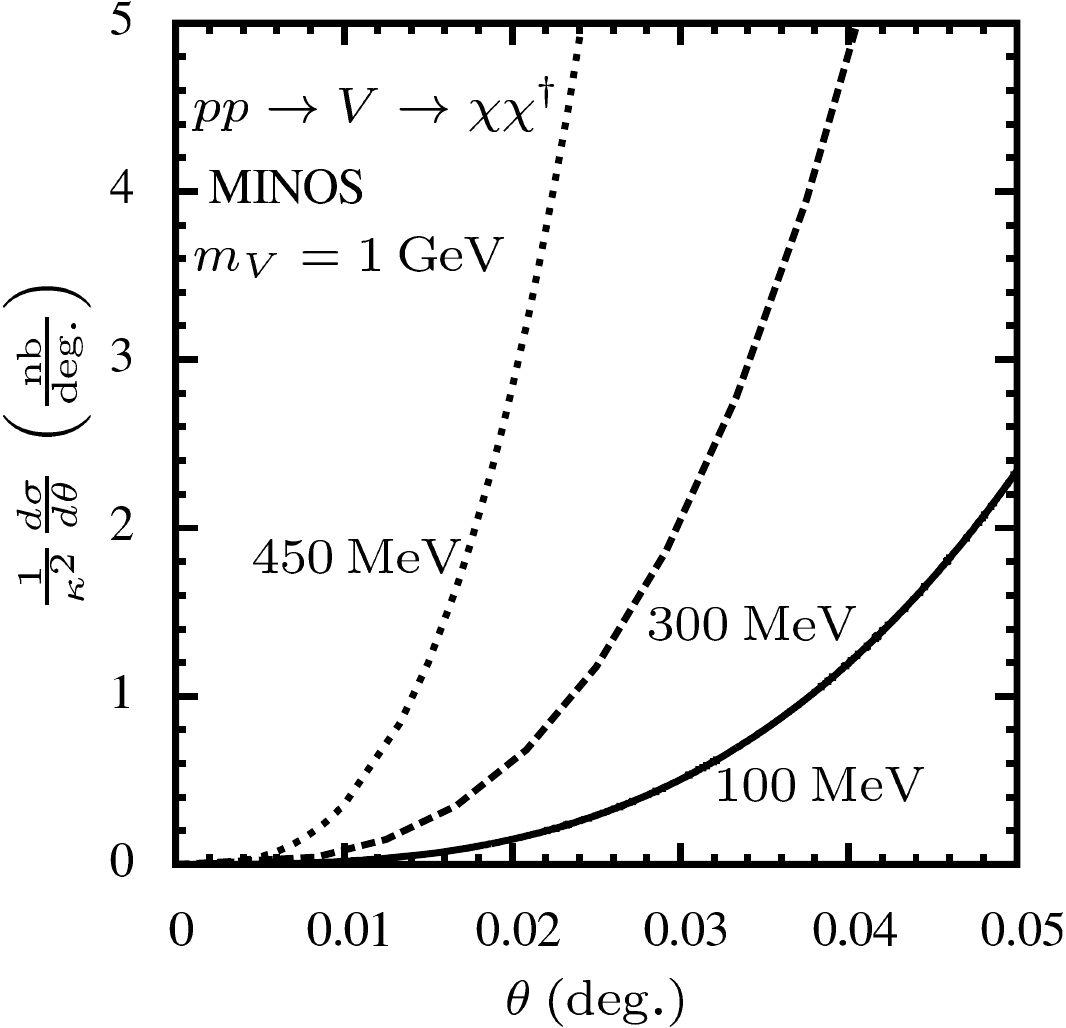}}
\caption{\footnotesize Top: production distributions of scalar DM as a function of lab frame angle with respect to the beam, normalized to unity, in the case of a vector mediator with $m_V=1~{\rm GeV}$.  The solid curves indicate $pp$ collisions and the dashed ones $pn$ collisions.  The set of curves that peak at $\theta\sim1^\circ$ corresponds to a $p$ beam with an energy equal to that of the MINOS experiment ($E_{\rm beam}=120~{\rm GeV}$, $\sqrt s=15~{\rm GeV}$) while the curves that peak at $\theta\sim4^\circ$ correspond to a $p$ beam with an energy equal to that of T2K ND280 ($E_{\rm beam}=30~{\rm GeV}$, $\sqrt s=7.6~{\rm GeV}$). Bottom left: production cross sections in the case of a vector mediator ($m_V=1~{\rm GeV}$) and scalar DM for $m_\chi=100,\ 300,\ 450~{\rm MeV}$ (solid, dashed, dotted) as functions of the DM angle with the beam in the lab frame in the case of $pp$ collisions at an energy corresponding to the T2K experiment.  The range of angles shown coincides with those covered by the off-axis ND280 near detector at T2K.  Bottom right: the same at MINOS beam energy.  The angles shown here are those that the MINOS near detector covers.}
\label{fig:ang}
\end{figure*}

\begin{figure*}[t]
\centerline{\includegraphics[scale=0.55]{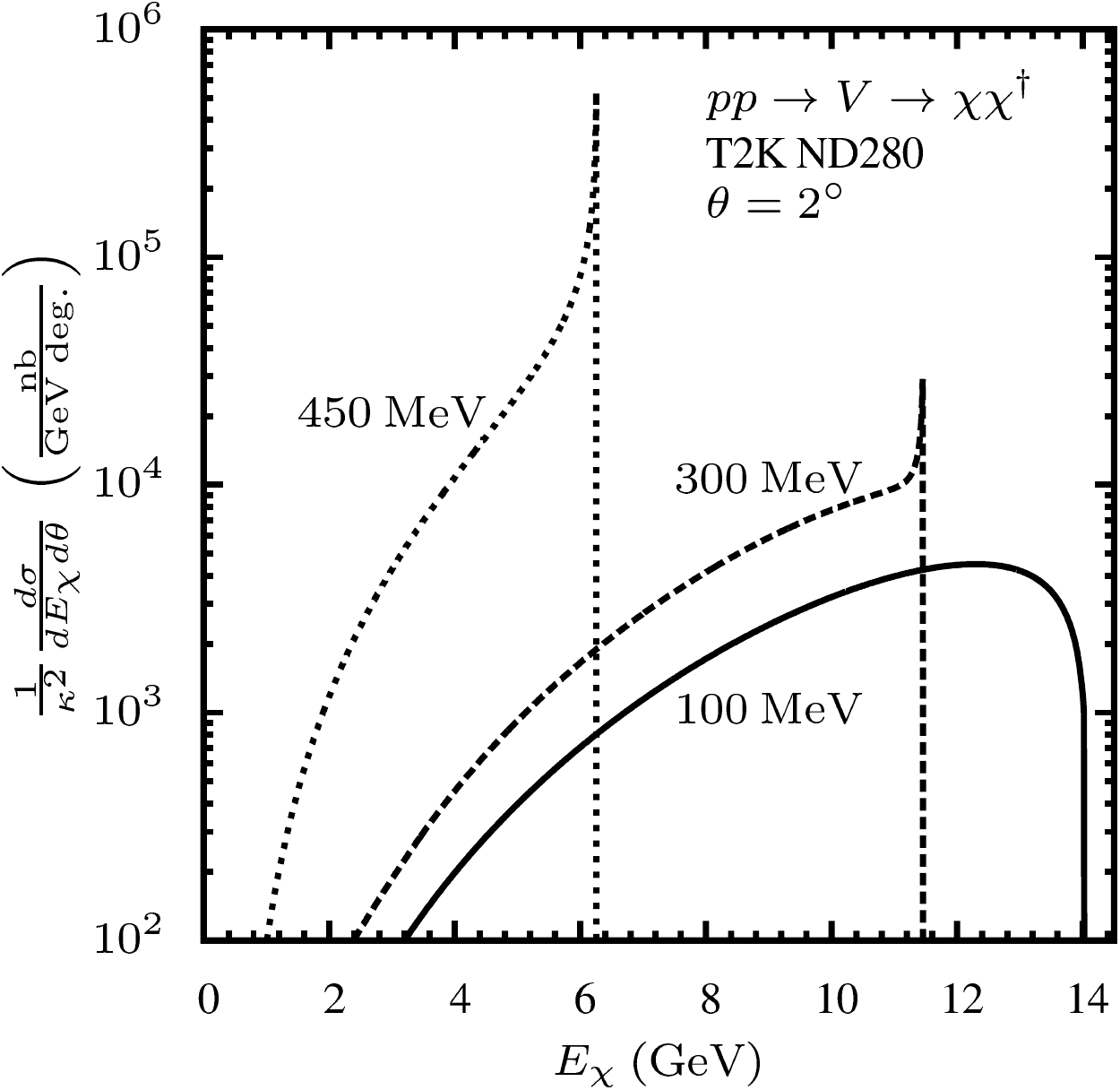}\hspace*{0.75cm}\includegraphics[scale=0.55]{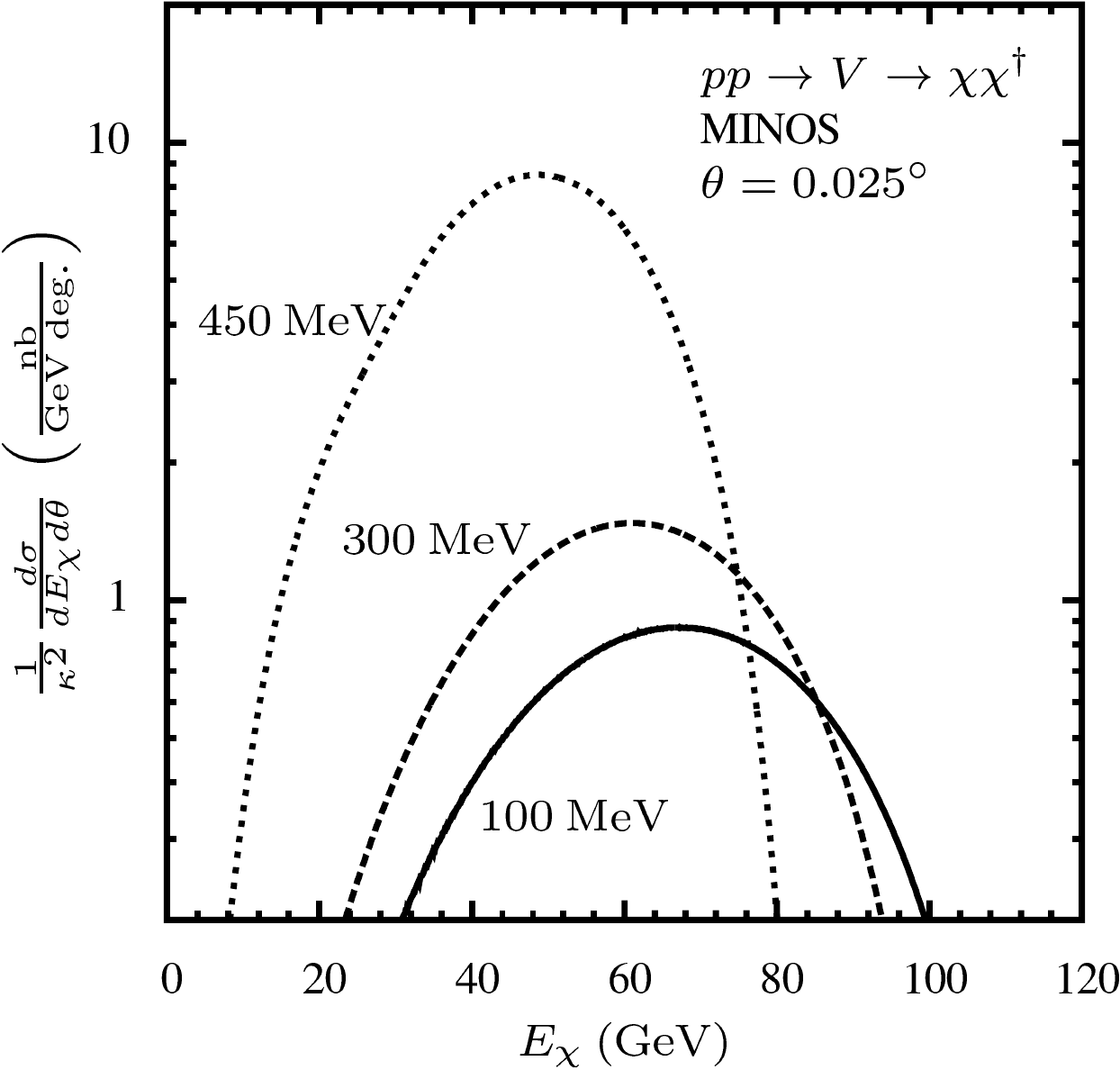}}
\caption{\footnotesize Left: $d\sigma\left(pp\to V\to\chi\chi^\dagger\right)/dE_\chi d\theta$ in the case of scalar $\chi$ and vector $V$ for $m_V=1~{\rm GeV}$ and $m_\chi=100,\ 300,\ 450~{\rm MeV}$ (solid, dashed, dotted) for $pp$ collisions at T2K ND280 at $\theta=2^\circ$.  The cusps at the kinematic limits for larger $m_\chi$ are the result of a degeneracy in the angle between $\chi$ and the beam direction in the lab frame, $\theta$, as a function of its value in the $V$ rest frame, $\hat\theta$, for relatively small DM velocities (e.g. in the limit that the DM is produced at threshold, $\theta=0$ for all $\hat\theta$).  Right: The same for $pp$ collisions at MINOS at $\theta=0.025^\circ$.}
\label{fig:dsigdE}
\end{figure*}

For illustration, in Figs.~\ref{fig:xsec}--\ref{fig:dsigdE} we present the resulting direct production distributions for a vector mediator that subsequently decays to scalar DM
at the T2K and MINOS experiments,  where $E_{\rm beam}=30,~120~{\rm GeV}$ ($\sqrt s\simeq7.6,~15.1~{\rm GeV}$), respectively; see Sec.~\ref{sec:Sens} for further details of these experiments. Fig.~\ref{fig:xsec} shows the total production cross section for $pp$ and $pn$ collisions at T2K and MINOS as a function of the vector mediator mass.
After integrating over energy, the angular distribution of scalar DM is shown in the top of Fig.~\ref{fig:ang} at T2K ND280 and MINOS in the case 
that $m_V=1~{\rm GeV}$ and $m_\chi=300~{\rm MeV}$.  We focus on the off-axis ND280 detector at T2K, to contrast with the on-axis detector at MINOS 
in sampling the angular production distribution. However, comparing ND280 to the on-axis INGRID detector at T2K would provide a similar contrast.
In the bottom left of Fig.~\ref{fig:ang}, we zoom in on the relevant angular region for the off-axis T2K ND280 near detector and show the scalar DM angular distribution for $m_V=1~{\rm GeV}$ and several DM masses produced in $pp$ collisions.  We do the same in the range of angles around the MINOS near detector in the bottom right of Fig.~\ref{fig:ang}.  As the mass of the DM is increased, it is produced in the more forward direction since its velocity in the $V$ rest frame decreases.  However, the angular distribution of scalar DM produced via a vector mediator, Eq.~(\ref{eq:sc_ang_dist}), suppresses the production of DM along the beam direction itself.  Thus, despite the smaller cross section for the production of vector mediators as a result of the lower energy of its beam, a larger number of DM particles may pass through the off-axis T2K ND280 near detector than the on-axis MINOS near detector.  This suppression along the beam axis is lessened somewhat when considering higher-order production mechanisms like the diagram on the right of Fig.~\ref{fig:Vprod}, which we do not include in this study.

We show the energy distribution of scalar DM for $m_V=1~{\rm GeV}$ and a range of $m_\chi$ in $pp$ collisions for T2K at $\theta=2^\circ$ and for MINOS at $\theta=0.025^\circ$ in Fig.~\ref{fig:dsigdE}.

For a scalar mediator, the leading-order direct production cross section is
\begin{align}
\sigma\left(pp(n)\to S\right)&=\frac{\alpha_s^2G_F N^2\theta^2}{288\sqrt{2}\pi}
\label{eq:sigmaS}
\\
&\quad\times\sum_q \int_{\tau}^1 \frac{dx}{x}\ \tau f_{g}\left(x\right)f_{g}\left(\frac{\tau}{x}\right).
\nonumber
\end{align}
Here, $\tau=m_S^2/s$ and the PDF $f_{g}\left(x\right)$ is the probability of finding a gluon with momentum fraction $x$ in a nucleon.  Up to threshold effect corrections, 
$N$ counts the number of quarks with a mass greater than $\sim 0.2 m_S$ \cite{ggmn}.  

The DM distributions in the lab frame can be related to the differential production cross section in the same way as in the vector mediator case in Eq.~(\ref{eq:com_to_lab}).  In the $S$ rest frame, the DM is simply produced isotropically,
\begin{align}
g(\cos\hat\theta)=\frac 12.
\end{align}

Because of the weak scale and loop factor suppressions in Eq.~(\ref{eq:sigmaS}), scalar mediator production is extremely small compared to the that of a vector at GeV scales.  Thus, current neutrino experiments are much less sensitive to DM scenarios involving a GeV-scale spin-0 mediator than they are to a spin-1 mediator.  Along with other factors to be discussed in the next section, this will lead us to focus only on the direct production of vector mediators.

\item {\it Indirect production:} This corresponds to production of $X$ via the decay of hadronic states (generically denoted $\ph$) produced in the primary $pp$ and $pn$ interactions,
\begin{align}
  p +p(n) \rightarrow & \;\ph + \cdots \nonumber\\
   &\;\;\mbox{\raisebox{6pt}{$\searrow$}} \;X + \cdots \nonumber\\
    & \qquad \;\mbox{\raisebox{4pt}{$\searrow$}}\; \ch\ch
\end{align}
Depending on the beam energy and form of the target, the relevant decay lengths ensure that this entire sequence of events will occur either inside the target itself or in the subsequent decay volume.

\begin{figure*}[t]
\centerline{\includegraphics[scale=0.55]{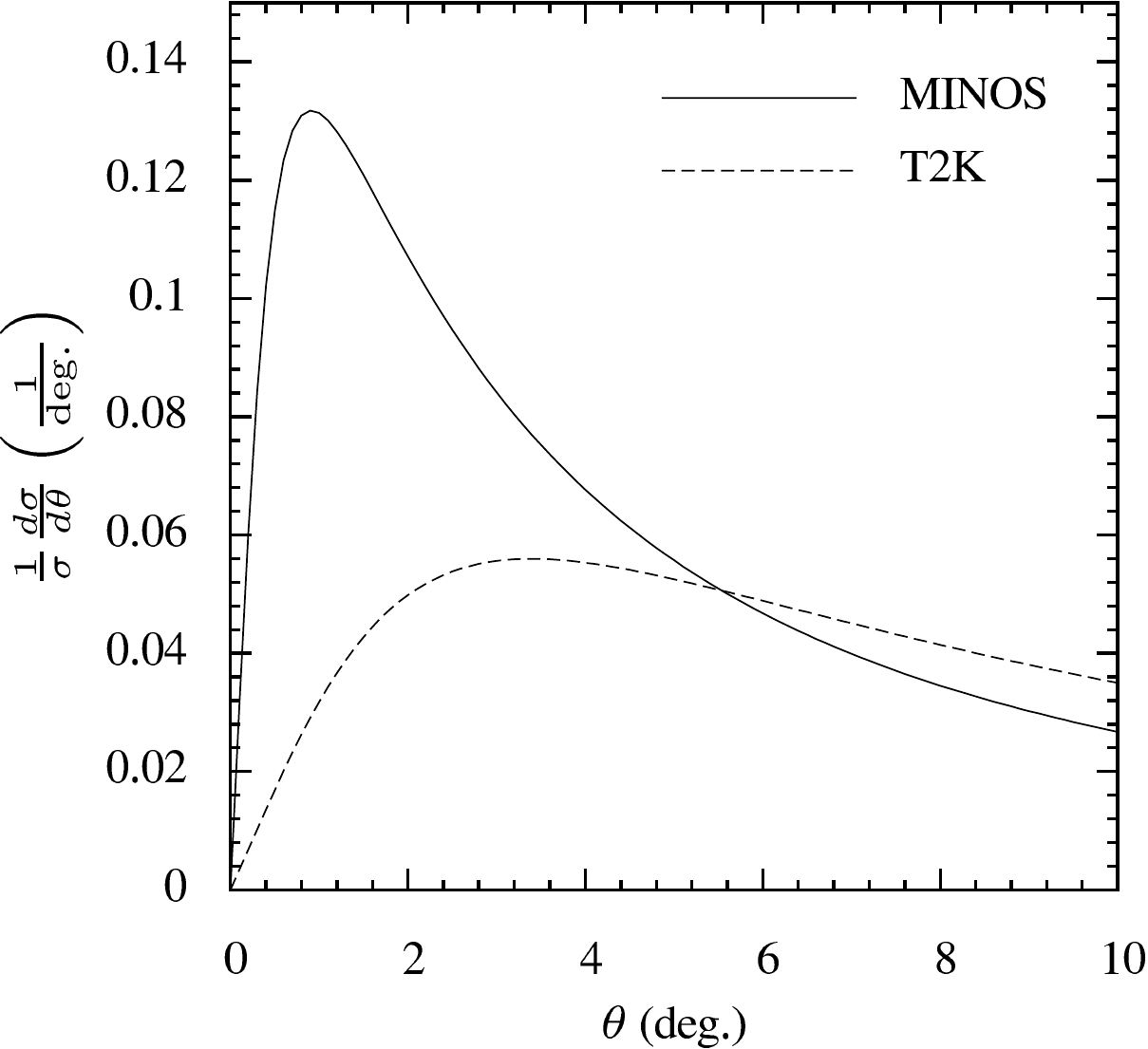}\hspace*{0.75cm}\includegraphics[scale=0.55]{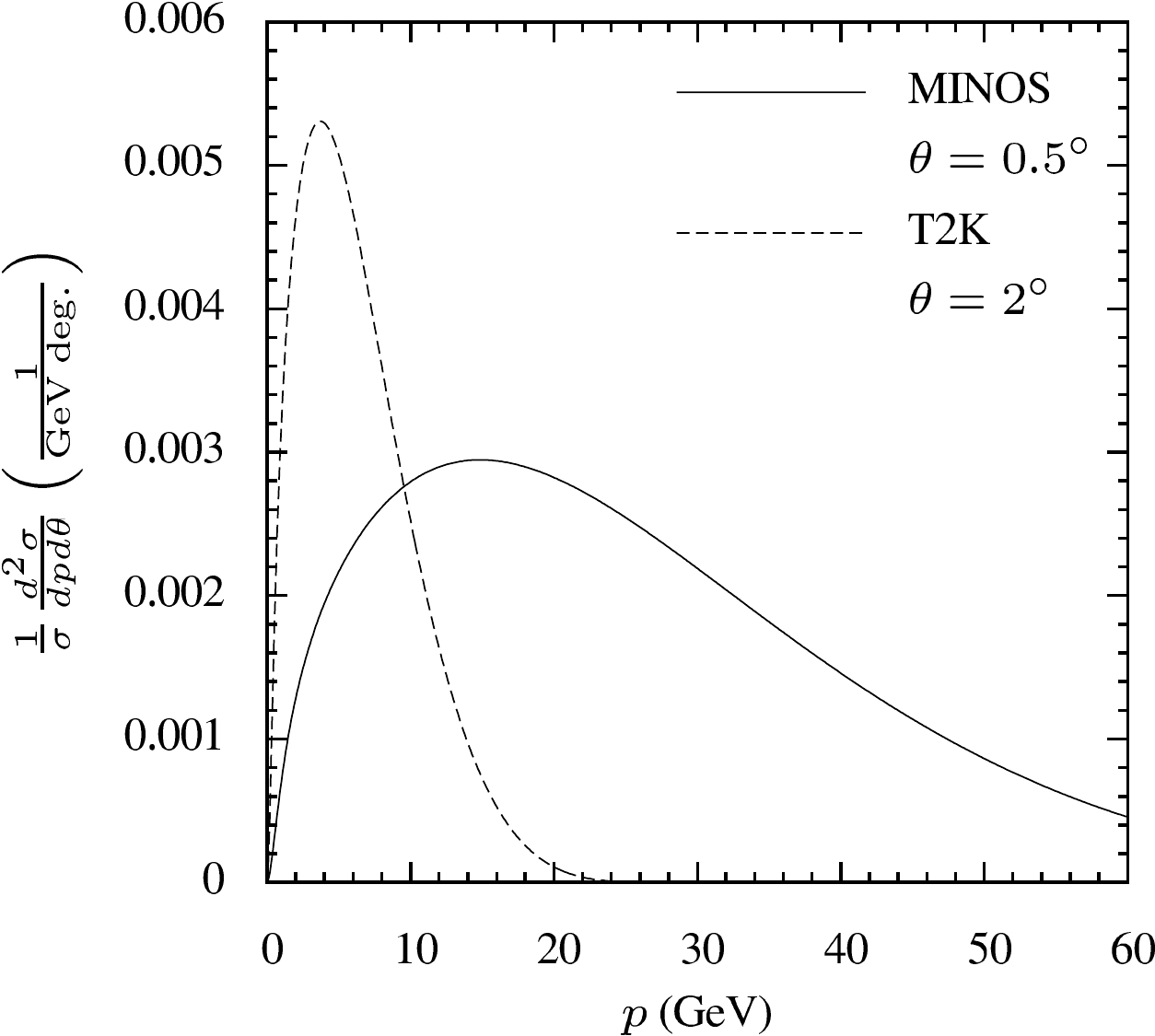}}
\caption{\footnotesize Pseudoscalar meson production distributions $f^{\rm BMPT}(\theta,p)$ in angle (left) and momentum (right), according to the 
fit \cite{bmpt}, scaled to the beam energy
and target composition for MINOS and T2K. These distributions determine the indirect production of vectors via e.g. $\et \rightarrow V\gamma$. Note that the
fitted distributions have an unphysical low-energy tail at large angles which we exclude in the analysis.}
\label{fig:bmpt}
\end{figure*}

In practice, this process is most important  in the low mass range where, for example, the large production rate of neutral pseudoscalar mesons $\ph = \pi^0, \eta$ 
can dominate the overall production of $V$'s in particular.  The meson production distribution at MiniBooNE is well-described by the Sanford and Wang fit 
$f^{\rm SW}(\theta,p)$ as
described in \cite{mininucl}, and utilized previously in \cite{den2011}. To estimate the meson production distribution for MINOS and T2K, 
we make use of an analytic fit $f^{\rm BMPT}(\theta,p)$ \cite{bmpt}
to data for (averaged $\pi^+$ and $\pi^-$) pion production obtained over a range of energies, which can be scaled to cover the target materials for the experiments of interest. Example distributions for MINOS and T2K configurations are shown in Fig.~\ref{fig:bmpt}. We have also tested this distribution against existing data published by NA61 \cite{na61_2011} for the T2K target configuration, and found good agreement. The aforementioned NA61 data is used to estimate the total pion flux at T2K. As there is currently no equivalent pion production dataset for MINOS, the pion production cross section was estimated by scaling the NA61 measured total cross section using the relative magnitudes of the BMPT distributions for T2K and MINOS.\footnote{The total $\pi^+$ flux could also be determined by working backwards from the measured neutrino flux, accounting for the angular acceptance of the detector \cite{bpr99c}. However, this reconstruction is more complex when the majority of pions decay 
in flight and are affected by magnetic focusing horns.}
The production rates of negatively and positively charged pions are averaged, and differ by ${\cal O}(1)$ factors (see e.g. \cite{bmpt}). To estimate the
$\et$ production rate, we use this averaged distribution and make use of some early experimental data \cite{etacross,hera84}, which indicates that in the 
appropriate energy range
\be
  \sigma_{pp \to pp\pi} \approx (25-30)\, \sigma_{pp \to pp\eta}.
\ee

This production mode is most relevant for $V$, and in the case $\ph=\pi^0$ or $\eta$ the branching ratio to $V$ is proportional to that of the radiative decay to two photons, though suppressed by coupling and phase space factors,
\begin{equation}
 \label{eq:brV}
 \textnormal{Br}_{\phi \to \gamma V} \simeq 2 \kappa^2 \left (1 - \frac{m_V^2}{m_\phi^2} \right)^3 \textnormal{Br}_{\phi \to \gamma\gamma}.
\end{equation}
For $\eta$ decays, as will be relevant here, $\textnormal{Br}_{\eta \to \gamma\gamma}\simeq0.39$.
This process becomes less competitive for higher mediator masses. We also explored the rate of $V$ production due to radiative decays of $c\bar{c}$ mesons such 
as $J/\psi \rightarrow V\eta$, but the overall rate is well below that of direct production discussed above. 

\end{itemize}

After either direct or indirect production, the suppressed portal couplings ensure that the real or virtual $X$ has an order-one branching to the hidden sector, Br$_{X\rightarrow 2\ch} \simeq 1$.
For completeness, we note that away from thresholds, the ratio of the decay rate of V to a single SM state \cite{BPR} relative to the hidden sector is 
of order $\ka^2 \al/\al'$, but is enhanced near resonances in hadronic channels and by the larger number of final states. In practice, Br$(V\rightarrow \bar\ch\ch)$ = 1 to a good approximation, apart from a small region near threshold of size $(1-4m_\ch^2/m_V^2)^n \lsim \ka^2 \al/\al'$
where the exponent $n=1/2$ (fermionic $\ch$) and $n=3/2$ (scalar $\ch$). We will excise this near-threshold region from the mass range.

The dark matter beam then propagates along with the neutrinos. For the couplings considered here, it has a weak-scale scattering rate with normal matter  and is
detectable  through neutral current-like elastic scattering processes with electrons or, of most relevance here, with nucleons. We will utilize the parameters and datasets of MiniBooNE, MINOS and T2K to probe this scenario. 
Importantly, MiniBooNE  has published an analysis on neutrino elastic scattering, which DM scattering will closely mimic, which allows some estimate of backgrounds
and efficiencies. MINOS and
T2K allow access to a higher mass range. We employed a simulation to determine the dark matter flux incident on the detector, which will be described in 
more detail in Sec.~\ref{sec:Sens}, after we have considered the viable model scenarios.

\section{Model scenarios}
\label{sec:Con}

The preceding analysis is applicable to generic scenarios of hidden sector states, coupled through the vector and scalar portals. In this section, we will
study more concrete models of sub-GeV dark matter, where the mediator mass satisfies $m_X > 2 m_\ch$, focusing on thermal relics for which the abundance provides a constraint on the annihilation cross section. 

\subsection{Constraints}

The constraints that we will take into account include:

\begin{itemize}
\item {\it Relic abundance}: If $\ch$ is a thermal relic, the WMAP constraint on the relic density $\Om_{\rm DM}h^2 \sim 0.1 \sim (0.1\,{\rm pb})/\langle \si v\rangle_{\rm fo}$ constrains the annihilation cross section at freeze-out to be $\langle \sigma v \rangle_{\rm fo} \sim 1\, {\rm pb}$. Even if the hidden sector state does not provide the dominant contribution 
to dark matter, the overclosure constraint $\langle \sigma v \rangle_{\rm fo} \gsim 1\, {\rm pb}$ applies more generally.
\item {\it Impact on the CMB}: In order not to distort the CMB due to energy injection into the IGM through annihilation, there are again restrictions on the annihilation
cross section that become particularly severe for light DM. The constraint takes the form $ f(z) \langle \si v \rangle_{\rm CMB} \lsim 0.1 \,(m_{\rm DM}/{\rm GeV})\, {\rm pb}$, where $f(z)$ is a redshift-dependent efficiency factor which for low mass varies from $f\sim 0.2$  for pion to $f\sim 1$ for electron final states \cite{cmb,*cmb2,*cmb3,*cmb4}.
These limits essentially exclude a thermal relic below a few GeV with an abundance fixed
via $s$-channel annihilation. Even for asymmetric DM, the requirement that the symmetric component annihilate away efficiently prior to decoupling
leads to a lower bound on the annihilation rate of the same order \cite{lyz}. This narrows down the field of viable models to those with 
(velocity-suppressed) annihilation since $v\sim 10^{-8}$ in this epoch. Other indirect signatures of annihilation in the galaxy 
(where $v\sim 10^{-3}$) are then necessarily suppressed as well.

\item {\it Visible decays}: Focusing first on the vector portal, we note that models where $V$ decays predominantly to the dark sector
are less constrained than those in which $V$ is metastable and decays mainly to the SM. For example, the fixed target constraints on dark
forces via leptonic $V$ decays \cite{best,*others,*others2,*others3,bpr99c} are avoided here for this reason as $V$ decays promptly
to the dark sector.  However, there are  constraints from high-luminosity colliders (particularly the $B$-factories in the case of GeV-scale vectors) 
which are sensitive to rare, but prompt, $V$ decays to the SM. In comparison to dark force searches where in the appropriate mass range
Br$(V\rightarrow l^+l^-)\sim {\cal O}(1)$, this SM branching is suppressed here by an additional factor of ${\cal O}(\al\ka^2/\al')$. 
A dedicated analysis for the higgs$'$strahlung signature \cite{BPR} was recently 
carried out at BaBar \cite{babarh'}, leading to limits translating to $\ka^4 \lsim {\rm few}\times 10^{-7}$ in the present scenario. 
However, this only applies when the dark Higgs is 
heavy enough to decay predominantly to two $V's$.  Although there are no specific analyses, one can infer $B$-factory (and $\ph$-factory)
limits on generic continuum processes, $e^+e^- \rightarrow V^*\gamma \rightarrow l^+l^-\gamma$ (and any exclusive decays, e.g of $\Up(nS)$ or $\ph$, not 
forbidden by $C$-parity \cite{BaBar,*Kloe}). The lack of significant peaks in similar rare-decay analyses suggests that limits,
which in the present case translate to $(\al/\al')\ka^4 \lsim 10^{-6}$, apply more generally \cite{Drees,slac,*Reece}. 

For the Higgs portal, there are significant $B$-factory limits on rare $B$ decays, Br$(B\rightarrow K + \slash{\!E})\lsim 10^{-5}$, which directly constrain
DM coupled via the Higgs portal \cite{bkp} due to decays of the form $B \rightarrow K + \ch + \bar{\ch}$.  

\item {\it Invisible decays}: The scenarios considered here allow for the mediator to decay on-shell to dark matter, leading to new invisible decay
channels \cite{McElrath,Fayet,*Fayet2,*Fayet3,Fayet4}.  Searches for rare radiative decays with missing energy, such as $J/\ps \rightarrow \gamma + \slash{\!E}$ \cite{CLEO}, are limited
in the present case by $C$-parity but there are also generic limits on purely invisible decays  of  $J/\ps$ and $\Up(1S)$ \cite{pdg} that 
constrain e.g. $J/\ps \rightarrow V^* \rightarrow \ch\ch^\dagger$. Off-resonance, the limit is of order $\al' \ka^2 \lsim {\rm few} \times 10^{-4}$ \cite{Fayet4}, which is
weaker than the visible decay constraint for $\al'\sim \al$ but becomes more significant for larger values of $\al'\sim 1$. Moving close to the 
resonance, where e.g. $m_V \approx m_{J/\ps}$, the limit becomes particularly stringent, $\ka^2/\al' \lsim {\rm few} \times 10^{-6}$. However, since the $V$ is still quite 
narrow for perturbative values of $\al'$, this only applies in a small $V$ mass range, which for $J/\ps$ and $\Up(1S)$ decays is somewhat above the 
scale considered here. We note that future limits on invisible decays, e.g. of $\ph$, would be sensitive to these scenarios.

\item {\it Energy injection during BBN}: Energy injection from decays of GeV-scale mediators in the early universe can be problematic if it occurs through 
hadronic channels
before about 10$^{-2}$~s. In the present case, most decays will occur to the hidden sector. Furthermore,  the couplings are sufficiently large to ensure
that decays occur well before BBN, so there are no significant constraints from this source.
\item {\it Self-interactions}: The models we are considering here will become nonperturbative when the couplings exceed the naive dimensional analysis (NDA) scale
of order $\al' \sim 4\pi$ (for the vector portal) and $\beta \sim 1$ (for the scalar portal). These rough limits characterize the point beyond which our
perturbative analysis breaks down, with a transfer cross section of order $\si_{\rm trans} \sim 4\pi (\al')^2 m_\ch^2/m_X^4 \sim 10^{-25}$cm$^2$ for a GeV-scale mediator. 
However, physical constraints on self-interaction are comparable in the low mass range. Limits
on halo ellipticity generally require that the transfer cross section for scattering is below about $10^{-23}-10^{-25}$~cm$^2$ \cite{dssw}; some
limits down to $10^{-26}$cm$^{2}$ do appear in the literature (see \cite{fkty,lyz} for recent discussions). This leads
to similar limits on $\al'$ and $\beta$ as the NDA limits quoted above. 
\end{itemize}

\subsection{Models and annihilation rates}

In this subsection we briefly outline the parameter space for a couple of DM scenarios, with the above constraints in mind. We will focus on thermal WIMPs, 
so that the annihilation rate determines the relic density. With sub-GeV masses, there are stringent constraints on models with $s$-wave annihilation cross sections, so we will focus on the cases that are $p$-wave suppressed, namely scalar DM coupled to the vector portal and Majorana DM coupled to the scalar portal.

\begin{itemize}

\item {\it Scalar dark matter with a U(1) mediator}:
The model contains four parameters; the masses $m_\ch$ and $m_V$ of the dark matter candidate and the vector mediator, 
the U(1)$'$ gauge coupling $e'$, and the kinetic mixing coefficient $\ka$. On requiring that $\ch$ comprises the majority of dark matter, the constraint on its relic abundance allows us to fix one relation between these four parameters. The primary quantity here is the annihilation rate, which for the GeV mass range of interest
is given by $s$-channel diagrams with $e^+e^-$, $\mu^+\mu^-$, and light hadronic final states. In the limit of small mixing, we can approximate this
rate by
\be
 \langle \si v\rangle_{\rm ann,V} \simeq \langle \si v\rangle_{e}+ \langle \si v\rangle_{\mu}(1+ R(s=4m_\ch^2)),
\ee
where $R = \si_{e^+e^-\rightarrow {\rm hadrons}}/\si_{e^+e^- \rightarrow \mu^+\mu^-}$, and the leptonic annihilation rate is \cite{BF},
\begin{align}
& \langle \si v\rangle_{l} = \frac{16\pi \ka^2 \al \al'}{3}  \langle v^2 \rangle 
 \\
 &\quad\times\frac{2m_\ch^2 + m_l^2}{(m_V^2 - 4 m_\ch^2)^2+m_V^2\Ga_V^2}\sqrt{1-\frac{m_l^2}{m_\ch^2}}.
 \nonumber
\end{align}
For $m_\ch \ll m_V \sim 1$~GeV, this rate scales as $\langle \si v \rangle_{l} \sim 10^{-33}\ka^2 (\al'/\al)(m_\ch/{\rm 100\,MeV})^2\,{\rm cm}^2$ using $v\sim 0.3$ at freeze-out. Accounting
for all annihilation channels, the observed relic density $\Om_{\rm DM}h^2 \sim 0.1 \sim (0.1\,{\rm pb})/\langle \si v\rangle_{\rm fo}$ reduces the number of free parameters to three via a constraint
that fixes $\al'=\al'(m_\ch,m_V,\ka)$.

The $p$-wave suppression of annihilation for low velocities allows this process to satisfy the CMB constraints \cite{cmb,*cmb2,*cmb3,*cmb4} alluded to above, 
as well as the galactic annihilation flux limits.

\item {\it Majorana dark matter with a scalar mediator}:
This model is a natural hidden sector generalization of the minimal model of scalar dark matter, and there are again four parameters: the masses $m_\ch$ and 
$m_S$, the hidden sector coupling $\beta$ and the mixing angle $\theta$. The abundance constraint will again allow us to determine e.g. $\beta=\beta(m_\ch,m_S,\theta)$.  Annihilation proceeds in the $p$-wave via mixing with the Higgs in the $s$-channel, and thus the rate is dictated by the light Higgs width. Given that $m_S \ll m_h$, the 
cross section scales as  $\langle \si v\rangle_{\rm ann,S} \sim \beta^2 \theta^2 m_\ch^2 \langle v^2 \rangle (\Gamma_{h^*}/m_{h^*})/(m_S^2-4m_\ch^2)^2$, where $h^*$ refers
to a virtual Higgs of mass $2m_\ch$. This rate is Yukawa-suppressed by the Higgs width $\Gamma_{h^*}$ unless the Higgs mixing angle $\theta \sim Av/m_h^2$ and $\beta$ are relatively large. However, these couplings are in turn constrained by the $B$-factory limits on rare $B$-decays with missing energy. The limits of \cite{bkp} imply
that $\beta^2 \theta^2 m_\ch^2 v_{\rm EW}^2 /m_S^4 \lsim {\cal O}(1)$, which we see is quite stringent for mediator and DM masses in the GeV range. These
$B$-decay limits make this scenario quite problematic as a model of thermal relic dark matter.

\end{itemize}

\subsection{Summary}

We can summarize the conclusions as follows for the four scenarios covered here: 

\begin{itemize}
\item {\it Vector portal, scalar $\ch$}: This DM candidate exhibits $p$-wave annihilation, which is crucial to satisfy galactic and CMB annihilation limits, and thus is viable for sub-percent mixing via the portal coupling. The sensitivity of neutrino facilities is significant in this case.
\item {\it Vector portal, fermionic $\ch$}: This implies $s$-wave annihilation, and the CMB annihilation limits can only be satisfied if $\ch$ is a  
highly subdominant component of WIMP dark matter, or has a more complex thermal history.
\item {\it Scalar portal, scalar $\ch$}: This again implies $s$-wave annihilation, suppressed in this case by the small Yukawa couplings. Thermal freezeout would
necessitate large mixing, which would rule out this scenario either due to the limits on rare $B$-decays, or its impact on the CMB.
\item {\it Scalar portal, fermionic $\ch$}: This DM candidate exhibits $p$-wave annihilation, which can avoid the CMB constraints, but to ensure the correct relic abundance the mixing must again be large, which is strongly constrained by rare $B$-decays.
\end{itemize}

Combining this information with the knowledge that DM coupled via the Higgs portal has a suppressed production rate, we conclude that rare $B$-decays
provide a more sensitive probe of the Higgs portal than dark matter beams. However, scalar DM coupled via the vector portal is a viable model
and we will focus on this scenario in the next section, where we outline the sensitivity of neutrino facilities to a GeV-scale dark matter beam.

\section{Sensitivity to a GeV dark matter beam}
\label{sec:Sens}
\subsection{Scattering}

The detection strategy studied here uses elastic scattering of the DM beam in the (near-)detector. We outline below the relevant
cross sections, focusing on the vector mediator for the reasons discussed in the previous section.

\begin{figure*}[t]
\centerline{\includegraphics[width=0.6\textwidth]{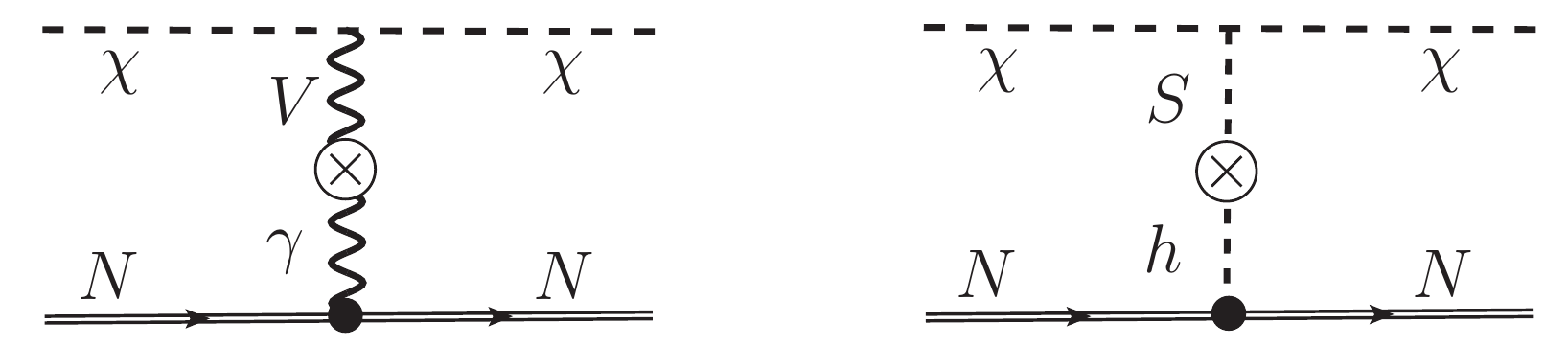}}
  \caption{\footnotesize Tree-level dark matter scattering off nucleons mediated by the vector and scalar portals.}
  \label{fig:scatt}
\end{figure*}

\begin{itemize}

\item {\it Vector-mediated scattering}:
For a vector mediator, the scattering of scalar DM on nucleons shown in Fig.~\ref{fig:scatt} is similar to neutrino-nucleon 
elastic scattering (see e.g. \cite{Ahrens:1986xe}), and the cross section takes the form
\begin{widetext} 
\begin{equation}
\label{eq:nscatterV}
\frac{d\sigma^V_{\ch N \to \ch N}}{dE_\ch} = \frac{\alpha' \kappa^2}{\alpha} \times \frac{4 \pi \alpha^2 \left[F^2_{1,N}(Q^2)A(E,E_\ch)-\frac{1}{4}F^2_{2,N}(Q^2)B(E,E_\ch)\right]}{{\left(m_V^2 + 2 m_N (E - E_\ch)\right)^2 (E^2-m_\chi^2)}},
\end{equation}
\end{widetext}
where $E$ and $E_\ch$ are the energies of the incident and outgoing dark matter particles, respectively and $Q^2=2m_N (E-E_\ch)$ is the momentum transfer.
We use simple monopole and dipole form-factors, $F_{1,N} = q_N/(1+Q^2/m_N^2)^2$ and  $F_{2,N} = \kappa_N/(1+Q^2/m_N^2)^2$, where $q_p=1$, $q_n=0$, 
$\kappa_p=1.79$ and $\kappa_n=-1.9$. The functions $A$ and $B$ are defined as
\begin{align}
A(E,E_\ch) & =2 m_N E E_\ch - m_\chi^2 (E-E_\ch)  \label{eq:A}, \\
B(E,E_\ch) & =(E_\ch-E)\left[(E_\ch+E)^2\right. \label{eq:B}
\\
&\quad\left.+2 m_N(E_\ch-E)-4m_\chi^2\right]. \nonumber
\end{align}

 \item {\it Scalar-mediated scattering}:
 For a scalar mediator, the $t$-channel scattering cross section of Majorana fermion DM on nucleons takes a similar form
 to the monopole contribution to (\ref{eq:nscatterV}), but in place of $\ka^2 \al\al'$ the cross section is suppressed by a factor 
 $\beta^2\theta^2 m_N^2 f_{Ts}^2/v_{\rm EW}^2$, where we have dropped isospin-violating corrections to the Higgs-nucleon coupling, and 
retained just the dominant contribution from $f_{Ts} \sim 0.118$ where $m_N f_{Tq} \equiv \langle N | m_q \bar{q}q| N\rangle$. Up to scalar mixing, 
this is analogous to conventional Higgs-mediated scattering and thus is quite suppressed relative the vector case above. Given
the suppressed production rate, we will not consider this case further in this section.
 \end{itemize}
 
 To account for the isotopic content of the detector material, we will use an effective differential cross section, given by
\begin{equation}
\label{eq:ch2scatter}
 \frac{d \sigma_{\chi N}^{\rm eff}}{dE_f}\simeq\frac{Z}{A}\frac{d\sigma_{\ch p \to \ch p}}{dE_\ch}+\frac{A-Z}{A}\frac{d\sigma_{\ch n \to \ch n}}{dE_\ch}.
\end{equation}
This expression is an approximation which ignores the differing detection efficiencies for scattering of bound nucleons. However, the $Q^2$-dependent efficiency
factors quoted by MiniBooNE \cite{mininucl}, for example, are close to one. Thus the error introduced by this simplification is small relative to the precision of our computation.

\subsection{Simulation}

A Monte Carlo simulation was employed to determine the kinematics of both the directly and indirectly produced dark matter beams. With the kinematics in hand, it is possible to calculate the expected sensitivity of MiniBooNE, MINOS and T2K to the hidden sector scenarios discussed in Sec.~\ref{sec:Con}, and the production channels of 
Sec.~\ref{sec:DMProd}. A detailed description of the simulation of the indirect production channel at MiniBooNE can be found in \cite{den2011}. Some of the pertinent parameters for each experiment are listed in Table~\ref{tab:numbers}, and we include some additional remarks below:

\begin{itemize}
\item {\it MINOS}: The MINOS experiment utilizes 120 GeV protons from the NuMI beamline impacting a graphite target. The near-detector has a large overall mass, but
 only part of the cross sectional area of the detector is instrumented and the near-detector itself is nearly 1~km from the target. Nonetheless, the large boost 
 provided by the 120 GeV proton beam leads to a significant event rate for dark matter scattering. We use an estimate for the total POT prior to the 2012 
 shutdown for the NOvA upgrade.
\item {\it T2K}: The T2K experiment has only been operating since 2010, and rather than use the current dataset we have
taken into account the final number of POT expected for the current run. 
T2K utilizes a 30 GeV proton beam impacting a graphite target, and has two near detectors, ND280 and INGRID, both located in a complex 
280~m from the target. ND280 is about 2 degrees off-axis and is well-instrumented with TPCs for tracking and analysis, while 
INGRID is on-axis and contains significantly more fiducial mass. 
\item {\it MiniBooNE}: The MiniBooNE experiment utilizes an 8.9 GeV proton beam impacting a Germanium target and, distinct from MINOS and T2K, has 
a single spherical mineral oil detector located 541~m from the target. This detector has a large fiducial mass, and importantly MiniBooNE has already
published a full neutrino elastic scattering analysis \cite{mininucl} with ${\cal O}(10^5)$ events and a measured energy spectrum, which provides the natural 
background for any dark matter beam search. We use an estimate for the total POT prior to the 2012 
 shutdown for the NOvA upgrade.
\end{itemize}

\begin{table*}
\begin{center}
\begin{tabular}{|c|c|c|c|c|c|c|c|c|c|c|c|c|c|} \hline
  & Target & $l_{T}$ & POT & $E_{\rm beam}$ & $L$  & $A_{\rm det}({\rm cm}^2)$ & $L_{\rm det}$ & $n_N$(cm$^{-3}$) & Fiducial Mass &$\ep_{\rm eff}$ \\ \hline
 MiniBooNE & Be & 71~cm & $1.2\times10^{21}$ & 8.9 GeV & 541~m  & $1.2 \times 10^{6}$ & 11.5 m & 9$\times 10^{23}$ & $\sim$650 tons & 0.6 \\
 MINOS & C & 94~cm & $1.5\times10^{21}$ & 120 GeV & 965~m & $7.1 \times 10^{4}$ & 1.3~m & 5$\times 10^{24}$ & 27 tons & 0.8 \\
 T2K ND280 & C & 90~cm & $5\times10^{21}$ & 30 GeV & 280~m & $5.5\times10^{4}$ & 0.7~m & 4$\times 10^{23}$ & 1.67 tons & $\sim$ \\ 
 T2K INGRID & C & 90~cm & $5\times10^{21}$ & 30 GeV & 280~m & $2.2\times10^{5}$ & 0.585~m & 5$\times 10^{24}$ & $\sim$110 tons & $\sim$ \\
 \hline
\end{tabular}
\caption{\footnotesize A summary of the parameters used for the three experiments considered in this work; see e.g. \cite{miniflux,mininucl,perevalov,mbPOT} for MiniBooNE, \cite{minos1,minos2,minos3,minos4,minosPOT} for MINOS, and \cite{t2k,nd280,ingrid,t2kpot} for T2K. Further details are in the text including a description of the notation. Note that in the 
absence of published analyses focusing on neutral currents, the overall efficiency $\ep_{\rm eff}$ for T2K is not known; we take it to be of the
same order as $\ep_{\rm eff}$  for MINOS. 
\label{tab:numbers}}
\end{center}
\end{table*}

The simulation of the dark matter beam used a re-weighting technique, first determining the dark matter trajectories that intersect the detector, and subsequently
weighting them according to the production distributions discussed earlier in Sec.~\ref{sec:DMProd}. We will describe these two steps in more detail
below, starting with the generation of the dark matter trajectories.

For direct production at either T2K or MINOS, the $V$'s were generated over an array of kinematically allowed momenta, and each $V$ was 
decayed isotropically into a random pair of $\chi$'s in the $V$'s center of mass frame. The lifetime of the $V$ is short enough for the parameter space considered that it will decay before escaping the target, and so the propagation of the $V$ through the target is ignored in the simulation. The trajectories of each of the $\chi$ particles are then checked to determine if they pass through the fiducial volume of the corresponding near detector. These trajectories are recorded along with the 
energy of the $\chi$. The treatment of indirect production at T2K, MINOS and MiniBooNE was similar (see \cite{den2011}), but required the extra
initial step of first generating kinematically allowed meson trajectories, with each then decayed isotropically into a $V$ and a $\gamma$ in the meson rest frame.
The newly produced $V$ is then treated in the same manner as in the direct production simulation.

With the trajectories in hand, for each point in parameter space the expected number of events could be determined by weighting them 
according to the production distribution $f(\theta,p)$, the scattering cross section $\sigma_{N\chi}^{\rm eff}(E)$, and the distance $R$ which 
$\ch$ propagates through the detector. There is also  an overall measure factor:  $\Delta=\delta p \delta \theta \delta \phi / (2 \pi)$ for indirect production, or $\Delta=\delta p$ for direct production, where the $\delta$ quantities refer to the step sizes used in the simulation for $\phi$ or $V$ production. Note that the distance $R$ travelled through the MINOS near detector and ND280 will almost always equal the length of the detector $L_{\rm det}$ shown in Table~\ref{tab:numbers}. For INGRID, it will occasionally be twice the listed number if it passes through the center of the detector, where two of the detector's modules overlap. MiniBooNE uses a spherical detector, and so $R$ can vary significantly in this case.

The final expression for the expected number of elastic nucleon dark matter scattering events is given by
\begin{align}
\label{eq:NeventsNucleon}
&N_{N\chi \to N\chi} = n_N \times \ep_{\rm eff} 
\\
&\quad\times \sum_{\substack{{\rm prod.}\\{\rm chans.}}} \left( N_{\ch} \sum_{{\rm trajec.}~i} R_i \sigma^{\rm eff}_{N\chi}(E_i) f(\theta_i,p_i) \Delta_i \right),
\nonumber
\end{align}
where $n_N$ is the nucleon density in the detector, while $\ep_{\rm eff}$ is the detection efficiency for events within the specified fiducial volume and cuts on momentum 
transfer. We will assume that lower cuts are above the range for coherent elastic scattering, so that our nucleon-level treatment in (\ref{eq:ch2scatter}) should be reliable. We
will also assume that the detection efficiencies do not deteriorate significantly for the full range of momentum transfer relevant for DM scattering.
The production quantities are given by
\ba
 N_\ch &=& \left\{ \begin{array}{ll} 2 N_{\rm POT} \times n_T  l_T \sigma_{PT} & {\rm direct} \\ 2 N_{\varphi} \times {\rm Br}(\varphi \rightarrow X+\cdots) & {\rm indirect}\end{array}\right., \\
 f(\theta,p) &=& \left\{ \begin{array}{ll} f_V(p) \times \frac{3}{4}(1-\cos^2 \theta) & {\rm direct} \\ f_{\varphi}^{\rm IND}(\theta,p) & {\rm indirect} \end{array}\right..
\ea
The distributions for direct ($f_{V}(p)$) and indirect ($f^{\rm IND}(\theta,p) = f^{\rm BMPT}_{\varphi}(\theta,p)$ or $f^{\rm SW}(\theta,p)$) production were discussed 
in Sec.~\ref{sec:DMProd}.\footnote{For T2K, rather
than $f^{\rm BMPT}(\theta,p)$, the indirect production distribution used was a parametrization of data from NA61 \cite{na61_2011}, using a replica T2K target. 
However, the results are  consistent with those using the BMPT parametrization.} Note that the meaning of $p$ and $\theta$ varies depending on the context. For direct production, $p$ is the $V$ momentum, and $\theta$ is the angle between the dark matter and the beam in the $V$ rest frame. For indirect production, both $p$ and $\theta$ refer to those of the original meson $\phi$ in the lab frame. The direct production parameters in $N_\ch$ are the number of 
protons on target $N_{\rm POT}$, the target length $l_T$ and density $n_T$, and the total cross section $\si_{\rm PT}$. The experimental quantities are listed 
in Table~\ref{tab:numbers}, while our treatment of $\si_{\rm PT}$ was discussed in Sec.~\ref{sec:DMProd}. For
indirect production, we use an estimate of the total $\phi=\eta$ yield $N_\eta$ 
and the branching ratio to  the mediator.

\subsection{Results}
\label{ssec:results}

\begin{figure*}[t]
  \centerline{\includegraphics[width=0.5\textwidth]{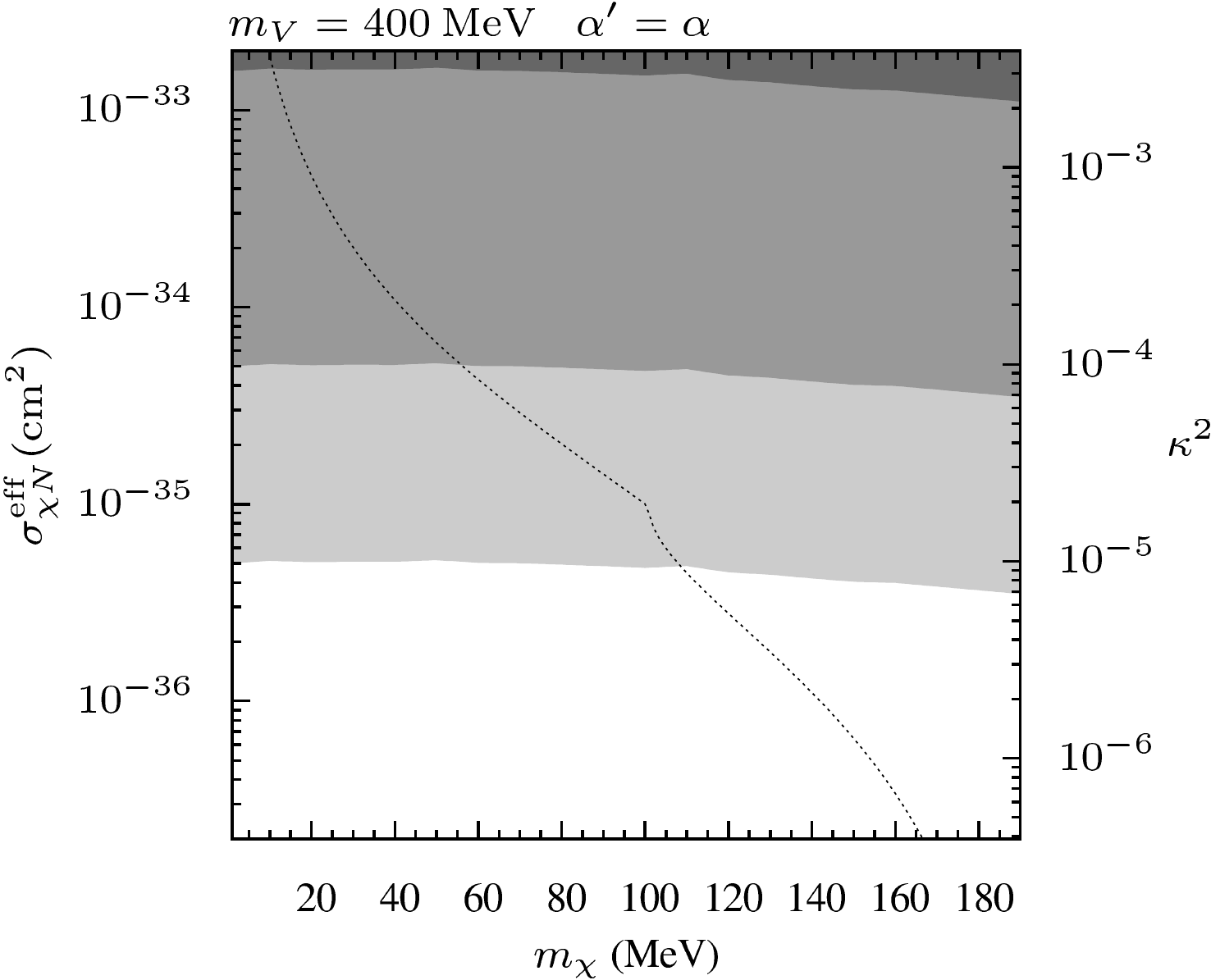} \hspace*{0.3cm} \includegraphics[width=0.5\textwidth]{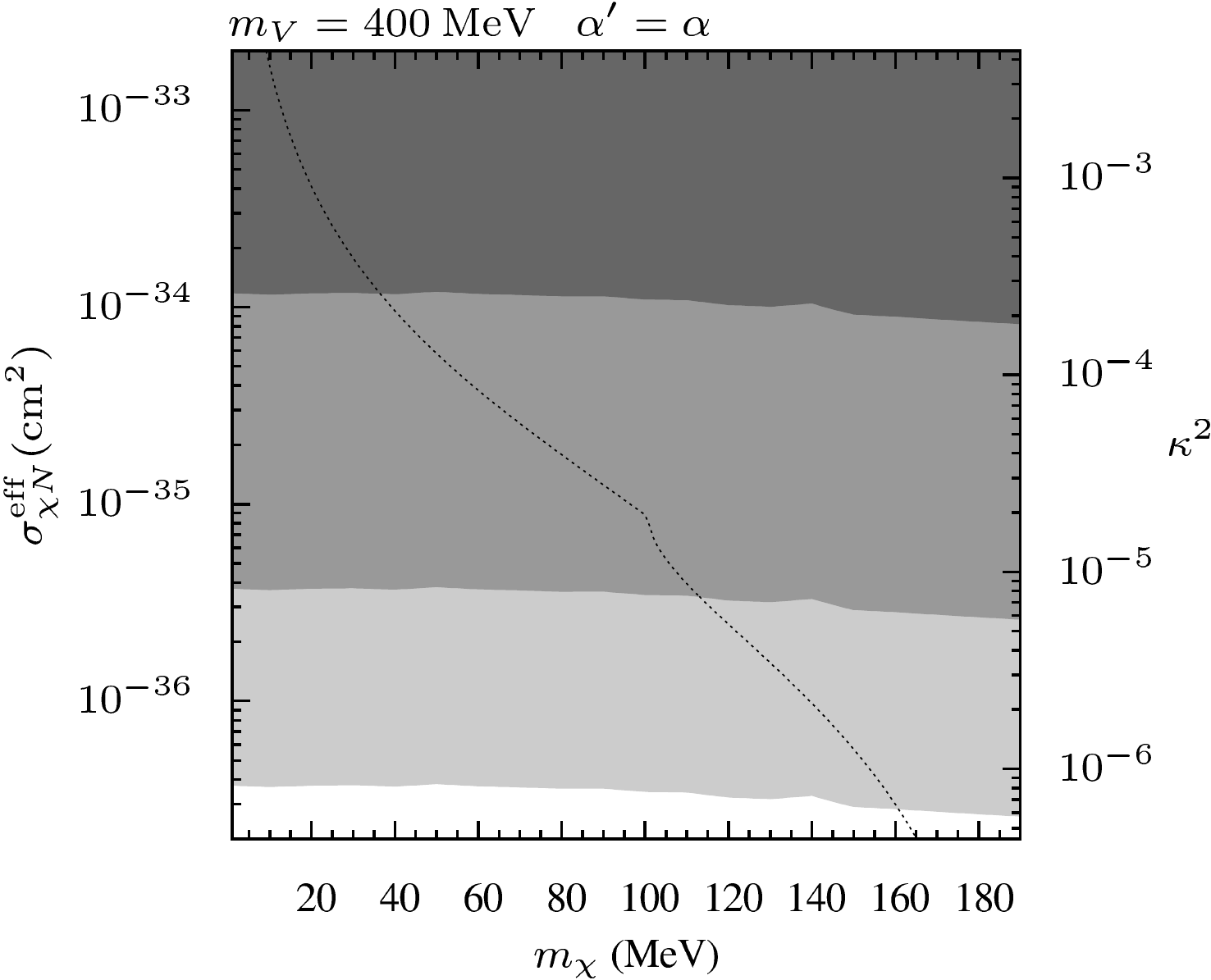}}
  \caption{\footnotesize Expected number of neutral current-like dark matter nucleon scattering events from V's produced through $\eta$ decays for the ND280 (left) and INGRID (right) detectors at T2K with $m_V=400$~MeV. The regions show greater than 10 (light), 1000 (medium) and $10^6$ (dark) expected events. The dashed curve indicates the value of $\kappa$ required for the dark matter annihilation cross section in the early universe to equal 1~pb.}
  \label{fig:eta1}
\end{figure*}

\begin{figure*}[t]
  \centerline{\includegraphics[width=0.5\textwidth]{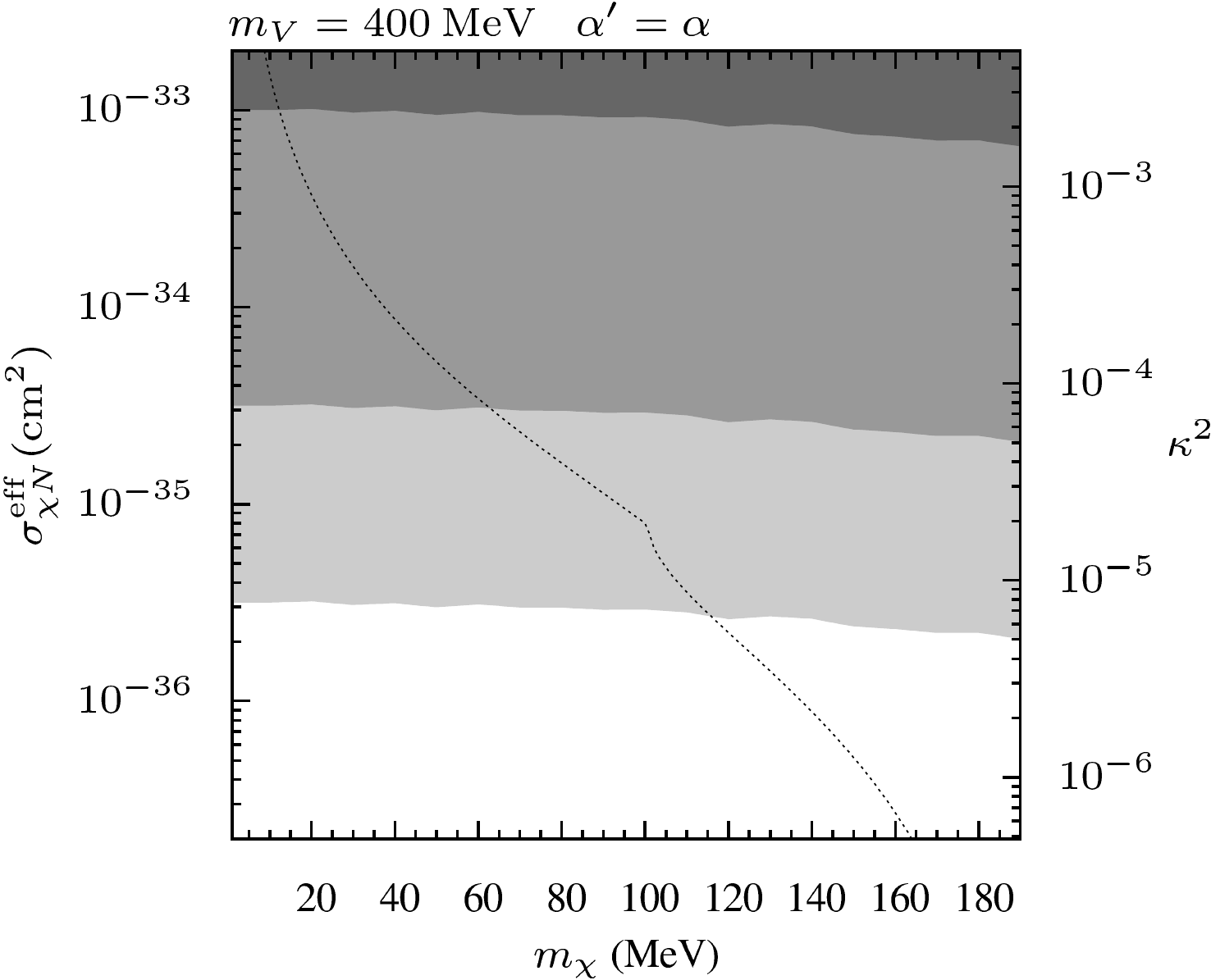} \hspace*{0.0cm} \includegraphics[width=0.5\textwidth]{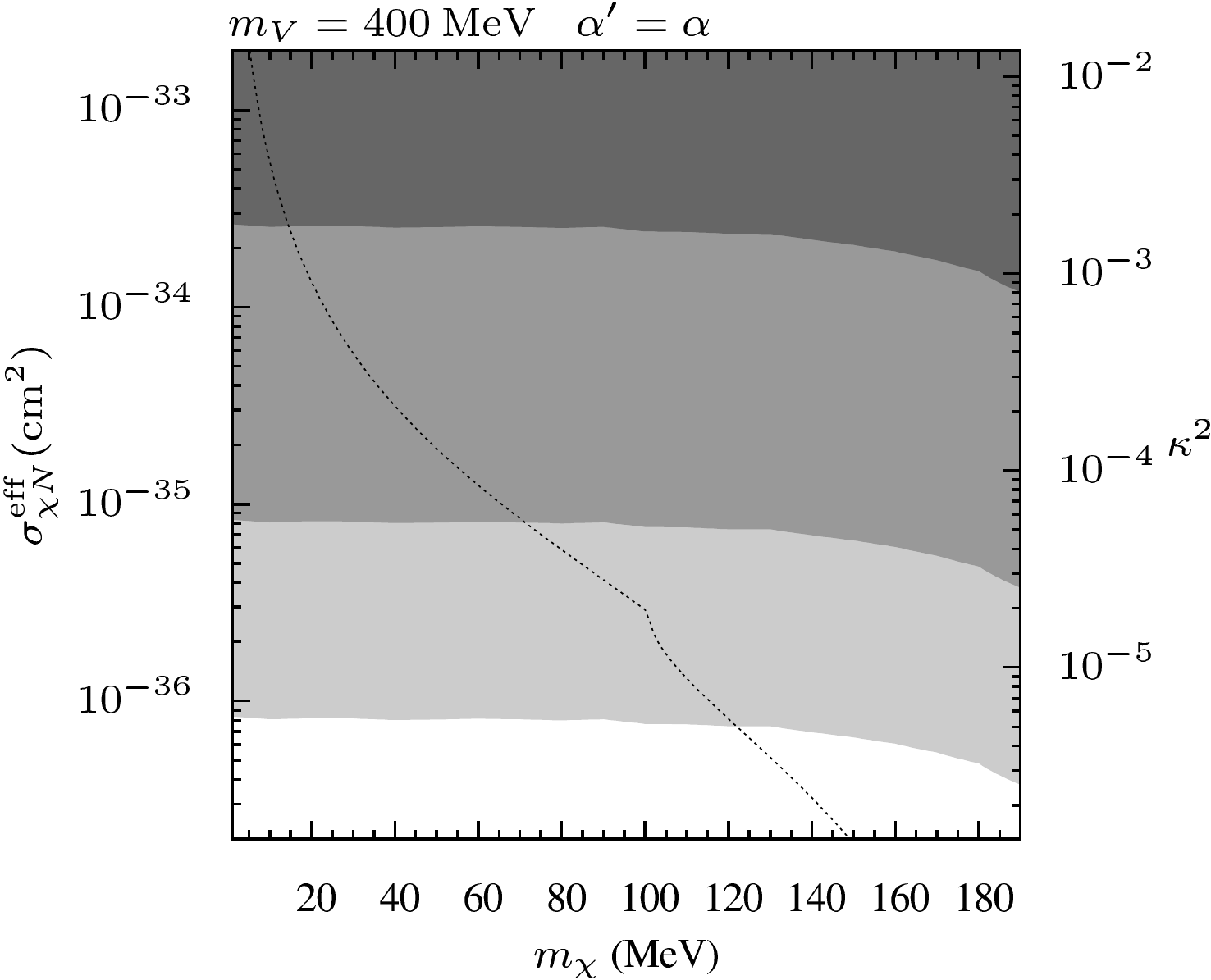}}
  \caption{\footnotesize Expected number of neutral current-like dark matter nucleon scattering events from V's produced through $\eta$ decays for the MINOS near detector (left) and MiniBooNE (right) with $m_V=400$~MeV. The contours are described in Fig.~\ref{fig:eta1}.}
  \label{fig:eta2}
\end{figure*}

\begin{figure*}[t]  
  \centerline{\includegraphics[width=0.5\textwidth]{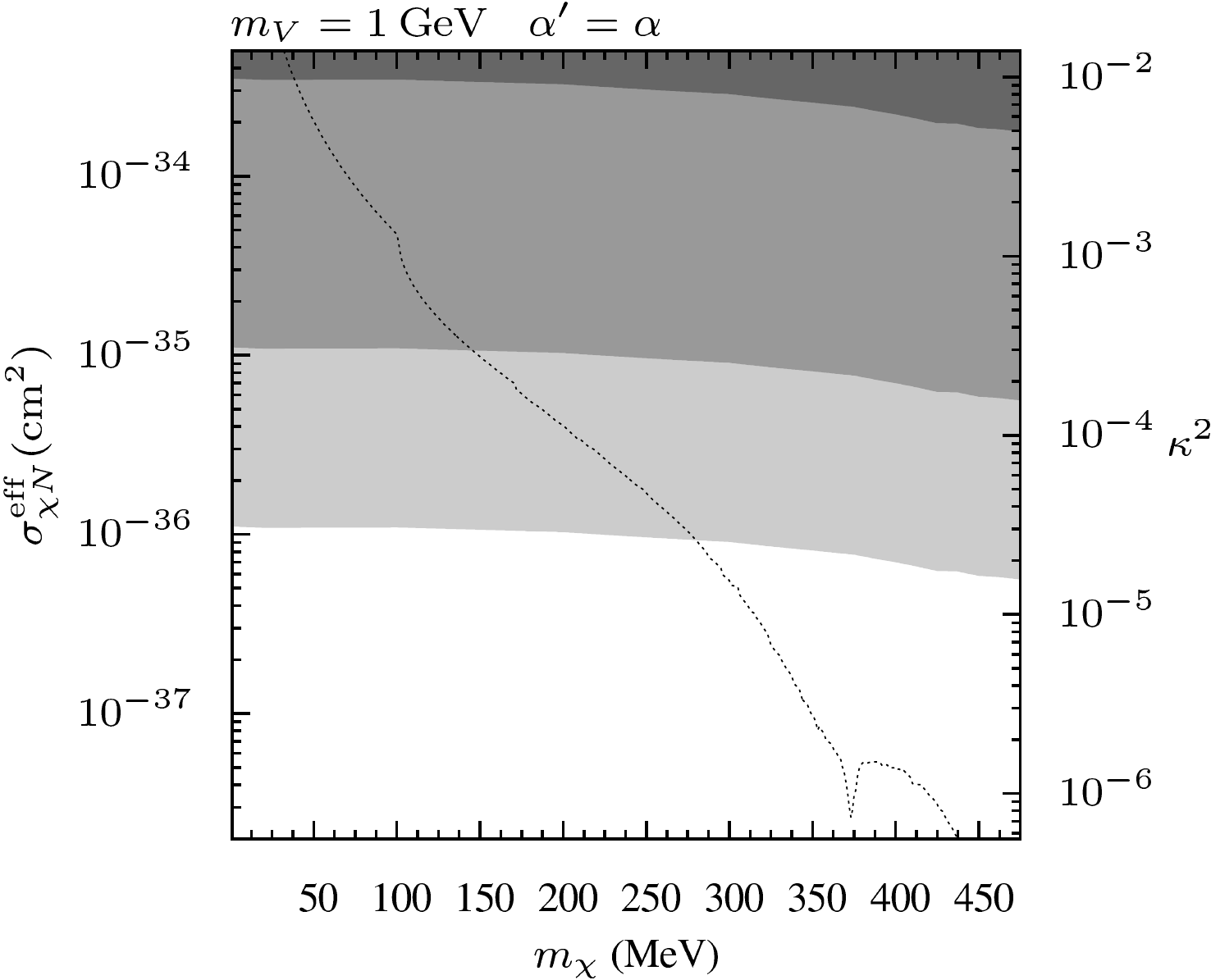} \hspace*{0.3cm} \includegraphics[width=0.5\textwidth]{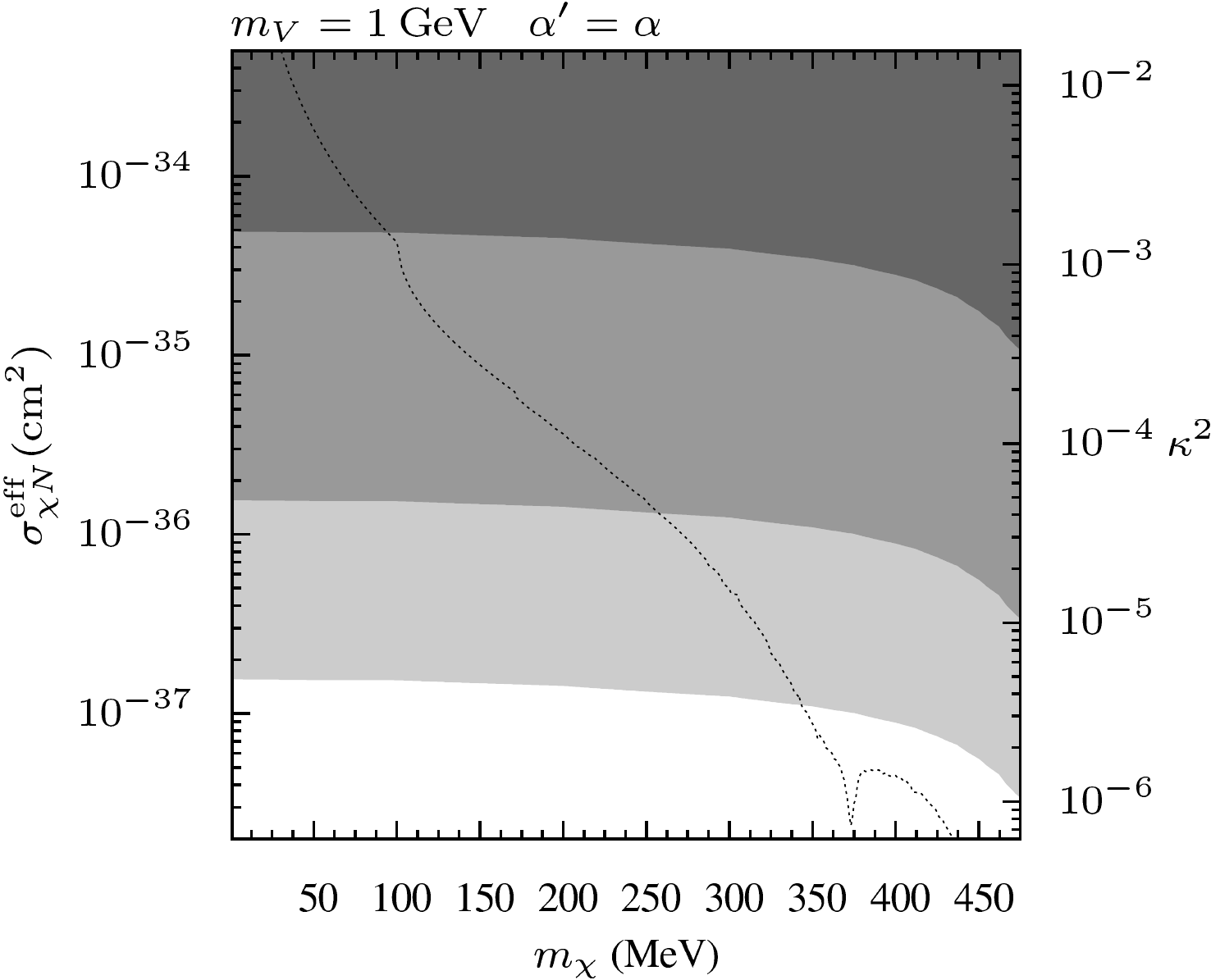}}
  \caption{\footnotesize Expected number of neutral current-like dark matter nucleon scattering events from direct V production for the ND280 (left) and INGRID (right) detectors at T2K with $m_V=1$~GeV. The contours are described in Fig.~\ref{fig:eta1}.}
  \label{fig:t2k2}
\end{figure*}

\begin{figure*}[t]
  \centerline{\includegraphics[width=0.5\textwidth]{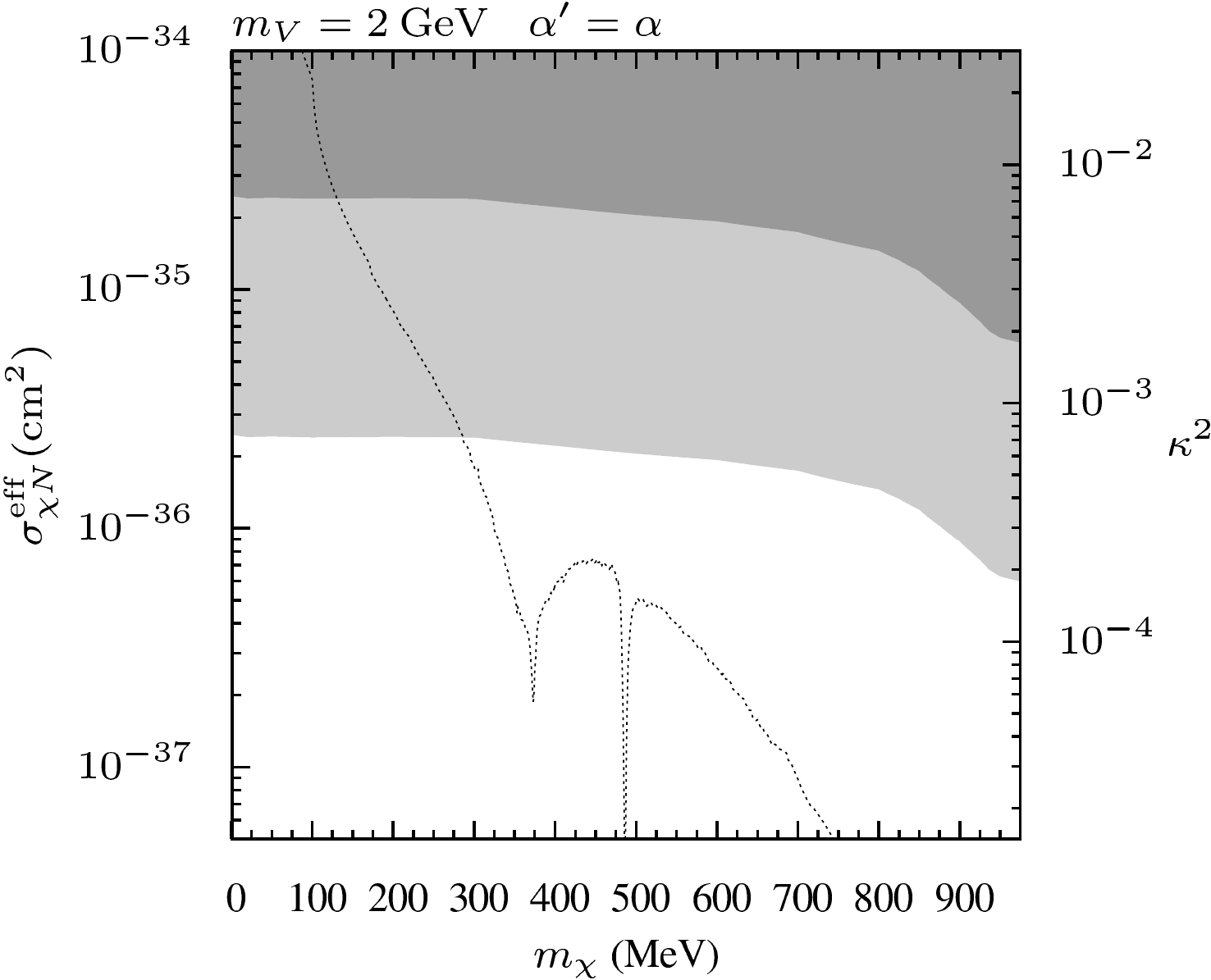} \hspace*{0.3cm} \includegraphics[width=0.5\textwidth]{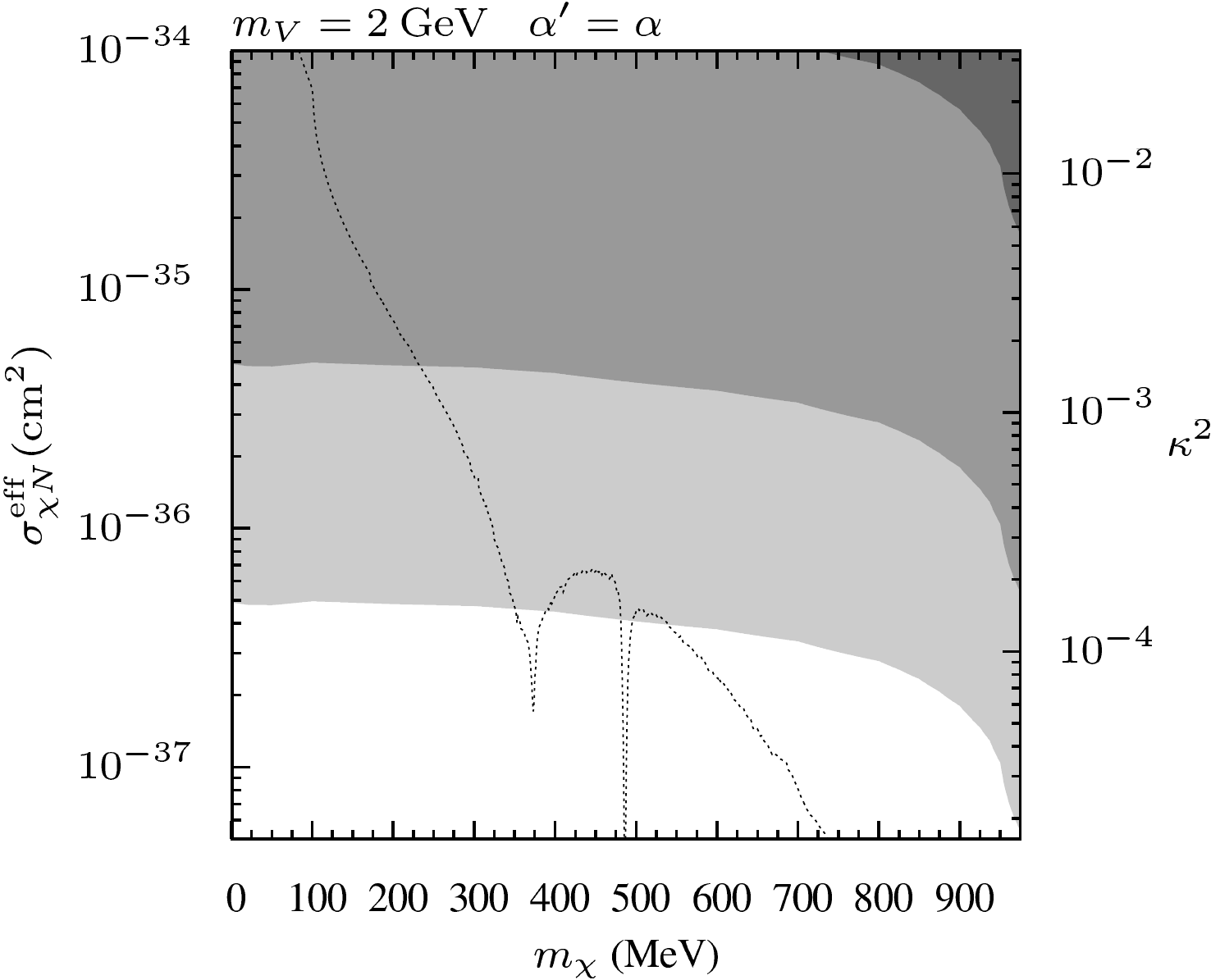}}
 \caption{\footnotesize Expected number of neutral current-like dark matter nucleon scattering events from direct V production for the ND280 (left) and INGRID (right) detectors at T2K with $m_V=2$~GeV. The contours are described in Fig.~\ref{fig:eta1}.}
  \label{fig:t2k3}
\end{figure*}

\begin{figure*}[t]
  \centerline{\includegraphics[width=0.5\textwidth]{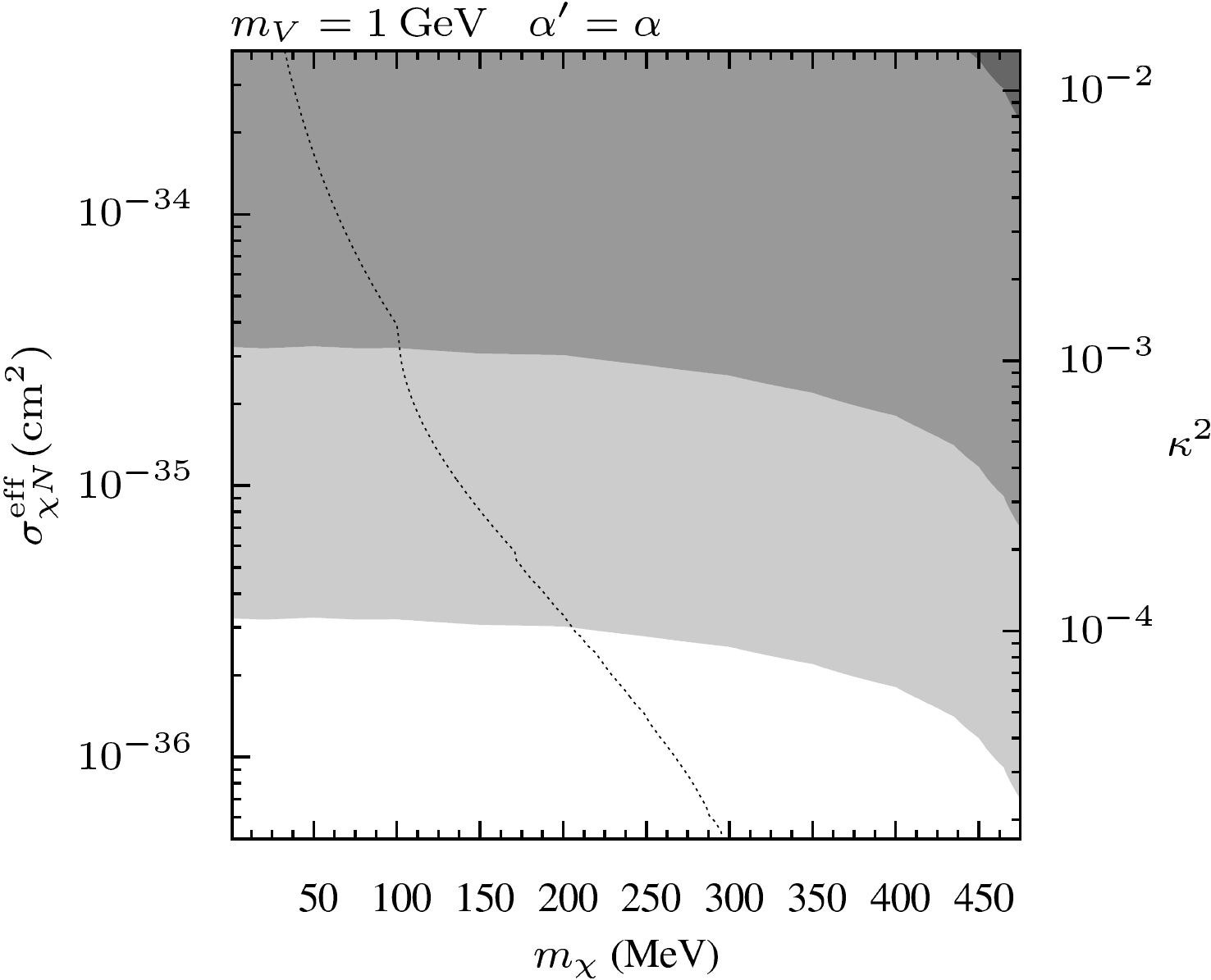} \hspace*{0.3cm} \includegraphics[width=0.5\textwidth]{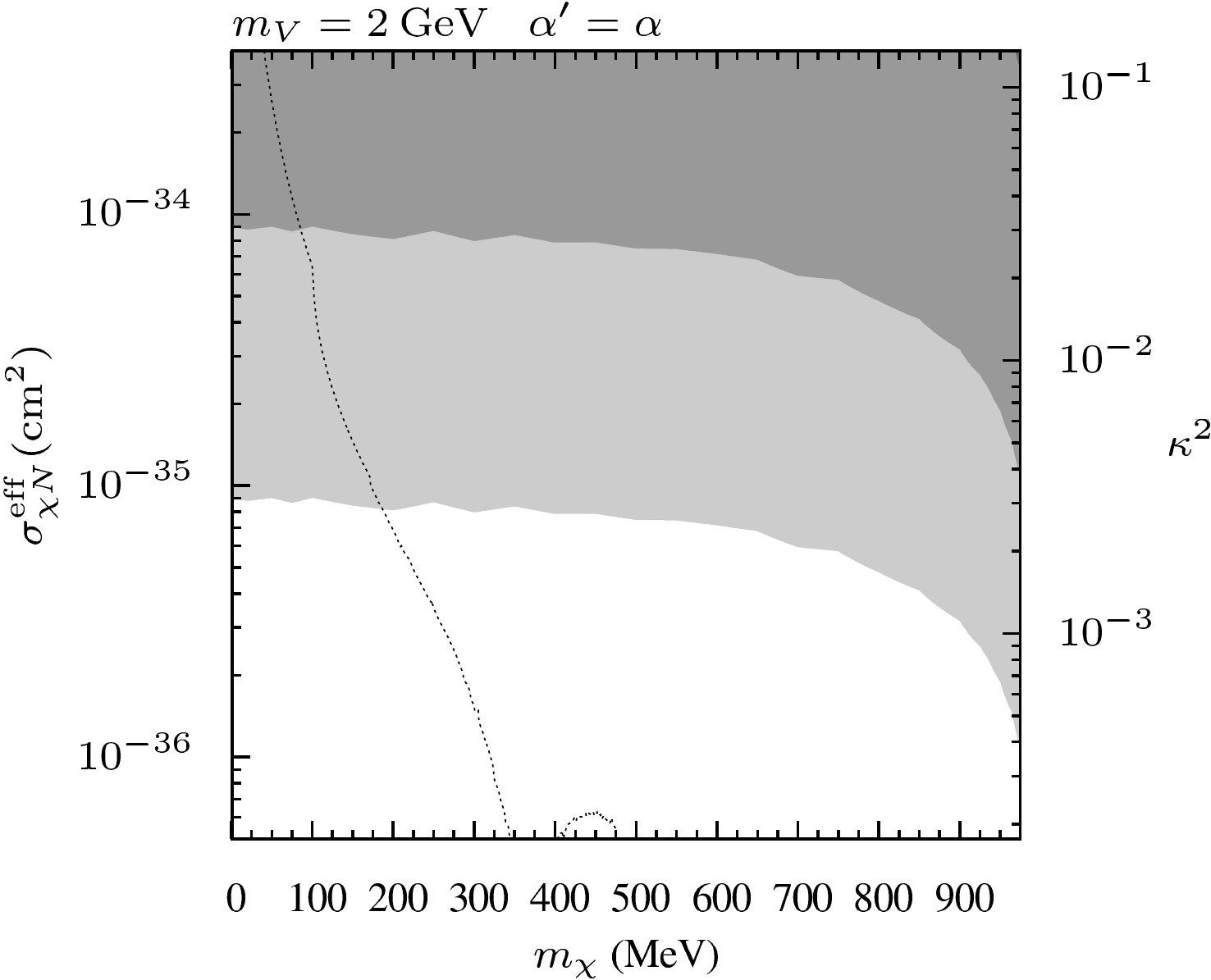}}
  \caption{\footnotesize Expected number of neutral current-like dark matter nucleon scattering events through direct V production for the MINOS near detector with two different vector mediator masses ($m_V=1$ GeV on the left and $m_V=2$ GeV on the right). The contours are  described in Fig.~\ref{fig:eta1}.}
  \label{fig:minos1}
\end{figure*}

\begin{figure*}[t]
  \centerline{\includegraphics[width=0.5\textwidth]{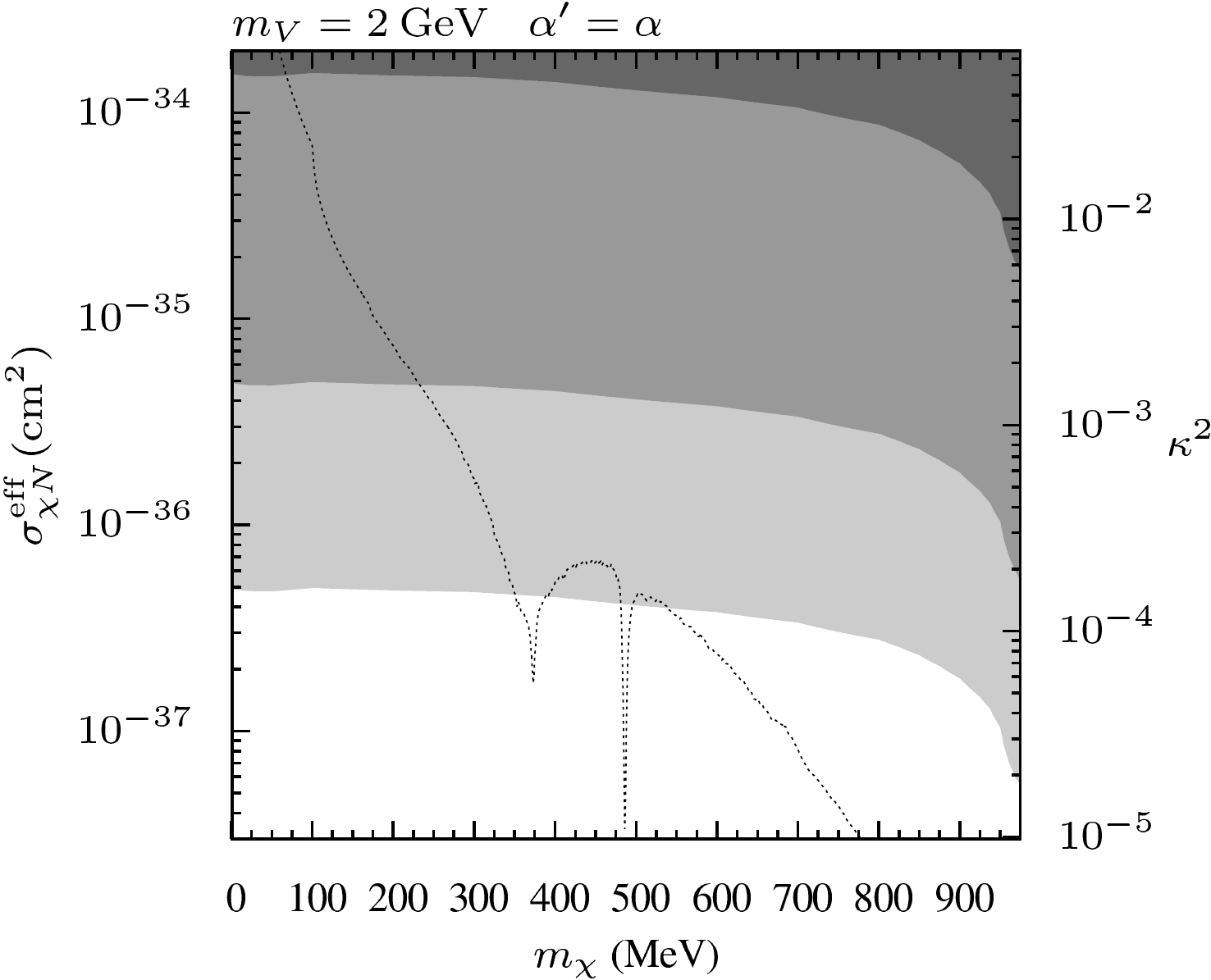} \hspace*{0.3cm} \includegraphics[width=0.5\textwidth]{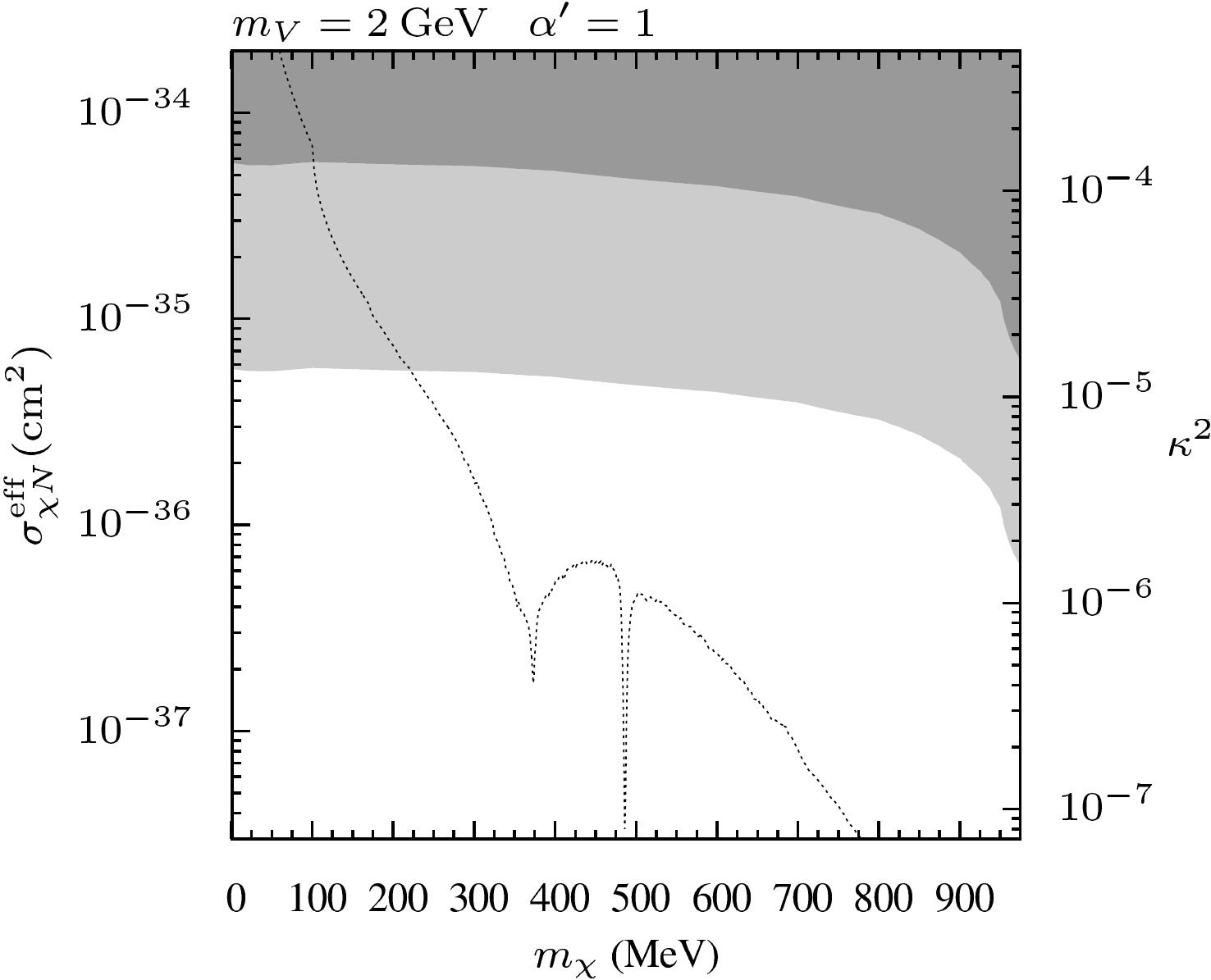}}
  \caption{\footnotesize Expected number of neutral current-like dark matter nucleon scattering events from direct V production with the INGRID detector at T2K, comparing two different $\alpha$ values ($\alpha'=\alpha$ on the left and $\alpha'=1$ on the right) for a 2 GeV Vector mediator. The contours are described in Fig.~\ref{fig:eta1}. }
  \label{fig:t2k1}
\end{figure*}

The results for DM nucleon scattering at MINOS, T2K, and MiniBooNE are shown for 
various parameter choices in Figs.~\ref{fig:eta1}-\ref{fig:t2k1}. All the plots show
contours of the number of events (10, 1000, or $10^6$) in the plane of nucleon scattering cross section (or kinetic mixing $\ka^2$) 
versus dark matter mass. The sensitivity tends to be fairly flat as a function of $m_\ch$, as the momentum transfer in the scattering tends to
be much larger than the mass and thus $m_\ch$ drops out of the kinematics. The exception to this general rule is that when $m_\ch$ approaches the decay threshold,
$m_\ch \sim m_V/2$, there is an enhancement in sensitivity as the dark matter has a small transverse boost from the $V$ decay and thus a 
larger fraction of trajectories will intersect on-axis detectors. 
The overlayed dark line denotes the parameter choices consistent with $\ch$ having the correct thermal relic density
to form WIMP dark matter. The structure in this curve for higher mass reflects the $\rh/\om$ and $\ph$ hadronic resonances that play a role in annihilation. Note that
the position of the thresholds and resonances is shifted down slightly from $2m_\ch$ due to the kinetic energy of the WIMPs. Assuming
an initial thermal abundance, the WIMP relic density would be too large in regions of parameter space below this curve. 

Existing particle physics limits on the parameter space would, as discussed in Sec.~\ref{sec:Con}, require that $\ka$ and $\al'$ satisfy 
both $(\al/\al')\kappa^4 \lsim 10^{-6}$ and $\al' \ka^2 \lsim {\rm few} \times 10^{-4}$ for models
of this type. For the perturbative values of $\al'\sim \al$ that will primarily be used in what follows, the former limit
is more restrictive implying $\ka^2 \lsim 10^{-3}$. However, this constraint is inferred from $B$-factory analyses of somewhat different models 
 rather than dedicated searches, so we refrain from showing any explicit exclusion curves.  Nonetheless, with this benchmark in mind we 
 observe from the plots that interesting sensitivity 
emerges with the ability to distinguish 
${\cal O}(10^3-10^4)$ dark matter scattering events from the neutrino background. 
The characteristic scattering cross section per nucleon to achieve ${\cal O}(1000)$ events ranges from $1-10$~pb, which is an
impressive level of sensitivity. Since this is not coherent scattering, it is more useful to contrast it with the best low-mass sensitivity achieved for
spin-dependent scattering in underground detectors \cite{Coupp2012} which is around 0.1pb for $m_\ch\sim 10$~GeV, but drops
off rapidly for lower masses. Of course, to implement a search for ${\cal O}(1000)$ events would require the ability to separate
this from the neutrino background of ${\cal O}(10^5-10^6)$ events. We will comment further on possible search strategies and
means for background rejection in the next section.

The individual plots reveal a number of other features summarized below:

\begin{itemize}
\item In Figs.~\ref{fig:eta1} and \ref{fig:eta2} we exhibit the sensitivity to dark matter in the 100-200~MeV mass range, with $m_V=400$~MeV. The direct parton-level approximation
for on-shell production of such low mass vectors is questionable with the PDF scale set to $Q=m_V$, so we use indirect production 
via $\et$-decays. As there is no significant resonant enhancement in $\eta$ production, one should bear in mind that this may be
an under-estimate for total $V$-production in this mass range. The ensuing sensitivity is shown for T2K in Fig.~\ref{fig:eta1}, and for 
MiniBooNE and MINOS in Fig.~\ref{fig:eta2}. Note that the lower momentum transfer in scattering at MiniBooNE leads to
an enhanced sensitivity to the cross section for a given sensitivity in $\ka$, relative to MINOS.
\item In Figs.~\ref{fig:t2k2} and \ref{fig:t2k3} for T2K and Fig.~\ref{fig:minos1} for MINOS we exhibit the sensitivity to higher mass dark matter, for $m_V=1$ and 2 GeV 
using direct parton-level production. We expect the narrow width approximation to work fairly well for 2 GeV vectors with this mass setting
the PDF scale. The precision of the estimate will certainly be lower using this method for 1 GeV vectors, due to the uncertainties in
the PDFs and the importance of higher-order QCD corrections. Nonetheless, we see that the tree-level sensitivity for 1 GeV vectors
is only marginally enhanced relative to $m_V=2$~GeV.
\item Using Figs.~\ref{fig:t2k2} and \ref{fig:t2k3}, it is interesting to compare the sensitivity of the two near-detectors at T2K. Given the suppression of 
direct $\ch$-production in the forward direction, the off-axis ND280 detector at T2K is ideally positioned to capture a comparatively 
large flux of dark matter, as compared to 
the on-axis detector INGRID. However, the much larger active mass of INGRID more than counteracts this effect, leading to 
an enhanced sensitivity. We are not aware if T2K has plans to use INGRID for analyses unrelated to diagnostics of the neutrino beam, 
but we see that there is considerable intrinsic sensitivity to light dark matter.
\item In Fig.~\ref{fig:t2k1} we compare the direct production sensitivity of the INGRID detector at T2K for two values of the dark U(1) coupling 
$\al'=\al$ and $\al'=1$. The latter value implies a self-interaction cross section for dark matter 
$\si \sim 4\pi m_\ch^2/m_V^4$ that can reach ${\cal O}(0.1~{\rm mb})$ for a 1 GeV mediator. This is close to, but somewhat below, 
the scale that would lead to detectable effects on halo structure, which may be relevant to the understanding of the inner regions of
dwarf spheroidal halos. As is apparent from the plot, increasing $\al'$ has the effect of enhancing the annihilation rate and
thus moving the relic density curve to lower values of $\ka$. Consequently, this increases the intrinsic sensitivity to $\ka$ while 
effectively lowering the sensitivity to the scattering cross section. Analogous sensitivity to $\al'$ applies to the other parameter regimes
and experiments shown in the earlier plots which all assume $\al'=\al$.
\item The plots all indicate that very light WIMPs with masses below about 100 MeV are problematic as thermal relics. Even for
larger values of $\al'$, the event rate along the measured relic density curve grows as $m_\ch$ decreases, and reaches levels which
are well above the elastic scattering rate for neutrinos. Thus, the measured elastic scattering of neutrinos at these facilities, and its
consistency with the Standard Model, serves to exclude
a large class of models of MeV scale dark matter. This strong tension with models of MeV-scale dark matter was already
exhibited in more detail using data from LSND and MiniBooNE in \cite{den2011}.
\end{itemize}

\section{Concluding Remarks}
\label{sec:Conc}
The direct search for dark matter is above all a search for weakly-interacting degrees of freedom, and at this stage the mass range
is relatively unconstrained. The LHC has as yet revealed little sign of new weak-scale physics, so it is important to keep in mind that the simple thermal relic
paradigm is broad enough to encompass a large mass range, extending well below the weak scale. Thus it is crucial to utilize all the available
experimental tools to explore the viable dark matter parameter space. The weak nature of DM-SM interactions
means that fixed target neutrino experiments provide a very natural source of low mass sensitivity. This goes both ways, as the next generation of underground dark matter
direct detection experiments may in turn be able to detect various astrophysical and cosmological sources of neutrinos. Rather than being simply an irreducible background,
it seems clear that these experiments will have to become observatories for all types of cosmic weakly-interacting degrees of freedom. In this
paper, we have explored another aspect of this convergence, namely the use of neutrino beam experiments to probe light dark matter that
can be produced in the target, and undergo elastic scattering in the detector. 

The challenge in developing a search strategy for this dark matter signal will be in disentangling the event spectrum
from the neutrino background at a level of maybe 1-10\%. This would allow sensitivity to kinetic mixing in the $\ka^2 \lsim 10^{-4}-10^{-5}$ range,
which is the most viable regime given the level of indirect constraints. This level of sensitivity to sub-GeV WIMPs, at the pb-level in terms
of per-nucleon scattering cross section, is only attained in spin-dependent direct detection for WIMPs with much larger masses, 
exceeding 10 GeV. Thus neutrino experiments could provide 
an important means of probing dark matter nucleon scattering below the mass range accessible via direct detection. In terms of
 isolating dark matter beam scattering events from the large background of neutrino elastic scattering, we note that there are
several distinctive characteristics. Firstly, the dark matter beam has a higher average energy (shown in Fig.~\ref{fig:dsigdE}) than the 
neutrino beam, and in particular a much higher cutoff that approaches the energy of the primary proton beam. This would permit
a relatively high cut in momentum transfer in scattering, provided such events are retained in the full sample. Secondly, the 
dark matter beam will be relatively unaffected by turning off or switching the polarity of the magnetic focusing horns, which would alter the neutrino beam significantly. Finally, there may also be useful information in the (nanosecond-scale) timing structure  
as the production mechanisms for vector-portal-coupled DM and the neutrino beam are quite distinct. Determining
whether one or more of these features could be put to practical use in a search strategy would require a dedicated analysis.

In concluding, we would also like to comment on some alternative approaches to explore the light WIMP regime.
\begin{itemize}
\item Direct detection in the low mass range could be feasible using electron scattering, as explored in recent work \cite{semicond,*semicond2}. This
approach is quite complementary to the neutrino beam analysis considered here. While the beam analysis requires relatively 
heavy vectors with $m_V> 2m_\ch$, the sensitivity for  electron scattering is enhanced when $m_V \ll m_\ch$. For comparison,
the projected electron scattering sensitivity to the vector portal model considered here is relatively weak for $m_V\sim 1$~GeV, 
but becomes significant for $m_V\sim 1$~MeV \cite{semicond}. Future progress using Ge crystals seems promising and may allow strong sensitivity to 
sub-GeV WIMPs with very light MeV-scale mediators, provided techniques are available to deal with all 
the backgrounds at such low recoil energies. 
\item Direct collider searches are also possible, utilizing missing energy signatures such as monophotons or
monojets \cite{Goodman:2010ku,*Fox:2011fx,*Rajaraman:2011wf,*Fox:2011pm,*Shoemaker:2011vi}. However, the 
sensitivity weakens significantly with light mediators. Collider searches can also pursue signatures of the light mediators 
directly, e.g. via subleading SM decays which may produce signatures
such as lepton jets at high energy. In the scenarios considered here, the decays
of the mediator are all prompt so there are no displaced vertices.
\end{itemize}

Looking to the future, the continued development of long-baseline neutrino facilities provides an ideal setting for expanding the
search for light hidden sector states. 
This sensitivity extends beyond the models of light dark matter discussed here to other classes of new physics, such as the scenarios discussed
in \cite{p1,*pp1}, that could impact the neutrino sector more directly.

\section*{Acknowledgements}

We would like to thank A.~Gaudin, C.~Polly and M.~Pospelov for helpful discussions. This work was supported in part by NSERC, Canada. 

\bibliography{GeVbeams}

\begin{thebibliography}{86}%
\makeatletter
\providecommand \@ifxundefined [1]{%
 \@ifx{#1\undefined}
}%
\providecommand \@ifnum [1]{%
 \ifnum #1\expandafter \@firstoftwo
 \else \expandafter \@secondoftwo
 \fi
}%
\providecommand \@ifx [1]{%
 \ifx #1\expandafter \@firstoftwo
 \else \expandafter \@secondoftwo
 \fi
}%
\providecommand \natexlab [1]{#1}%
\providecommand \enquote  [1]{``#1''}%
\providecommand \bibnamefont  [1]{#1}%
\providecommand \bibfnamefont [1]{#1}%
\providecommand \citenamefont [1]{#1}%
\providecommand \href@noop [0]{\@secondoftwo}%
\providecommand \href [0]{\begingroup \@sanitize@url \@href}%
\providecommand \@href[1]{\@@startlink{#1}\@@href}%
\providecommand \@@href[1]{\endgroup#1\@@endlink}%
\providecommand \@sanitize@url [0]{\catcode `\\12\catcode `\$12\catcode
  `\&12\catcode `\#12\catcode `\^12\catcode `\_12\catcode `\%12\relax}%
\providecommand \@@startlink[1]{}%
\providecommand \@@endlink[0]{}%
\providecommand \url  [0]{\begingroup\@sanitize@url \@url }%
\providecommand \@url [1]{\endgroup\@href {#1}{\urlprefix }}%
\providecommand \urlprefix  [0]{URL }%
\providecommand \Eprint [0]{\href }%
\providecommand \doibase [0]{http://dx.doi.org/}%
\providecommand \selectlanguage [0]{\@gobble}%
\providecommand \bibinfo  [0]{\@secondoftwo}%
\providecommand \bibfield  [0]{\@secondoftwo}%
\providecommand \translation [1]{[#1]}%
\providecommand \BibitemOpen [0]{}%
\providecommand \bibitemStop [0]{}%
\providecommand \bibitemNoStop [0]{.\EOS\space}%
\providecommand \EOS [0]{\spacefactor3000\relax}%
\providecommand \BibitemShut  [1]{\csname bibitem#1\endcsname}%
\let\auto@bib@innerbib\@empty
\bibitem [{\citenamefont {Boehm}\ \emph {et~al.}(2004)\citenamefont {Boehm},
  \citenamefont {Hooper}, \citenamefont {Silk}, \citenamefont {Casse},\ and\
  \citenamefont {Paul}}]{Boehm}%
  \BibitemOpen
  \bibfield  {author} {\bibinfo {author} {\bibfnamefont {C.}~\bibnamefont
  {Boehm}}, \bibinfo {author} {\bibfnamefont {D.}~\bibnamefont {Hooper}},
  \bibinfo {author} {\bibfnamefont {J.}~\bibnamefont {Silk}}, \bibinfo {author}
  {\bibfnamefont {M.}~\bibnamefont {Casse}}, \ and\ \bibinfo {author}
  {\bibfnamefont {J.}~\bibnamefont {Paul}},\ }\href {\doibase
  10.1103/PhysRevLett.92.101301} {\bibfield  {journal} {\bibinfo  {journal}
  {Phys. Rev. Lett.}\ }\textbf {\bibinfo {volume} {92}},\ \bibinfo {pages}
  {101301} (\bibinfo {year} {2004})},\ \Eprint
  {http://arxiv.org/abs/astro-ph/0309686} {arXiv:astro-ph/0309686 [astro-ph]}
  \BibitemShut {NoStop}%
\bibitem [{\citenamefont {Gondolo}\ and\ \citenamefont
  {Gelmini}(2005)}]{light-chi}%
  \BibitemOpen
  \bibfield  {author} {\bibinfo {author} {\bibfnamefont {P.}~\bibnamefont
  {Gondolo}}\ and\ \bibinfo {author} {\bibfnamefont {G.}~\bibnamefont
  {Gelmini}},\ }\href {\doibase 10.1103/PhysRevD.71.123520} {\bibfield
  {journal} {\bibinfo  {journal} {Phys. Rev. D}\ }\textbf {\bibinfo {volume}
  {71}},\ \bibinfo {pages} {123520} (\bibinfo {year} {2005})},\ \Eprint
  {http://arxiv.org/abs/hep-ph/0504010} {arXiv:hep-ph/0504010 [hep-ph]}
  \BibitemShut {NoStop}%
\bibitem [{\citenamefont {Finkbeiner}\ and\ \citenamefont
  {Weiner}(2007)}]{Neil}%
  \BibitemOpen
  \bibfield  {author} {\bibinfo {author} {\bibfnamefont {D.~P.}\ \bibnamefont
  {Finkbeiner}}\ and\ \bibinfo {author} {\bibfnamefont {N.}~\bibnamefont
  {Weiner}},\ }\href {\doibase 10.1103/PhysRevD.76.083519} {\bibfield
  {journal} {\bibinfo  {journal} {Phys. Rev. D}\ }\textbf {\bibinfo {volume}
  {76}},\ \bibinfo {pages} {083519} (\bibinfo {year} {2007})},\ \Eprint
  {http://arxiv.org/abs/astro-ph/0702587} {arXiv:astro-ph/0702587 [astro-ph]}
  \BibitemShut {NoStop}%
\bibitem [{\citenamefont {Hooper}\ and\ \citenamefont {Zurek}(2008)}]{HZ}%
  \BibitemOpen
  \bibfield  {author} {\bibinfo {author} {\bibfnamefont {D.}~\bibnamefont
  {Hooper}}\ and\ \bibinfo {author} {\bibfnamefont {K.~M.}\ \bibnamefont
  {Zurek}},\ }\href {\doibase 10.1103/PhysRevD.77.087302} {\bibfield  {journal}
  {\bibinfo  {journal} {Phys. Rev. D}\ }\textbf {\bibinfo {volume} {77}},\
  \bibinfo {pages} {087302} (\bibinfo {year} {2008})},\ \Eprint
  {http://arxiv.org/abs/0801.3686} {arXiv:0801.3686 [hep-ph]} \BibitemShut
  {NoStop}%
\bibitem [{\citenamefont {Arkani-Hamed}\ \emph {et~al.}(2009)\citenamefont
  {Arkani-Hamed}, \citenamefont {Finkbeiner}, \citenamefont {Slatyer},\ and\
  \citenamefont {Weiner}}]{AFSW}%
  \BibitemOpen
  \bibfield  {author} {\bibinfo {author} {\bibfnamefont {N.}~\bibnamefont
  {Arkani-Hamed}}, \bibinfo {author} {\bibfnamefont {D.~P.}\ \bibnamefont
  {Finkbeiner}}, \bibinfo {author} {\bibfnamefont {T.~R.}\ \bibnamefont
  {Slatyer}}, \ and\ \bibinfo {author} {\bibfnamefont {N.}~\bibnamefont
  {Weiner}},\ }\href {\doibase 10.1103/PhysRevD.79.015014} {\bibfield
  {journal} {\bibinfo  {journal} {Phys. Rev. D}\ }\textbf {\bibinfo {volume}
  {79}},\ \bibinfo {pages} {015014} (\bibinfo {year} {2009})},\ \Eprint
  {http://arxiv.org/abs/0810.0713} {arXiv:0810.0713 [hep-ph]} \BibitemShut
  {NoStop}%
\bibitem [{\citenamefont {Pospelov}\ and\ \citenamefont {Ritz}(2009)}]{PR}%
  \BibitemOpen
  \bibfield  {author} {\bibinfo {author} {\bibfnamefont {M.}~\bibnamefont
  {Pospelov}}\ and\ \bibinfo {author} {\bibfnamefont {A.}~\bibnamefont
  {Ritz}},\ }\href {\doibase 10.1016/j.physletb.2008.12.012} {\bibfield
  {journal} {\bibinfo  {journal} {Phys. Lett. B}\ }\textbf {\bibinfo {volume}
  {671}},\ \bibinfo {pages} {391} (\bibinfo {year} {2009})},\ \Eprint
  {http://arxiv.org/abs/0810.1502} {arXiv:0810.1502 [hep-ph]} \BibitemShut
  {NoStop}%
\bibitem [{\citenamefont {Fayet}(2004)}]{Fayet}%
  \BibitemOpen
  \bibfield  {author} {\bibinfo {author} {\bibfnamefont {P.}~\bibnamefont
  {Fayet}},\ }\href {\doibase 10.1103/PhysRevD.70.023514} {\bibfield  {journal}
  {\bibinfo  {journal} {Phys. Rev. D}\ }\textbf {\bibinfo {volume} {70}},\
  \bibinfo {pages} {023514} (\bibinfo {year} {2004})},\ \Eprint
  {http://arxiv.org/abs/hep-ph/0403226} {arXiv:hep-ph/0403226 [hep-ph]}
  \BibitemShut {NoStop}%
\bibitem [{\citenamefont {Fayet}(2006)}]{Fayet2}%
  \BibitemOpen
  \bibfield  {author} {\bibinfo {author} {\bibfnamefont {P.}~\bibnamefont
  {Fayet}},\ }\href {\doibase 10.1103/PhysRevD.74.054034} {\bibfield  {journal}
  {\bibinfo  {journal} {Phys. Rev. D}\ }\textbf {\bibinfo {volume} {74}},\
  \bibinfo {pages} {054034} (\bibinfo {year} {2006})},\ \Eprint
  {http://arxiv.org/abs/hep-ph/0607318} {arXiv:hep-ph/0607318 [hep-ph]}
  \BibitemShut {NoStop}%
\bibitem [{\citenamefont {Fayet}(2007)}]{Fayet3}%
  \BibitemOpen
  \bibfield  {author} {\bibinfo {author} {\bibfnamefont {P.}~\bibnamefont
  {Fayet}},\ }\href {\doibase 10.1103/PhysRevD.75.115017} {\bibfield  {journal}
  {\bibinfo  {journal} {Phys. Rev. D}\ }\textbf {\bibinfo {volume} {75}},\
  \bibinfo {pages} {115017} (\bibinfo {year} {2007})},\ \Eprint
  {http://arxiv.org/abs/hep-ph/0702176} {arXiv:hep-ph/0702176 [hep-ph]}
  \BibitemShut {NoStop}%
\bibitem [{\citenamefont {Boehm}\ and\ \citenamefont {Fayet}(2004)}]{BF}%
  \BibitemOpen
  \bibfield  {author} {\bibinfo {author} {\bibfnamefont {C.}~\bibnamefont
  {Boehm}}\ and\ \bibinfo {author} {\bibfnamefont {P.}~\bibnamefont {Fayet}},\
  }\href {\doibase 10.1016/j.nuclphysb.2004.01.015} {\bibfield  {journal}
  {\bibinfo  {journal} {Nucl. Phys. B}\ }\textbf {\bibinfo {volume} {683}},\
  \bibinfo {pages} {219} (\bibinfo {year} {2004})},\ \Eprint
  {http://arxiv.org/abs/hep-ph/0305261} {arXiv:hep-ph/0305261 [hep-ph]}
  \BibitemShut {NoStop}%
\bibitem [{\citenamefont {Pospelov}\ \emph {et~al.}(2008)\citenamefont
  {Pospelov}, \citenamefont {Ritz},\ and\ \citenamefont {Voloshin}}]{PRV}%
  \BibitemOpen
  \bibfield  {author} {\bibinfo {author} {\bibfnamefont {M.}~\bibnamefont
  {Pospelov}}, \bibinfo {author} {\bibfnamefont {A.}~\bibnamefont {Ritz}}, \
  and\ \bibinfo {author} {\bibfnamefont {M.~B.}\ \bibnamefont {Voloshin}},\
  }\href {\doibase 10.1016/j.physletb.2008.02.052} {\bibfield  {journal}
  {\bibinfo  {journal} {Phys. Lett. B}\ }\textbf {\bibinfo {volume} {662}},\
  \bibinfo {pages} {53} (\bibinfo {year} {2008})},\ \Eprint
  {http://arxiv.org/abs/0711.4866} {arXiv:0711.4866 [hep-ph]} \BibitemShut
  {NoStop}%
\bibitem [{\citenamefont {Lee}\ and\ \citenamefont {Weinberg}(1977)}]{LW}%
  \BibitemOpen
  \bibfield  {author} {\bibinfo {author} {\bibfnamefont {B.~W.}\ \bibnamefont
  {Lee}}\ and\ \bibinfo {author} {\bibfnamefont {S.}~\bibnamefont {Weinberg}},\
  }\href {\doibase 10.1103/PhysRevLett.39.165} {\bibfield  {journal} {\bibinfo
  {journal} {Phys. Rev. Lett.}\ }\textbf {\bibinfo {volume} {39}},\ \bibinfo
  {pages} {165} (\bibinfo {year} {1977})}\BibitemShut {NoStop}%
\bibitem [{\citenamefont {Borodatchenkova}\ \emph {et~al.}(2006)\citenamefont
  {Borodatchenkova}, \citenamefont {Choudhury},\ and\ \citenamefont
  {Drees}}]{Drees}%
  \BibitemOpen
  \bibfield  {author} {\bibinfo {author} {\bibfnamefont {N.}~\bibnamefont
  {Borodatchenkova}}, \bibinfo {author} {\bibfnamefont {D.}~\bibnamefont
  {Choudhury}}, \ and\ \bibinfo {author} {\bibfnamefont {M.}~\bibnamefont
  {Drees}},\ }\href {\doibase 10.1103/PhysRevLett.96.141802} {\bibfield
  {journal} {\bibinfo  {journal} {Phys. Rev. Lett.}\ }\textbf {\bibinfo
  {volume} {96}},\ \bibinfo {pages} {141802} (\bibinfo {year} {2006})},\
  \Eprint {http://arxiv.org/abs/hep-ph/0510147} {arXiv:hep-ph/0510147 [hep-ph]}
  \BibitemShut {NoStop}%
\bibitem [{\citenamefont {Pospelov}(2009)}]{tests}%
  \BibitemOpen
  \bibfield  {author} {\bibinfo {author} {\bibfnamefont {M.}~\bibnamefont
  {Pospelov}},\ }\href {\doibase 10.1103/PhysRevD.80.095002} {\bibfield
  {journal} {\bibinfo  {journal} {Phys. Rev. D}\ }\textbf {\bibinfo {volume}
  {80}},\ \bibinfo {pages} {095002} (\bibinfo {year} {2009})},\ \bibinfo {note}
  {14 pages, 2 figures},\ \Eprint {http://arxiv.org/abs/0811.1030}
  {arXiv:0811.1030 [hep-ph]} \BibitemShut {NoStop}%
\bibitem [{\citenamefont {Bjorken}\ \emph {et~al.}(2009)\citenamefont
  {Bjorken}, \citenamefont {Essig}, \citenamefont {Schuster},\ and\
  \citenamefont {Toro}}]{best}%
  \BibitemOpen
  \bibfield  {author} {\bibinfo {author} {\bibfnamefont {J.~D.}\ \bibnamefont
  {Bjorken}}, \bibinfo {author} {\bibfnamefont {R.}~\bibnamefont {Essig}},
  \bibinfo {author} {\bibfnamefont {P.}~\bibnamefont {Schuster}}, \ and\
  \bibinfo {author} {\bibfnamefont {N.}~\bibnamefont {Toro}},\ }\href {\doibase
  10.1103/PhysRevD.80.075018} {\bibfield  {journal} {\bibinfo  {journal} {Phys.
  Rev. D}\ }\textbf {\bibinfo {volume} {80}},\ \bibinfo {pages} {075018}
  (\bibinfo {year} {2009})},\ \Eprint {http://arxiv.org/abs/0906.0580}
  {arXiv:0906.0580 [hep-ph]} \BibitemShut {NoStop}%
\bibitem [{\citenamefont {Schuster}\ \emph {et~al.}(2010)\citenamefont
  {Schuster}, \citenamefont {Toro},\ and\ \citenamefont {Yavin}}]{others}%
  \BibitemOpen
  \bibfield  {author} {\bibinfo {author} {\bibfnamefont {P.}~\bibnamefont
  {Schuster}}, \bibinfo {author} {\bibfnamefont {N.}~\bibnamefont {Toro}}, \
  and\ \bibinfo {author} {\bibfnamefont {I.}~\bibnamefont {Yavin}},\ }\href
  {\doibase 10.1103/PhysRevD.81.016002} {\bibfield  {journal} {\bibinfo
  {journal} {Phys. Rev. D}\ }\textbf {\bibinfo {volume} {81}},\ \bibinfo
  {pages} {016002} (\bibinfo {year} {2010})},\ \Eprint
  {http://arxiv.org/abs/0910.1602} {arXiv:0910.1602 [hep-ph]} \BibitemShut
  {NoStop}%
\bibitem [{\citenamefont {Essig}\ \emph
  {et~al.}(2011{\natexlab{a}})\citenamefont {Essig}, \citenamefont {Schuster},
  \citenamefont {Toro},\ and\ \citenamefont {Wojtsekhowski}}]{others2}%
  \BibitemOpen
  \bibfield  {author} {\bibinfo {author} {\bibfnamefont {R.}~\bibnamefont
  {Essig}}, \bibinfo {author} {\bibfnamefont {P.}~\bibnamefont {Schuster}},
  \bibinfo {author} {\bibfnamefont {N.}~\bibnamefont {Toro}}, \ and\ \bibinfo
  {author} {\bibfnamefont {B.}~\bibnamefont {Wojtsekhowski}},\ }\href {\doibase
  10.1007/JHEP02(2011)009} {\bibfield  {journal} {\bibinfo  {journal} {JHEP}\
  }\textbf {\bibinfo {volume} {1102}},\ \bibinfo {pages} {009} (\bibinfo {year}
  {2011}{\natexlab{a}})},\ \Eprint {http://arxiv.org/abs/1001.2557}
  {arXiv:1001.2557 [hep-ph]} \BibitemShut {NoStop}%
\bibitem [{\citenamefont {Essig}\ \emph {et~al.}(2010)\citenamefont {Essig},
  \citenamefont {Harnik}, \citenamefont {Kaplan},\ and\ \citenamefont
  {Toro}}]{others3}%
  \BibitemOpen
  \bibfield  {author} {\bibinfo {author} {\bibfnamefont {R.}~\bibnamefont
  {Essig}}, \bibinfo {author} {\bibfnamefont {R.}~\bibnamefont {Harnik}},
  \bibinfo {author} {\bibfnamefont {J.}~\bibnamefont {Kaplan}}, \ and\ \bibinfo
  {author} {\bibfnamefont {N.}~\bibnamefont {Toro}},\ }\href {\doibase
  10.1103/PhysRevD.82.113008} {\bibfield  {journal} {\bibinfo  {journal} {Phys.
  Rev. D}\ }\textbf {\bibinfo {volume} {82}},\ \bibinfo {pages} {113008}
  (\bibinfo {year} {2010})},\ \Eprint {http://arxiv.org/abs/1008.0636}
  {arXiv:1008.0636 [hep-ph]} \BibitemShut {NoStop}%
\bibitem [{\citenamefont {Essig}\ \emph {et~al.}(2009)\citenamefont {Essig},
  \citenamefont {Schuster},\ and\ \citenamefont {Toro}}]{slac}%
  \BibitemOpen
  \bibfield  {author} {\bibinfo {author} {\bibfnamefont {R.}~\bibnamefont
  {Essig}}, \bibinfo {author} {\bibfnamefont {P.}~\bibnamefont {Schuster}}, \
  and\ \bibinfo {author} {\bibfnamefont {N.}~\bibnamefont {Toro}},\ }\href
  {\doibase 10.1103/PhysRevD.80.015003} {\bibfield  {journal} {\bibinfo
  {journal} {Phys. Rev. D}\ }\textbf {\bibinfo {volume} {80}},\ \bibinfo
  {pages} {015003} (\bibinfo {year} {2009})},\ \Eprint
  {http://arxiv.org/abs/0903.3941} {arXiv:0903.3941 [hep-ph]} \BibitemShut
  {NoStop}%
\bibitem [{\citenamefont {Reece}\ and\ \citenamefont {Wang}(2009)}]{Reece}%
  \BibitemOpen
  \bibfield  {author} {\bibinfo {author} {\bibfnamefont {M.}~\bibnamefont
  {Reece}}\ and\ \bibinfo {author} {\bibfnamefont {L.-T.}\ \bibnamefont
  {Wang}},\ }\href {\doibase 10.1088/1126-6708/2009/07/051} {\bibfield
  {journal} {\bibinfo  {journal} {JHEP}\ }\textbf {\bibinfo {volume} {0907}},\
  \bibinfo {pages} {051} (\bibinfo {year} {2009})},\ \Eprint
  {http://arxiv.org/abs/0904.1743} {arXiv:0904.1743 [hep-ph]} \BibitemShut
  {NoStop}%
\bibitem [{\citenamefont {Batell}\ \emph
  {et~al.}(2009{\natexlab{a}})\citenamefont {Batell}, \citenamefont
  {Pospelov},\ and\ \citenamefont {Ritz}}]{BPR}%
  \BibitemOpen
  \bibfield  {author} {\bibinfo {author} {\bibfnamefont {B.}~\bibnamefont
  {Batell}}, \bibinfo {author} {\bibfnamefont {M.}~\bibnamefont {Pospelov}}, \
  and\ \bibinfo {author} {\bibfnamefont {A.}~\bibnamefont {Ritz}},\ }\href
  {\doibase 10.1103/PhysRevD.79.115008} {\bibfield  {journal} {\bibinfo
  {journal} {Phys. Rev. D}\ }\textbf {\bibinfo {volume} {79}},\ \bibinfo
  {pages} {115008} (\bibinfo {year} {2009}{\natexlab{a}})},\ \Eprint
  {http://arxiv.org/abs/0903.0363} {arXiv:0903.0363 [hep-ph]} \BibitemShut
  {NoStop}%
\bibitem [{\citenamefont {Batell}\ \emph
  {et~al.}(2009{\natexlab{b}})\citenamefont {Batell}, \citenamefont
  {Pospelov},\ and\ \citenamefont {Ritz}}]{bpr99c}%
  \BibitemOpen
  \bibfield  {author} {\bibinfo {author} {\bibfnamefont {B.}~\bibnamefont
  {Batell}}, \bibinfo {author} {\bibfnamefont {M.}~\bibnamefont {Pospelov}}, \
  and\ \bibinfo {author} {\bibfnamefont {A.}~\bibnamefont {Ritz}},\ }\href
  {\doibase 10.1103/PhysRevD.80.095024} {\bibfield  {journal} {\bibinfo
  {journal} {Phys. Rev. D}\ }\textbf {\bibinfo {volume} {80}},\ \bibinfo
  {pages} {095024} (\bibinfo {year} {2009}{\natexlab{b}})},\ \Eprint
  {http://arxiv.org/abs/0906.5614} {arXiv:0906.5614 [hep-ph]} \BibitemShut
  {NoStop}%
\bibitem [{\citenamefont {deNiverville}\ \emph {et~al.}(2011)\citenamefont
  {deNiverville}, \citenamefont {Pospelov},\ and\ \citenamefont
  {Ritz}}]{den2011}%
  \BibitemOpen
  \bibfield  {author} {\bibinfo {author} {\bibfnamefont {P.}~\bibnamefont
  {deNiverville}}, \bibinfo {author} {\bibfnamefont {M.}~\bibnamefont
  {Pospelov}}, \ and\ \bibinfo {author} {\bibfnamefont {A.}~\bibnamefont
  {Ritz}},\ }\href@noop {} {\bibfield  {journal} {\bibinfo  {journal} {Phys.
  Rev. D}\ }\textbf {\bibinfo {volume} {84}},\ \bibinfo {pages} {075020}
  (\bibinfo {year} {2011})},\ \Eprint {http://arxiv.org/abs/1107.4580}
  {arXiv:1107.4580 [hep-ph]} \BibitemShut {NoStop}%
\bibitem [{\citenamefont {Rosenberg}\ and\ \citenamefont {van
  Bibber}(2000)}]{axions_exp}%
  \BibitemOpen
  \bibfield  {author} {\bibinfo {author} {\bibfnamefont {L.~J.}\ \bibnamefont
  {Rosenberg}}\ and\ \bibinfo {author} {\bibfnamefont {K.~A.}\ \bibnamefont
  {van Bibber}},\ }\href {\doibase 10.1016/S0370-1573(99)00045-9} {\bibfield
  {journal} {\bibinfo  {journal} {Phys. Rept.}\ }\textbf {\bibinfo {volume}
  {325}},\ \bibinfo {pages} {1} (\bibinfo {year} {2000})}\BibitemShut {NoStop}%
\bibitem [{\citenamefont {Carosi}\ and\ \citenamefont {van
  Bibber}(2008)}]{axions_exp2}%
  \BibitemOpen
  \bibfield  {author} {\bibinfo {author} {\bibfnamefont {G.}~\bibnamefont
  {Carosi}}\ and\ \bibinfo {author} {\bibfnamefont {K.}~\bibnamefont {van
  Bibber}},\ }\href@noop {} {\bibfield  {journal} {\bibinfo  {journal} {Lect.
  Notes Phys.}\ }\textbf {\bibinfo {volume} {741}},\ \bibinfo {pages} {135}
  (\bibinfo {year} {2008})},\ \Eprint {http://arxiv.org/abs/hep-ex/0701025}
  {arXiv:hep-ex/0701025 [hep-ex]} \BibitemShut {NoStop}%
\bibitem [{\citenamefont {Gallas}\ \emph {et~al.}(1995)\citenamefont {Gallas}
  \emph {et~al.}}]{neutrino_exp}%
  \BibitemOpen
  \bibfield  {author} {\bibinfo {author} {\bibfnamefont {E.}~\bibnamefont
  {Gallas}} \emph {et~al.} (\bibinfo {collaboration} {FMMF Collaboration}),\
  }\href {\doibase 10.1103/PhysRevD.52.6} {\bibfield  {journal} {\bibinfo
  {journal} {Phys. Rev. D}\ }\textbf {\bibinfo {volume} {52}},\ \bibinfo
  {pages} {6} (\bibinfo {year} {1995})}\BibitemShut {NoStop}%
\bibitem [{\citenamefont {Bernardi}\ \emph {et~al.}(1988)\citenamefont
  {Bernardi}, \citenamefont {Carugno}, \citenamefont {Chauveau}, \citenamefont
  {Dicarlo}, \citenamefont {Dris} \emph {et~al.}}]{neutrino_exp2}%
  \BibitemOpen
  \bibfield  {author} {\bibinfo {author} {\bibfnamefont {G.}~\bibnamefont
  {Bernardi}}, \bibinfo {author} {\bibfnamefont {G.}~\bibnamefont {Carugno}},
  \bibinfo {author} {\bibfnamefont {J.}~\bibnamefont {Chauveau}}, \bibinfo
  {author} {\bibfnamefont {F.}~\bibnamefont {Dicarlo}}, \bibinfo {author}
  {\bibfnamefont {M.}~\bibnamefont {Dris}},  \emph {et~al.},\ }\href {\doibase
  10.1016/0370-2693(88)90563-1} {\bibfield  {journal} {\bibinfo  {journal}
  {Phys. Lett. B}\ }\textbf {\bibinfo {volume} {203}},\ \bibinfo {pages} {332}
  (\bibinfo {year} {1988})}\BibitemShut {NoStop}%
\bibitem [{\citenamefont {Adams}\ \emph {et~al.}(1997)\citenamefont {Adams}
  \emph {et~al.}}]{gluino_exp}%
  \BibitemOpen
  \bibfield  {author} {\bibinfo {author} {\bibfnamefont {J.}~\bibnamefont
  {Adams}} \emph {et~al.} (\bibinfo {collaboration} {KTeV Collaboration}),\
  }\href {\doibase 10.1103/PhysRevLett.79.4083} {\bibfield  {journal} {\bibinfo
   {journal} {Phys. Rev. Lett.}\ }\textbf {\bibinfo {volume} {79}},\ \bibinfo
  {pages} {4083} (\bibinfo {year} {1997})},\ \Eprint
  {http://arxiv.org/abs/hep-ex/9709028} {arXiv:hep-ex/9709028 [hep-ex]}
  \BibitemShut {NoStop}%
\bibitem [{\citenamefont {Badier}\ \emph {et~al.}(1986)\citenamefont {Badier}
  \emph {et~al.}}]{unstable}%
  \BibitemOpen
  \bibfield  {author} {\bibinfo {author} {\bibfnamefont {J.}~\bibnamefont
  {Badier}} \emph {et~al.} (\bibinfo {collaboration} {NA3 Collaboration}),\
  }\href {\doibase 10.1007/BF01559588} {\bibfield  {journal} {\bibinfo
  {journal} {Z. Phys. C}\ }\textbf {\bibinfo {volume} {31}},\ \bibinfo {pages}
  {21} (\bibinfo {year} {1986})}\BibitemShut {NoStop}%
\bibitem [{\citenamefont {LoSecco}\ \emph {et~al.}(1981)\citenamefont
  {LoSecco}, \citenamefont {Sulak}, \citenamefont {Galik}, \citenamefont
  {Horstkotte}, \citenamefont {Knauer} \emph {et~al.}}]{losecco}%
  \BibitemOpen
  \bibfield  {author} {\bibinfo {author} {\bibfnamefont {J.}~\bibnamefont
  {LoSecco}}, \bibinfo {author} {\bibfnamefont {L.}~\bibnamefont {Sulak}},
  \bibinfo {author} {\bibfnamefont {R.}~\bibnamefont {Galik}}, \bibinfo
  {author} {\bibfnamefont {J.}~\bibnamefont {Horstkotte}}, \bibinfo {author}
  {\bibfnamefont {J.}~\bibnamefont {Knauer}},  \emph {et~al.},\ }\href
  {\doibase 10.1016/0370-2693(81)91064-9} {\bibfield  {journal} {\bibinfo
  {journal} {Phys. Lett. B}\ }\textbf {\bibinfo {volume} {102}},\ \bibinfo
  {pages} {209} (\bibinfo {year} {1981})}\BibitemShut {NoStop}%
\bibitem [{\citenamefont {Foot}\ \emph {et~al.}(1991)\citenamefont {Foot},
  \citenamefont {Lew},\ and\ \citenamefont {Volkas}}]{portal1}%
  \BibitemOpen
  \bibfield  {author} {\bibinfo {author} {\bibfnamefont {R.}~\bibnamefont
  {Foot}}, \bibinfo {author} {\bibfnamefont {H.}~\bibnamefont {Lew}}, \ and\
  \bibinfo {author} {\bibfnamefont {R.}~\bibnamefont {Volkas}},\ }\href
  {\doibase 10.1016/0370-2693(91)91013-L} {\bibfield  {journal} {\bibinfo
  {journal} {Phys. Lett. B}\ }\textbf {\bibinfo {volume} {272}},\ \bibinfo
  {pages} {67} (\bibinfo {year} {1991})}\BibitemShut {NoStop}%
\bibitem [{\citenamefont {Foot}\ and\ \citenamefont {He}(1991)}]{portal2}%
  \BibitemOpen
  \bibfield  {author} {\bibinfo {author} {\bibfnamefont {R.}~\bibnamefont
  {Foot}}\ and\ \bibinfo {author} {\bibfnamefont {X.-G.}\ \bibnamefont {He}},\
  }\href {\doibase 10.1016/0370-2693(91)90901-2} {\bibfield  {journal}
  {\bibinfo  {journal} {Phys. Lett. B}\ }\textbf {\bibinfo {volume} {267}},\
  \bibinfo {pages} {509} (\bibinfo {year} {1991})}\BibitemShut {NoStop}%
\bibitem [{\citenamefont {Patt}\ and\ \citenamefont {Wilczek}(2006)}]{pw}%
  \BibitemOpen
  \bibfield  {author} {\bibinfo {author} {\bibfnamefont {B.}~\bibnamefont
  {Patt}}\ and\ \bibinfo {author} {\bibfnamefont {F.}~\bibnamefont {Wilczek}},\
  }\href@noop {} {\  (\bibinfo {year} {2006})},\ \Eprint
  {http://arxiv.org/abs/hep-ph/0605188} {arXiv:hep-ph/0605188 [hep-ph]}
  \BibitemShut {NoStop}%
\bibitem [{\citenamefont {Schabinger}\ and\ \citenamefont
  {Wells}(2005)}]{higgsportal}%
  \BibitemOpen
  \bibfield  {author} {\bibinfo {author} {\bibfnamefont {R.}~\bibnamefont
  {Schabinger}}\ and\ \bibinfo {author} {\bibfnamefont {J.~D.}\ \bibnamefont
  {Wells}},\ }\href {\doibase 10.1103/PhysRevD.72.093007} {\bibfield  {journal}
  {\bibinfo  {journal} {Phys. Rev. D}\ }\textbf {\bibinfo {volume} {72}},\
  \bibinfo {pages} {093007} (\bibinfo {year} {2005})},\ \Eprint
  {http://arxiv.org/abs/hep-ph/0509209} {arXiv:hep-ph/0509209 [hep-ph]}
  \BibitemShut {NoStop}%
\bibitem [{\citenamefont {Cerdeno}\ \emph {et~al.}(2006)\citenamefont
  {Cerdeno}, \citenamefont {Dedes},\ and\ \citenamefont
  {Underwood}}]{higgsportal2}%
  \BibitemOpen
  \bibfield  {author} {\bibinfo {author} {\bibfnamefont {D.~G.}\ \bibnamefont
  {Cerdeno}}, \bibinfo {author} {\bibfnamefont {A.}~\bibnamefont {Dedes}}, \
  and\ \bibinfo {author} {\bibfnamefont {T.~E.~J.}\ \bibnamefont {Underwood}},\
  }\href {\doibase 10.1088/1126-6708/2006/09/067} {\bibfield  {journal}
  {\bibinfo  {journal} {JHEP}\ }\textbf {\bibinfo {volume} {0609}},\ \bibinfo
  {pages} {067} (\bibinfo {year} {2006})},\ \Eprint
  {http://arxiv.org/abs/hep-ph/0607157} {arXiv:hep-ph/0607157 [hep-ph]}
  \BibitemShut {NoStop}%
\bibitem [{\citenamefont {Espinosa}\ and\ \citenamefont
  {Quiros}(2007)}]{higgsportal3}%
  \BibitemOpen
  \bibfield  {author} {\bibinfo {author} {\bibfnamefont {J.~R.}\ \bibnamefont
  {Espinosa}}\ and\ \bibinfo {author} {\bibfnamefont {M.}~\bibnamefont
  {Quiros}},\ }\href {\doibase 10.1103/PhysRevD.76.076004} {\bibfield
  {journal} {\bibinfo  {journal} {Phys. Rev. D}\ }\textbf {\bibinfo {volume}
  {76}},\ \bibinfo {pages} {076004} (\bibinfo {year} {2007})},\ \Eprint
  {http://arxiv.org/abs/hep-ph/0701145} {arXiv:hep-ph/0701145 [hep-ph]}
  \BibitemShut {NoStop}%
\bibitem [{\citenamefont {March-Russell}\ \emph {et~al.}(2008)\citenamefont
  {March-Russell}, \citenamefont {West}, \citenamefont {Cumberbatch},\ and\
  \citenamefont {Hooper}}]{higgsportal4}%
  \BibitemOpen
  \bibfield  {author} {\bibinfo {author} {\bibfnamefont {J.}~\bibnamefont
  {March-Russell}}, \bibinfo {author} {\bibfnamefont {S.~M.}\ \bibnamefont
  {West}}, \bibinfo {author} {\bibfnamefont {D.}~\bibnamefont {Cumberbatch}}, \
  and\ \bibinfo {author} {\bibfnamefont {D.}~\bibnamefont {Hooper}},\ }\href
  {\doibase 10.1088/1126-6708/2008/07/058} {\bibfield  {journal} {\bibinfo
  {journal} {JHEP}\ }\textbf {\bibinfo {volume} {0807}},\ \bibinfo {pages}
  {058} (\bibinfo {year} {2008})},\ \Eprint {http://arxiv.org/abs/0801.3440}
  {arXiv:0801.3440 [hep-ph]} \BibitemShut {NoStop}%
\bibitem [{\citenamefont {Ahlers}\ \emph {et~al.}(2008)\citenamefont {Ahlers},
  \citenamefont {Jaeckel}, \citenamefont {Redondo},\ and\ \citenamefont
  {Ringwald}}]{DMportal}%
  \BibitemOpen
  \bibfield  {author} {\bibinfo {author} {\bibfnamefont {M.}~\bibnamefont
  {Ahlers}}, \bibinfo {author} {\bibfnamefont {J.}~\bibnamefont {Jaeckel}},
  \bibinfo {author} {\bibfnamefont {J.}~\bibnamefont {Redondo}}, \ and\
  \bibinfo {author} {\bibfnamefont {A.}~\bibnamefont {Ringwald}},\ }\href
  {\doibase 10.1103/PhysRevD.78.075005} {\bibfield  {journal} {\bibinfo
  {journal} {Phys. Rev. D}\ }\textbf {\bibinfo {volume} {78}},\ \bibinfo
  {pages} {075005} (\bibinfo {year} {2008})},\ \Eprint
  {http://arxiv.org/abs/0807.4143} {arXiv:0807.4143 [hep-ph]} \BibitemShut
  {NoStop}%
\bibitem [{\citenamefont {Feng}\ \emph {et~al.}(2008)\citenamefont {Feng},
  \citenamefont {Tu},\ and\ \citenamefont {Yu}}]{DMportal2}%
  \BibitemOpen
  \bibfield  {author} {\bibinfo {author} {\bibfnamefont {J.~L.}\ \bibnamefont
  {Feng}}, \bibinfo {author} {\bibfnamefont {H.}~\bibnamefont {Tu}}, \ and\
  \bibinfo {author} {\bibfnamefont {H.-B.}\ \bibnamefont {Yu}},\ }\href
  {\doibase 10.1088/1475-7516/2008/10/043} {\bibfield  {journal} {\bibinfo
  {journal} {JCAP}\ }\textbf {\bibinfo {volume} {0810}},\ \bibinfo {pages}
  {043} (\bibinfo {year} {2008})},\ \Eprint {http://arxiv.org/abs/0808.2318}
  {arXiv:0808.2318 [hep-ph]} \BibitemShut {NoStop}%
\bibitem [{\citenamefont {Kohri}\ \emph {et~al.}(2010)\citenamefont {Kohri},
  \citenamefont {McDonald},\ and\ \citenamefont {Sahu}}]{DMportal3}%
  \BibitemOpen
  \bibfield  {author} {\bibinfo {author} {\bibfnamefont {K.}~\bibnamefont
  {Kohri}}, \bibinfo {author} {\bibfnamefont {J.}~\bibnamefont {McDonald}}, \
  and\ \bibinfo {author} {\bibfnamefont {N.}~\bibnamefont {Sahu}},\ }\href
  {\doibase 10.1103/PhysRevD.81.023530} {\bibfield  {journal} {\bibinfo
  {journal} {Phys. Rev. D}\ }\textbf {\bibinfo {volume} {81}},\ \bibinfo
  {pages} {023530} (\bibinfo {year} {2010})},\ \Eprint
  {http://arxiv.org/abs/0905.1312} {arXiv:0905.1312 [hep-ph]} \BibitemShut
  {NoStop}%
\bibitem [{\citenamefont {Feng}\ \emph
  {et~al.}(2009{\natexlab{a}})\citenamefont {Feng}, \citenamefont {Kaplinghat},
  \citenamefont {Tu},\ and\ \citenamefont {Yu}}]{DMportal4}%
  \BibitemOpen
  \bibfield  {author} {\bibinfo {author} {\bibfnamefont {J.~L.}\ \bibnamefont
  {Feng}}, \bibinfo {author} {\bibfnamefont {M.}~\bibnamefont {Kaplinghat}},
  \bibinfo {author} {\bibfnamefont {H.}~\bibnamefont {Tu}}, \ and\ \bibinfo
  {author} {\bibfnamefont {H.-B.}\ \bibnamefont {Yu}},\ }\href {\doibase
  10.1088/1475-7516/2009/07/004} {\bibfield  {journal} {\bibinfo  {journal}
  {JCAP}\ }\textbf {\bibinfo {volume} {0907}},\ \bibinfo {pages} {004}
  (\bibinfo {year} {2009}{\natexlab{a}})},\ \Eprint
  {http://arxiv.org/abs/0905.3039} {arXiv:0905.3039 [hep-ph]} \BibitemShut
  {NoStop}%
\bibitem [{\citenamefont {Nadolsky}\ \emph {et~al.}(2008)\citenamefont
  {Nadolsky}, \citenamefont {Lai}, \citenamefont {Cao}, \citenamefont {Huston},
  \citenamefont {Pumplin} \emph {et~al.}}]{cteq}%
  \BibitemOpen
  \bibfield  {author} {\bibinfo {author} {\bibfnamefont {P.~M.}\ \bibnamefont
  {Nadolsky}}, \bibinfo {author} {\bibfnamefont {H.-L.}\ \bibnamefont {Lai}},
  \bibinfo {author} {\bibfnamefont {Q.-H.}\ \bibnamefont {Cao}}, \bibinfo
  {author} {\bibfnamefont {J.}~\bibnamefont {Huston}}, \bibinfo {author}
  {\bibfnamefont {J.}~\bibnamefont {Pumplin}},  \emph {et~al.},\ }\href
  {\doibase 10.1103/PhysRevD.78.013004} {\bibfield  {journal} {\bibinfo
  {journal} {Phys. Rev. D}\ }\textbf {\bibinfo {volume} {78}},\ \bibinfo
  {pages} {013004} (\bibinfo {year} {2008})},\ \Eprint
  {http://arxiv.org/abs/0802.0007} {arXiv:0802.0007 [hep-ph]} \BibitemShut
  {NoStop}%
\bibitem [{\citenamefont {Georgi}\ \emph {et~al.}(1978)\citenamefont {Georgi},
  \citenamefont {Glashow}, \citenamefont {Machacek},\ and\ \citenamefont
  {Nanopoulos}}]{ggmn}%
  \BibitemOpen
  \bibfield  {author} {\bibinfo {author} {\bibfnamefont {H.~M.}\ \bibnamefont
  {Georgi}}, \bibinfo {author} {\bibfnamefont {S.~L.}\ \bibnamefont {Glashow}},
  \bibinfo {author} {\bibfnamefont {M.~E.}\ \bibnamefont {Machacek}}, \ and\
  \bibinfo {author} {\bibfnamefont {D.~V.}\ \bibnamefont {Nanopoulos}},\ }\href
  {\doibase 10.1103/PhysRevLett.40.692} {\bibfield  {journal} {\bibinfo
  {journal} {Phys. Rev. Lett.}\ }\textbf {\bibinfo {volume} {40}},\ \bibinfo
  {pages} {692} (\bibinfo {year} {1978})}\BibitemShut {NoStop}%
\bibitem [{\citenamefont {Bonesini}\ \emph {et~al.}(2001)\citenamefont
  {Bonesini}, \citenamefont {Marchionni}, \citenamefont {Pietropaolo},\ and\
  \citenamefont {Tabarelli~de Fatis}}]{bmpt}%
  \BibitemOpen
  \bibfield  {author} {\bibinfo {author} {\bibfnamefont {M.}~\bibnamefont
  {Bonesini}}, \bibinfo {author} {\bibfnamefont {A.}~\bibnamefont
  {Marchionni}}, \bibinfo {author} {\bibfnamefont {F.}~\bibnamefont
  {Pietropaolo}}, \ and\ \bibinfo {author} {\bibfnamefont {T.}~\bibnamefont
  {Tabarelli~de Fatis}},\ }\href {\doibase 10.1007/s100520100656} {\bibfield
  {journal} {\bibinfo  {journal} {Eur. Phys. J. C}\ }\textbf {\bibinfo {volume}
  {20}},\ \bibinfo {pages} {13} (\bibinfo {year} {2001})},\ \Eprint
  {http://arxiv.org/abs/hep-ph/0101163} {arXiv:hep-ph/0101163 [hep-ph]}
  \BibitemShut {NoStop}%
\bibitem [{\citenamefont {Aguilar-Arevalo}\ \emph {et~al.}(2010)\citenamefont
  {Aguilar-Arevalo} \emph {et~al.}}]{mininucl}%
  \BibitemOpen
  \bibfield  {author} {\bibinfo {author} {\bibfnamefont {A.}~\bibnamefont
  {Aguilar-Arevalo}} \emph {et~al.} (\bibinfo {collaboration} {MiniBooNE
  Collaboration}),\ }\href {\doibase 10.1103/PhysRevD.82.092005} {\bibfield
  {journal} {\bibinfo  {journal} {Phys. Rev. D}\ }\textbf {\bibinfo {volume}
  {82}},\ \bibinfo {pages} {092005} (\bibinfo {year} {2010})},\ \Eprint
  {http://arxiv.org/abs/1007.4730} {arXiv:1007.4730 [hep-ex]} \BibitemShut
  {NoStop}%
\bibitem [{\citenamefont {Abgrall}\ \emph {et~al.}(2011)\citenamefont {Abgrall}
  \emph {et~al.}}]{na61_2011}%
  \BibitemOpen
  \bibfield  {author} {\bibinfo {author} {\bibfnamefont {N.}~\bibnamefont
  {Abgrall}} \emph {et~al.} (\bibinfo {collaboration} {NA61/SHINE
  Collaboration}),\ }\href {\doibase 10.1103/PhysRevC.84.034604} {\bibfield
  {journal} {\bibinfo  {journal} {Phys. Rev. C}\ }\textbf {\bibinfo {volume}
  {84}},\ \bibinfo {pages} {034604} (\bibinfo {year} {2011})},\ \Eprint
  {http://arxiv.org/abs/1102.0983} {arXiv:1102.0983 [hep-ex]} \BibitemShut
  {NoStop}%
\bibitem [{\citenamefont {Teis}\ \emph {et~al.}(1997)\citenamefont {Teis},
  \citenamefont {Cassing}, \citenamefont {Effenberger}, \citenamefont
  {Hombach}, \citenamefont {Mosel} \emph {et~al.}}]{etacross}%
  \BibitemOpen
  \bibfield  {author} {\bibinfo {author} {\bibfnamefont {S.}~\bibnamefont
  {Teis}}, \bibinfo {author} {\bibfnamefont {W.}~\bibnamefont {Cassing}},
  \bibinfo {author} {\bibfnamefont {M.}~\bibnamefont {Effenberger}}, \bibinfo
  {author} {\bibfnamefont {A.}~\bibnamefont {Hombach}}, \bibinfo {author}
  {\bibfnamefont {U.}~\bibnamefont {Mosel}},  \emph {et~al.},\ }\href {\doibase
  10.1007/s002180050198} {\bibfield  {journal} {\bibinfo  {journal} {Z. Phys.
  A}\ }\textbf {\bibinfo {volume} {356}},\ \bibinfo {pages} {421} (\bibinfo
  {year} {1997})},\ \Eprint {http://arxiv.org/abs/nucl-th/9609009}
  {arXiv:nucl-th/9609009 [nucl-th]} \BibitemShut {NoStop}%
\bibitem [{\citenamefont {Flaminio}\ \emph {et~al.}(1984)\citenamefont
  {Flaminio}, \citenamefont {Moorhead}, \citenamefont {Morrison},\ and\
  \citenamefont {Rivoire}}]{hera84}%
  \BibitemOpen
  \bibfield  {author} {\bibinfo {author} {\bibfnamefont {V.}~\bibnamefont
  {Flaminio}}, \bibinfo {author} {\bibfnamefont {W.}~\bibnamefont {Moorhead}},
  \bibinfo {author} {\bibfnamefont {D.~R.}\ \bibnamefont {Morrison}}, \ and\
  \bibinfo {author} {\bibfnamefont {N.}~\bibnamefont {Rivoire}},\ }\href@noop
  {} {\emph {\bibinfo {title} {Compilation of cross-sections 3. p and anti-p
  induced reactions}}} (\bibinfo {year} {1984}),\ \bibinfo {note}
  {{CERN-HERA-84-01}}\BibitemShut {NoStop}%
\bibitem [{\citenamefont {Padmanabhan}\ and\ \citenamefont
  {Finkbeiner}(2005)}]{cmb}%
  \BibitemOpen
  \bibfield  {author} {\bibinfo {author} {\bibfnamefont {N.}~\bibnamefont
  {Padmanabhan}}\ and\ \bibinfo {author} {\bibfnamefont {D.~P.}\ \bibnamefont
  {Finkbeiner}},\ }\href {\doibase 10.1103/PhysRevD.72.023508} {\bibfield
  {journal} {\bibinfo  {journal} {Phys. Rev. D}\ }\textbf {\bibinfo {volume}
  {72}},\ \bibinfo {pages} {023508} (\bibinfo {year} {2005})},\ \Eprint
  {http://arxiv.org/abs/astro-ph/0503486} {arXiv:astro-ph/0503486 [astro-ph]}
  \BibitemShut {NoStop}%
\bibitem [{\citenamefont {Slatyer}\ \emph {et~al.}(2009)\citenamefont
  {Slatyer}, \citenamefont {Padmanabhan},\ and\ \citenamefont
  {Finkbeiner}}]{cmb2}%
  \BibitemOpen
  \bibfield  {author} {\bibinfo {author} {\bibfnamefont {T.~R.}\ \bibnamefont
  {Slatyer}}, \bibinfo {author} {\bibfnamefont {N.}~\bibnamefont
  {Padmanabhan}}, \ and\ \bibinfo {author} {\bibfnamefont {D.~P.}\ \bibnamefont
  {Finkbeiner}},\ }\href {\doibase 10.1103/PhysRevD.80.043526} {\bibfield
  {journal} {\bibinfo  {journal} {Phys. Rev. D}\ }\textbf {\bibinfo {volume}
  {80}},\ \bibinfo {pages} {043526} (\bibinfo {year} {2009})},\ \bibinfo {note}
  {16 pages, 6 figures},\ \Eprint {http://arxiv.org/abs/0906.1197}
  {arXiv:0906.1197 [astro-ph.CO]} \BibitemShut {NoStop}%
\bibitem [{\citenamefont {Galli}\ \emph {et~al.}(2011)\citenamefont {Galli},
  \citenamefont {Iocco}, \citenamefont {Bertone},\ and\ \citenamefont
  {Melchiorri}}]{cmb3}%
  \BibitemOpen
  \bibfield  {author} {\bibinfo {author} {\bibfnamefont {S.}~\bibnamefont
  {Galli}}, \bibinfo {author} {\bibfnamefont {F.}~\bibnamefont {Iocco}},
  \bibinfo {author} {\bibfnamefont {G.}~\bibnamefont {Bertone}}, \ and\
  \bibinfo {author} {\bibfnamefont {A.}~\bibnamefont {Melchiorri}},\ }\href
  {\doibase 10.1103/PhysRevD.84.027302} {\bibfield  {journal} {\bibinfo
  {journal} {Phys. Rev. D}\ }\textbf {\bibinfo {volume} {84}},\ \bibinfo
  {pages} {027302} (\bibinfo {year} {2011})},\ \Eprint
  {http://arxiv.org/abs/1106.1528} {arXiv:1106.1528 [astro-ph.CO]} \BibitemShut
  {NoStop}%
\bibitem [{\citenamefont {Finkbeiner}\ \emph {et~al.}(2012)\citenamefont
  {Finkbeiner}, \citenamefont {Galli}, \citenamefont {Lin},\ and\ \citenamefont
  {Slatyer}}]{cmb4}%
  \BibitemOpen
  \bibfield  {author} {\bibinfo {author} {\bibfnamefont {D.~P.}\ \bibnamefont
  {Finkbeiner}}, \bibinfo {author} {\bibfnamefont {S.}~\bibnamefont {Galli}},
  \bibinfo {author} {\bibfnamefont {T.}~\bibnamefont {Lin}}, \ and\ \bibinfo
  {author} {\bibfnamefont {T.~R.}\ \bibnamefont {Slatyer}},\ }\href {\doibase
  10.1103/PhysRevD.85.043522} {\bibfield  {journal} {\bibinfo  {journal} {Phys.
  Rev. D}\ }\textbf {\bibinfo {volume} {85}},\ \bibinfo {pages} {043522}
  (\bibinfo {year} {2012})},\ \bibinfo {note} {30 pages, 24 figures},\ \Eprint
  {http://arxiv.org/abs/1109.6322} {arXiv:1109.6322 [astro-ph.CO]} \BibitemShut
  {NoStop}%
\bibitem [{\citenamefont {Lin}\ \emph {et~al.}(2012)\citenamefont {Lin},
  \citenamefont {Yu},\ and\ \citenamefont {Zurek}}]{lyz}%
  \BibitemOpen
  \bibfield  {author} {\bibinfo {author} {\bibfnamefont {T.}~\bibnamefont
  {Lin}}, \bibinfo {author} {\bibfnamefont {H.-B.}\ \bibnamefont {Yu}}, \ and\
  \bibinfo {author} {\bibfnamefont {K.~M.}\ \bibnamefont {Zurek}},\ }\href
  {\doibase 10.1103/PhysRevD.85.063503} {\bibfield  {journal} {\bibinfo
  {journal} {Phys. Rev. D}\ }\textbf {\bibinfo {volume} {85}},\ \bibinfo
  {pages} {063503} (\bibinfo {year} {2012})},\ \bibinfo {note} {21 pages, 6
  figures},\ \Eprint {http://arxiv.org/abs/1111.0293} {arXiv:1111.0293
  [hep-ph]} \BibitemShut {NoStop}%
\bibitem [{\citenamefont {Lees}\ \emph {et~al.}(2012)\citenamefont {Lees} \emph
  {et~al.}}]{babarh'}%
  \BibitemOpen
  \bibfield  {author} {\bibinfo {author} {\bibfnamefont {J.}~\bibnamefont
  {Lees}} \emph {et~al.} (\bibinfo {collaboration} {The BABAR Collaboration}),\
  }\href@noop {} {\  (\bibinfo {year} {2012})},\ \Eprint
  {http://arxiv.org/abs/1202.1313} {arXiv:1202.1313 [hep-ex]} \BibitemShut
  {NoStop}%
\bibitem [{\citenamefont {Aubert}\ \emph {et~al.}(2009)\citenamefont {Aubert}
  \emph {et~al.}}]{BaBar}%
  \BibitemOpen
  \bibfield  {author} {\bibinfo {author} {\bibfnamefont {B.}~\bibnamefont
  {Aubert}} \emph {et~al.} (\bibinfo {collaboration} {BABAR Collaboration}),\
  }\href@noop {} {\  (\bibinfo {year} {2009})},\ \Eprint
  {http://arxiv.org/abs/0902.2176} {arXiv:0902.2176 [hep-ex]} \BibitemShut
  {NoStop}%
\bibitem [{\citenamefont {Archilli}\ \emph {et~al.}(2011)\citenamefont
  {Archilli} \emph {et~al.}}]{Kloe}%
  \BibitemOpen
  \bibfield  {author} {\bibinfo {author} {\bibfnamefont {F.}~\bibnamefont
  {Archilli}} \emph {et~al.},\ }\href@noop {} {\bibfield  {journal} {\bibinfo
  {journal} {J. Phys. Conf. Ser.}\ }\textbf {\bibinfo {volume} {335}},\
  \bibinfo {pages} {012067} (\bibinfo {year} {2011})},\ \Eprint
  {http://arxiv.org/abs/1107.2531} {arXiv:1107.2531 [hep-ex]} \BibitemShut
  {NoStop}%
\bibitem [{\citenamefont {Bird}\ \emph {et~al.}(2006)\citenamefont {Bird},
  \citenamefont {Kowalewski},\ and\ \citenamefont {Pospelov}}]{bkp}%
  \BibitemOpen
  \bibfield  {author} {\bibinfo {author} {\bibfnamefont {C.}~\bibnamefont
  {Bird}}, \bibinfo {author} {\bibfnamefont {R.~V.}\ \bibnamefont
  {Kowalewski}}, \ and\ \bibinfo {author} {\bibfnamefont {M.}~\bibnamefont
  {Pospelov}},\ }\href {\doibase 10.1142/S0217732306019852} {\bibfield
  {journal} {\bibinfo  {journal} {Mod. Phys. Lett. A}\ }\textbf {\bibinfo
  {volume} {21}},\ \bibinfo {pages} {457} (\bibinfo {year} {2006})},\ \Eprint
  {http://arxiv.org/abs/hep-ph/0601090} {arXiv:hep-ph/0601090 [hep-ph]}
  \BibitemShut {NoStop}%
\bibitem [{\citenamefont {McElrath}(2005)}]{McElrath}%
  \BibitemOpen
  \bibfield  {author} {\bibinfo {author} {\bibfnamefont {B.}~\bibnamefont
  {McElrath}},\ }\href {\doibase 10.1103/PhysRevD.72.103508} {\bibfield
  {journal} {\bibinfo  {journal} {Phys. Rev. D}\ }\textbf {\bibinfo {volume}
  {72}},\ \bibinfo {pages} {103508} (\bibinfo {year} {2005})},\ \Eprint
  {http://arxiv.org/abs/hep-ph/0506151} {arXiv:hep-ph/0506151 [hep-ph]}
  \BibitemShut {NoStop}%
\bibitem [{\citenamefont {Fayet}(2010)}]{Fayet4}%
  \BibitemOpen
  \bibfield  {author} {\bibinfo {author} {\bibfnamefont {P.}~\bibnamefont
  {Fayet}},\ }\href {\doibase 10.1103/PhysRevD.81.054025} {\bibfield  {journal}
  {\bibinfo  {journal} {Phys. Rev. D}\ }\textbf {\bibinfo {volume} {81}},\
  \bibinfo {pages} {054025} (\bibinfo {year} {2010})},\ \Eprint
  {http://arxiv.org/abs/0910.2587} {arXiv:0910.2587 [hep-ph]} \BibitemShut
  {NoStop}%
\bibitem [{\citenamefont {Insler}\ \emph {et~al.}(2010)\citenamefont {Insler}
  \emph {et~al.}}]{CLEO}%
  \BibitemOpen
  \bibfield  {author} {\bibinfo {author} {\bibfnamefont {J.}~\bibnamefont
  {Insler}} \emph {et~al.} (\bibinfo {collaboration} {CLEO Collaboration}),\
  }\href {\doibase 10.1103/PhysRevD.81.091101} {\bibfield  {journal} {\bibinfo
  {journal} {Phys. Rev. D}\ }\textbf {\bibinfo {volume} {81}},\ \bibinfo
  {pages} {091101} (\bibinfo {year} {2010})},\ \Eprint
  {http://arxiv.org/abs/1003.0417} {arXiv:1003.0417 [hep-ex]} \BibitemShut
  {NoStop}%
\bibitem [{\citenamefont {Nakamura}\ \emph {et~al.}(2010)\citenamefont
  {Nakamura} \emph {et~al.}}]{pdg}%
  \BibitemOpen
  \bibfield  {author} {\bibinfo {author} {\bibfnamefont {K.}~\bibnamefont
  {Nakamura}} \emph {et~al.} (\bibinfo {collaboration} {Particle Data Group}),\
  }\href {http://pdg.lbl.gov} {\bibfield  {journal} {\bibinfo  {journal}
  {{Journal of Physics G}}\ }\textbf {\bibinfo {volume} {37}},\ \bibinfo
  {pages} {075021} (\bibinfo {year} {2010})}\BibitemShut {NoStop}%
\bibitem [{\citenamefont {Dave}\ \emph {et~al.}(2001)\citenamefont {Dave},
  \citenamefont {Spergel}, \citenamefont {Steinhardt},\ and\ \citenamefont
  {Wandelt}}]{dssw}%
  \BibitemOpen
  \bibfield  {author} {\bibinfo {author} {\bibfnamefont {R.}~\bibnamefont
  {Dave}}, \bibinfo {author} {\bibfnamefont {D.~N.}\ \bibnamefont {Spergel}},
  \bibinfo {author} {\bibfnamefont {P.~J.}\ \bibnamefont {Steinhardt}}, \ and\
  \bibinfo {author} {\bibfnamefont {B.~D.}\ \bibnamefont {Wandelt}},\ }\href
  {\doibase 10.1086/318417} {\bibfield  {journal} {\bibinfo  {journal}
  {Astrophys.J.}\ }\textbf {\bibinfo {volume} {547}},\ \bibinfo {pages} {574}
  (\bibinfo {year} {2001})},\ \Eprint {http://arxiv.org/abs/astro-ph/0006218}
  {arXiv:astro-ph/0006218 [astro-ph]} \BibitemShut {NoStop}%
\bibitem [{\citenamefont {Feng}\ \emph
  {et~al.}(2009{\natexlab{b}})\citenamefont {Feng}, \citenamefont {Kaplinghat},
  \citenamefont {Tu},\ and\ \citenamefont {Yu}}]{fkty}%
  \BibitemOpen
  \bibfield  {author} {\bibinfo {author} {\bibfnamefont {J.~L.}\ \bibnamefont
  {Feng}}, \bibinfo {author} {\bibfnamefont {M.}~\bibnamefont {Kaplinghat}},
  \bibinfo {author} {\bibfnamefont {H.}~\bibnamefont {Tu}}, \ and\ \bibinfo
  {author} {\bibfnamefont {H.-B.}\ \bibnamefont {Yu}},\ }\href {\doibase
  10.1088/1475-7516/2009/07/004} {\bibfield  {journal} {\bibinfo  {journal}
  {JCAP}\ }\textbf {\bibinfo {volume} {0907}},\ \bibinfo {pages} {004}
  (\bibinfo {year} {2009}{\natexlab{b}})},\ \Eprint
  {http://arxiv.org/abs/0905.3039} {arXiv:0905.3039 [hep-ph]} \BibitemShut
  {NoStop}%
\bibitem [{\citenamefont {Ahrens}\ \emph {et~al.}(1987)\citenamefont {Ahrens},
  \citenamefont {Aronson}, \citenamefont {Connolly}, \citenamefont {Gibbard},
  \citenamefont {Murtagh} \emph {et~al.}}]{Ahrens:1986xe}%
  \BibitemOpen
  \bibfield  {author} {\bibinfo {author} {\bibfnamefont {L.}~\bibnamefont
  {Ahrens}}, \bibinfo {author} {\bibfnamefont {S.}~\bibnamefont {Aronson}},
  \bibinfo {author} {\bibfnamefont {P.}~\bibnamefont {Connolly}}, \bibinfo
  {author} {\bibfnamefont {B.}~\bibnamefont {Gibbard}}, \bibinfo {author}
  {\bibfnamefont {M.}~\bibnamefont {Murtagh}},  \emph {et~al.},\ }\href
  {\doibase 10.1103/PhysRevD.35.785} {\bibfield  {journal} {\bibinfo  {journal}
  {Phys. Rev. D}\ }\textbf {\bibinfo {volume} {35}},\ \bibinfo {pages} {785}
  (\bibinfo {year} {1987})}\BibitemShut {NoStop}%
\bibitem [{\citenamefont {Aguilar-Arevalo}\ \emph {et~al.}(2009)\citenamefont
  {Aguilar-Arevalo} \emph {et~al.}}]{miniflux}%
  \BibitemOpen
  \bibfield  {author} {\bibinfo {author} {\bibfnamefont {A.}~\bibnamefont
  {Aguilar-Arevalo}} \emph {et~al.} (\bibinfo {collaboration} {MiniBooNE
  Collaboration}),\ }\href {\doibase 10.1103/PhysRevD.79.072002} {\bibfield
  {journal} {\bibinfo  {journal} {Phys. Rev. D}\ }\textbf {\bibinfo {volume}
  {79}},\ \bibinfo {pages} {072002} (\bibinfo {year} {2009})},\ \Eprint
  {http://arxiv.org/abs/0806.1449} {arXiv:0806.1449 [hep-ex]} \BibitemShut
  {NoStop}%
\bibitem [{\citenamefont {Perevalov}(2009)}]{perevalov}%
  \BibitemOpen
  \bibfield  {author} {\bibinfo {author} {\bibfnamefont {D.}~\bibnamefont
  {Perevalov}},\ }\href@noop {} {\bibfield  {journal} {\bibinfo  {journal} {PhD
  Thesis, University of Alabama}\ } (\bibinfo {year} {2009})}\BibitemShut
  {NoStop}%
\bibitem [{\citenamefont {Karagiorgi}()}]{mbPOT}%
  \BibitemOpen
  \bibfield  {author} {\bibinfo {author} {\bibfnamefont {G.}~\bibnamefont
  {Karagiorgi}} (\bibinfo {collaboration} {MiniBooNE}),\ }\href@noop {} {\emph
  {\bibinfo {title} {{Appearance results from MiniBooNE}}}},\ \bibinfo {note}
  {{WIN 2012,
  \url{http://www-boone.fnal.gov/slides-talks/conf-talk/georgiak/WIN11_miniboone.pdf}}}\BibitemShut
  {NoStop}%
\bibitem [{\citenamefont {Pittam}(2010)}]{minos1}%
  \BibitemOpen
  \bibfield  {author} {\bibinfo {author} {\bibfnamefont {R.}~\bibnamefont
  {Pittam}},\ }\href@noop {} {\  (\bibinfo {year} {2010})},\ \bibinfo {note}
  {fermilab-thesis-2010-43}\BibitemShut {NoStop}%
\bibitem [{\citenamefont {Adamson}\ \emph {et~al.}(2011)\citenamefont {Adamson}
  \emph {et~al.}}]{minos2}%
  \BibitemOpen
  \bibfield  {author} {\bibinfo {author} {\bibfnamefont {P.}~\bibnamefont
  {Adamson}} \emph {et~al.} (\bibinfo {collaboration} {MINOS Collaboration}),\
  }\href {\doibase 10.1103/PhysRevLett.107.011802} {\bibfield  {journal}
  {\bibinfo  {journal} {Phys. Rev. Lett.}\ }\textbf {\bibinfo {volume} {107}},\
  \bibinfo {pages} {011802} (\bibinfo {year} {2011})},\ \Eprint
  {http://arxiv.org/abs/1104.3922} {arXiv:1104.3922 [hep-ex]} \BibitemShut
  {NoStop}%
\bibitem [{\citenamefont {Adamson}\ \emph {et~al.}(2010)\citenamefont {Adamson}
  \emph {et~al.}}]{minos3}%
  \BibitemOpen
  \bibfield  {author} {\bibinfo {author} {\bibfnamefont {P.}~\bibnamefont
  {Adamson}} \emph {et~al.} (\bibinfo {collaboration} {The MINOS
  Collaboration}),\ }\href {\doibase 10.1103/PhysRevD.81.052004} {\bibfield
  {journal} {\bibinfo  {journal} {Phys. Rev. D}\ }\textbf {\bibinfo {volume}
  {81}},\ \bibinfo {pages} {052004} (\bibinfo {year} {2010})},\ \Eprint
  {http://arxiv.org/abs/1001.0336} {arXiv:1001.0336 [hep-ex]} \BibitemShut
  {NoStop}%
\bibitem [{\citenamefont {Ambats}\ \emph {et~al.}()\citenamefont {Ambats} \emph
  {et~al.}}]{minos4}%
  \BibitemOpen
  \bibfield  {author} {\bibinfo {author} {\bibfnamefont {I.}~\bibnamefont
  {Ambats}} \emph {et~al.} (\bibinfo {collaboration} {MINOS}),\ }\href@noop {}
  {\emph {\bibinfo {title} {{The MINOS Detectors Technical Design Report}}}},\
  \bibinfo {note} {{NUMI-L-337}}\BibitemShut {NoStop}%
\bibitem [{\citenamefont {Messier}()}]{minosPOT}%
  \BibitemOpen
  \bibfield  {author} {\bibinfo {author} {\bibfnamefont {M.}~\bibnamefont
  {Messier}} (\bibinfo {collaboration} {MINOS}),\ }\href@noop {} {\emph
  {\bibinfo {title} {{MINOS, NOvA, Fermilab perspective}}}},\ \bibinfo {note}
  {{Recontres de Moriond EW 2012,
  \url{http://nova-docdb.fnal.gov/cgi-bin/ShowDocument?docid=7197}}}\BibitemShut
  {NoStop}%
\bibitem [{\citenamefont {Abe}\ \emph {et~al.}(2011{\natexlab{a}})\citenamefont
  {Abe} \emph {et~al.}}]{t2k}%
  \BibitemOpen
  \bibfield  {author} {\bibinfo {author} {\bibfnamefont {K.}~\bibnamefont
  {Abe}} \emph {et~al.} (\bibinfo {collaboration} {T2K Collaboration}),\ }\href
  {\doibase 10.1016/j.nima.2011.06.067} {\bibfield  {journal} {\bibinfo
  {journal} {Nucl. Instrum. Meth. A}\ }\textbf {\bibinfo {volume} {659}},\
  \bibinfo {pages} {106} (\bibinfo {year} {2011}{\natexlab{a}})},\ \Eprint
  {http://arxiv.org/abs/1106.1238} {arXiv:1106.1238 [physics.ins-det]}
  \BibitemShut {NoStop}%
\bibitem [{\citenamefont {Blaszczyk}(2011)}]{nd280}%
  \BibitemOpen
  \bibfield  {author} {\bibinfo {author} {\bibfnamefont {F.~d.}\ \bibnamefont
  {Blaszczyk}},\ }\href@noop {} {\emph {\bibinfo {title} {T2K off-axis near
  detector $nu_\mu$ flux measurement and absolute momentum scale calibration of
  the off-axis near detector tracker}}} (\bibinfo {year} {2011}),\ \bibinfo
  {note}
  {\url{http://laguna.ethz.ch/indico/getFile.py/access?contribId=1&sessionId=7&resId=0&materialId=slides&confId=1}}\BibitemShut
  {NoStop}%
\bibitem [{\citenamefont {Abe}\ \emph {et~al.}(2011{\natexlab{b}})\citenamefont
  {Abe}, \citenamefont {Abgrall}, \citenamefont {Ajima}, \citenamefont
  {Aihara}, \citenamefont {Albert} \emph {et~al.}}]{ingrid}%
  \BibitemOpen
  \bibfield  {author} {\bibinfo {author} {\bibfnamefont {K.}~\bibnamefont
  {Abe}}, \bibinfo {author} {\bibfnamefont {N.}~\bibnamefont {Abgrall}},
  \bibinfo {author} {\bibfnamefont {Y.}~\bibnamefont {Ajima}}, \bibinfo
  {author} {\bibfnamefont {H.}~\bibnamefont {Aihara}}, \bibinfo {author}
  {\bibfnamefont {J.}~\bibnamefont {Albert}},  \emph {et~al.},\ }\href@noop {}
  {\  (\bibinfo {year} {2011}{\natexlab{b}})},\ \Eprint
  {http://arxiv.org/abs/1111.3119} {arXiv:1111.3119 [physics.ins-det]}
  \BibitemShut {NoStop}%
\bibitem [{\citenamefont {Le}(2009)}]{t2kpot}%
  \BibitemOpen
  \bibfield  {author} {\bibinfo {author} {\bibfnamefont {T.}~\bibnamefont {Le}}
  (\bibinfo {collaboration} {T2K Collaboration}),\ }\href@noop {} {\  (\bibinfo
  {year} {2009})},\ \Eprint {http://arxiv.org/abs/0910.4211} {arXiv:0910.4211
  [hep-ex]} \BibitemShut {NoStop}%
\bibitem [{\citenamefont {Behnke}\ \emph {et~al.}(2012)\citenamefont {Behnke},
  \citenamefont {Behnke}, \citenamefont {Brice}, \citenamefont {Broemmelsiek},
  \citenamefont {Collar} \emph {et~al.}}]{Coupp2012}%
  \BibitemOpen
  \bibfield  {author} {\bibinfo {author} {\bibfnamefont {E.}~\bibnamefont
  {Behnke}}, \bibinfo {author} {\bibfnamefont {J.}~\bibnamefont {Behnke}},
  \bibinfo {author} {\bibfnamefont {S.}~\bibnamefont {Brice}}, \bibinfo
  {author} {\bibfnamefont {D.}~\bibnamefont {Broemmelsiek}}, \bibinfo {author}
  {\bibfnamefont {J.}~\bibnamefont {Collar}},  \emph {et~al.},\ }\href@noop {}
  {\  (\bibinfo {year} {2012})},\ \Eprint {http://arxiv.org/abs/1204.3094}
  {arXiv:1204.3094 [astro-ph.CO]} \BibitemShut {NoStop}%
\bibitem [{\citenamefont {Essig}\ \emph
  {et~al.}(2011{\natexlab{b}})\citenamefont {Essig}, \citenamefont {Mardon},\
  and\ \citenamefont {Volansky}}]{semicond}%
  \BibitemOpen
  \bibfield  {author} {\bibinfo {author} {\bibfnamefont {R.}~\bibnamefont
  {Essig}}, \bibinfo {author} {\bibfnamefont {J.}~\bibnamefont {Mardon}}, \
  and\ \bibinfo {author} {\bibfnamefont {T.}~\bibnamefont {Volansky}},\
  }\href@noop {} {\  (\bibinfo {year} {2011}{\natexlab{b}})},\ \Eprint
  {http://arxiv.org/abs/1108.5383} {arXiv:1108.5383 [hep-ph]} \BibitemShut
  {NoStop}%
\bibitem [{\citenamefont {Graham}\ \emph {et~al.}(2012)\citenamefont {Graham},
  \citenamefont {Kaplan}, \citenamefont {Rajendran},\ and\ \citenamefont
  {Walters}}]{semicond2}%
  \BibitemOpen
  \bibfield  {author} {\bibinfo {author} {\bibfnamefont {P.~W.}\ \bibnamefont
  {Graham}}, \bibinfo {author} {\bibfnamefont {D.~E.}\ \bibnamefont {Kaplan}},
  \bibinfo {author} {\bibfnamefont {S.}~\bibnamefont {Rajendran}}, \ and\
  \bibinfo {author} {\bibfnamefont {M.~T.}\ \bibnamefont {Walters}},\
  }\href@noop {} {\  (\bibinfo {year} {2012})},\ \Eprint
  {http://arxiv.org/abs/1203.2531} {arXiv:1203.2531 [hep-ph]} \BibitemShut
  {NoStop}%
\bibitem [{\citenamefont {Goodman}\ \emph {et~al.}(2010)\citenamefont
  {Goodman}, \citenamefont {Ibe}, \citenamefont {Rajaraman}, \citenamefont
  {Shepherd}, \citenamefont {Tait} \emph {et~al.}}]{Goodman:2010ku}%
  \BibitemOpen
  \bibfield  {author} {\bibinfo {author} {\bibfnamefont {J.}~\bibnamefont
  {Goodman}}, \bibinfo {author} {\bibfnamefont {M.}~\bibnamefont {Ibe}},
  \bibinfo {author} {\bibfnamefont {A.}~\bibnamefont {Rajaraman}}, \bibinfo
  {author} {\bibfnamefont {W.}~\bibnamefont {Shepherd}}, \bibinfo {author}
  {\bibfnamefont {T.~M.}\ \bibnamefont {Tait}},  \emph {et~al.},\ }\href
  {\doibase 10.1103/PhysRevD.82.116010} {\bibfield  {journal} {\bibinfo
  {journal} {Phys. Rev. D}\ }\textbf {\bibinfo {volume} {82}},\ \bibinfo
  {pages} {116010} (\bibinfo {year} {2010})},\ \Eprint
  {http://arxiv.org/abs/1008.1783} {arXiv:1008.1783 [hep-ph]} \BibitemShut
  {NoStop}%
\bibitem [{\citenamefont {Fox}\ \emph {et~al.}(2011)\citenamefont {Fox},
  \citenamefont {Harnik}, \citenamefont {Kopp},\ and\ \citenamefont
  {Tsai}}]{Fox:2011fx}%
  \BibitemOpen
  \bibfield  {author} {\bibinfo {author} {\bibfnamefont {P.~J.}\ \bibnamefont
  {Fox}}, \bibinfo {author} {\bibfnamefont {R.}~\bibnamefont {Harnik}},
  \bibinfo {author} {\bibfnamefont {J.}~\bibnamefont {Kopp}}, \ and\ \bibinfo
  {author} {\bibfnamefont {Y.}~\bibnamefont {Tsai}},\ }\href {\doibase
  10.1103/PhysRevD.84.014028} {\bibfield  {journal} {\bibinfo  {journal} {Phys.
  Rev. D}\ }\textbf {\bibinfo {volume} {84}},\ \bibinfo {pages} {014028}
  (\bibinfo {year} {2011})},\ \Eprint {http://arxiv.org/abs/1103.0240}
  {arXiv:1103.0240 [hep-ph]} \BibitemShut {NoStop}%
\bibitem [{\citenamefont {Rajaraman}\ \emph {et~al.}(2011)\citenamefont
  {Rajaraman}, \citenamefont {Shepherd}, \citenamefont {Tait},\ and\
  \citenamefont {Wijangco}}]{Rajaraman:2011wf}%
  \BibitemOpen
  \bibfield  {author} {\bibinfo {author} {\bibfnamefont {A.}~\bibnamefont
  {Rajaraman}}, \bibinfo {author} {\bibfnamefont {W.}~\bibnamefont {Shepherd}},
  \bibinfo {author} {\bibfnamefont {T.~M.}\ \bibnamefont {Tait}}, \ and\
  \bibinfo {author} {\bibfnamefont {A.~M.}\ \bibnamefont {Wijangco}},\ }\href
  {\doibase 10.1103/PhysRevD.84.095013} {\bibfield  {journal} {\bibinfo
  {journal} {Phys. Rev. D}\ }\textbf {\bibinfo {volume} {84}},\ \bibinfo
  {pages} {095013} (\bibinfo {year} {2011})},\ \Eprint
  {http://arxiv.org/abs/1108.1196} {arXiv:1108.1196 [hep-ph]} \BibitemShut
  {NoStop}%
\bibitem [{\citenamefont {Fox}\ \emph {et~al.}(2012)\citenamefont {Fox},
  \citenamefont {Harnik}, \citenamefont {Kopp},\ and\ \citenamefont
  {Tsai}}]{Fox:2011pm}%
  \BibitemOpen
  \bibfield  {author} {\bibinfo {author} {\bibfnamefont {P.~J.}\ \bibnamefont
  {Fox}}, \bibinfo {author} {\bibfnamefont {R.}~\bibnamefont {Harnik}},
  \bibinfo {author} {\bibfnamefont {J.}~\bibnamefont {Kopp}}, \ and\ \bibinfo
  {author} {\bibfnamefont {Y.}~\bibnamefont {Tsai}},\ }\href {\doibase
  10.1103/PhysRevD.85.056011} {\bibfield  {journal} {\bibinfo  {journal} {Phys.
  Rev. D}\ }\textbf {\bibinfo {volume} {85}},\ \bibinfo {pages} {056011}
  (\bibinfo {year} {2012})},\ \Eprint {http://arxiv.org/abs/1109.4398}
  {arXiv:1109.4398 [hep-ph]} \BibitemShut {NoStop}%
\bibitem [{\citenamefont {Shoemaker}\ and\ \citenamefont
  {Vecchi}(2011)}]{Shoemaker:2011vi}%
  \BibitemOpen
  \bibfield  {author} {\bibinfo {author} {\bibfnamefont {I.~M.}\ \bibnamefont
  {Shoemaker}}\ and\ \bibinfo {author} {\bibfnamefont {L.}~\bibnamefont
  {Vecchi}},\ }\href@noop {} {\  (\bibinfo {year} {2011})},\ \Eprint
  {http://arxiv.org/abs/1112.5457} {arXiv:1112.5457 [hep-ph]} \BibitemShut
  {NoStop}%
\bibitem [{\citenamefont {Pospelov}(2011)}]{p1}%
  \BibitemOpen
  \bibfield  {author} {\bibinfo {author} {\bibfnamefont {M.}~\bibnamefont
  {Pospelov}},\ }\href {\doibase 10.1103/PhysRevD.84.085008} {\bibfield
  {journal} {\bibinfo  {journal} {Phys. Rev. D}\ }\textbf {\bibinfo {volume}
  {84}},\ \bibinfo {pages} {085008} (\bibinfo {year} {2011})},\ \Eprint
  {http://arxiv.org/abs/1103.3261} {arXiv:1103.3261 [hep-ph]} \BibitemShut
  {NoStop}%
\bibitem [{\citenamefont {Pospelov}\ and\ \citenamefont {Pradler}(2012)}]{pp1}%
  \BibitemOpen
  \bibfield  {author} {\bibinfo {author} {\bibfnamefont {M.}~\bibnamefont
  {Pospelov}}\ and\ \bibinfo {author} {\bibfnamefont {J.}~\bibnamefont
  {Pradler}},\ }\href@noop {} {\  (\bibinfo {year} {2012})},\ \Eprint
  {http://arxiv.org/abs/1203.0545} {arXiv:1203.0545 [hep-ph]} \BibitemShut
  {NoStop}%
\end{thebibliography}%
\end{document}